\documentclass[prd,twocolumn,showpacs,preprintnumbers,nofootinbib]{revtex4}
\usepackage[dvips]{graphicx}
\usepackage{enumerate}
\usepackage{amsmath,amssymb}
\usepackage{mathrsfs}
\usepackage[dvips]{graphicx}
\usepackage{bm}

\begin{document}
\preprint{CHIBA-EP-211-v2, 2015}

\title{
Confinement--deconfinement phase transition
\\
and gauge-invariant gluonic mass
in Yang-Mills theory
}

\author{Kei-Ichi Kondo}
\email{kondok@faculty.chiba-u.jp}

\affiliation{Department of Physics,  
Graduate School of Science, 
Chiba University, Chiba 263-8522, Japan
}
\begin{abstract}
We give an analytical derivation of the confinement/deconfinement phase transition at finite temperature in the $SU(N)$ Yang-Mills theory in the $D$-dimensional space time for $D>2$. 
We elucidate what is the mechanism for quark confinement and deconfinement at finite temperature and why the phase transition  occurs at a certain temperature.
For this purpose, we use a novel reformulation of the Yang-Mills theory which allows the gauge-invariant gluonic mass term and calculate analytically the effective potential of the Polyakov loop average concretely for the $SU(2)$ and $SU(3)$ Yang-Mills theories by including the gauge-invariant dynamical gluonic mass.  
For $D=4$, we give an  estimate on the transition temperature $T_d$ as  the ratio to the gauge-invariant gluonic mass $M$ which has been measured on the lattice at zero temperature and is  calculable also at finite temperature. 
We show that the order of the phase transition at $T_d$ is the second  order for $SU(2)$ and weakly first order for $SU(3)$ Yang-Mills theory.
These initial results are obtained easily based on the analytical calculations of the ``one-loop type'' in the first approximation.
Then these results are identified with the initial condition in solving the flow equation of the Wetterich type to improve the initial results in a systematic way  in the framework of the functional renormalization group. 
But the improvements do not change the initial results in an essential way  except for some thermodynamic observables.  
We argue how the artifacts in the first approximation are eliminated to obtain the correct behaviors for such thermodynamic observables.

\end{abstract}


\maketitle

\section{Introduction and main results}


Quark confinement and chiral-symmetry breaking  are the main subjects to be investigated for understanding the various phases in the gauge theory for strong interactions, namely QCD at finite temperature and density, see e.g., \cite{QCDtexts} for recent texts and \cite{QCDreviews} for recent reviews. 
In a previous paper \cite{Kondo10}, we have proposed a theoretical framework to obtain a low-energy effective theory of QCD towards a first-principle derivation of confinement/deconfinement and chiral-symmetry breaking/restoration crossover transitions at finite temperature.
The basic ingredients are a novel reformulation \cite{KKSS15} of Yang-Mills theory and QCD based on new variables originated from \cite{Cho80,DG79,FN99,Shabanov99,Cho80c,FN99a,KMS05,KMS06,Kondo06,KSM08}, and the flow equation of the Wetterich type \cite{Wetterich93} in the framework of the functional renormalization group (FRG) \cite{FRG} as a realization of the Wilsonian renormalization group \cite{WRG}. 
In fact, we have demonstrated that an effective theory obtained in this framework enables us to treat both transitions simultaneously on equal footing from QCD. 
The resulting effective theory in  simple but non-trivial approximations  is regarded as a modified and improved version of  nonlocal Polyakov-loop extended Nambu-Jona-Lasinio (PNJL) models proposed in \cite{HRCW08,SFR06,BBRV07}
extending the original (local) PNJL model  \cite{Fukushima04}. 
See e.g., \cite{HRCW10,HKW11,RBBV11,SSKY10} for later developments of the nonlocal PNJL models. 


A novel feature of the resulting effective theory is that the nonlocal Nambu-Jona-Lasinio (NJL) coupling depends explicitly on the temperature and Polyakov loop, which affects the entanglement between confinement and chiral symmetry breaking,  in addition to the well-known coupling in the conventional PNJL model between the Polyakov loop and quarks coming from the cross term generated through the covariant derivative in the quark sector.  The chiral symmetry breaking/restoration transition is mainly controlled by the nonlocal NJL interaction.


On the other hand, the confinement/deconfinement transition in the pure gluon sector \cite{BS83} is described by the nonperturbative effective potential for the Polyakov loop average \cite{Polyakov78} which is obtained in a nonperturbative way put forward by \cite{MP08,BGP10}
in the framework of FRG (See also \cite{BEGP10,FP13,HPS11,HSBBPSB13}). 
At present, however, the FRG studies of the Yang-Mills theory and QCD rely heavily on hard numerical works and the outcome is obtained only in the numerical way. This fact unables everyone to reproduce the FRG results and to understand the results in a physically transparent manner. Therefore, a simple analytical derivation of the results is desired to understand such nonperturbative phenomena from the first principle.

In this paper, we focus on the confinement/deconfinement phase transition in the pure gluon sector for $SU(N)$ Yang-Mills theory.
We demonstrate that the essential features on the confinement/deconfinement phase transition summarized below can be obtained in a simple analytical way without hard numerical works, once we take into account a gauge-invariant and dynamical gluonic mass $M$ which is allowed to introduce in the reformulation of the Yang-Mills theory.
In fact, we have already emphasized the importance of such a gluonic mass in the previous papers \cite{KKSS15}, but have not exhausted the outcome yet. 
The results obtained in this paper will be applied to the chiral-symmetry breaking/restoration and its crossover to confinement/deconfinement in QCD in a subsequent paper.

For this purpose, we use the reformulation of the Yang-Mills theory which allows one to introduce a gauge-invariant ``mass term'' for a specific gluonic degree of freedom called the remaining field $\mathscr{X}_\mu(x)$, see e.g., \cite{KKSS15} for a review. 
Such a gluonic mass has already played the very important role in quark confinement at zero temperature to explain and understand the ``Abelian dominance'' in the Maximally Abelian gauge gauge%
\footnote{
The Abelian dominance in the low energy regime in the Yang-Mills theory was first proposed in \cite{EI82} and was confirmed for the string tension by the numerical simulations on the lattice \cite{SY90} in the Maximally Abelian gauge \cite{KLSW87}, which is a realization of the Abelian projection \cite{tHooft81}.
The magnetic monopole dominance was also confirmed for the string tension by the numerical simulations on the lattice in the Maximally Abelian gauge \cite{SNW94}.
The Abelian dominance was also shown for the gluon field propagators 
\cite{AS99,BCGMP03}.
These results suggest the dual superconductor hypothesis as a promising mechanism for quark confinement \cite{dualsuper}.
}
which is replaced by the gauge-independent restricted field dominance \cite{KKMSSI05,IKKMSS06,SKKMSI07,Shibata-lattice2007,SKKS13} in our terminology 
\cite{Kondo06}.

First of all, notice that the mechanism of the dynamical mass generation for the gluon field has been already proposed and that the dynamical gluonic mass generation has been shown to occur at zero temperature in \cite{Kondo06,Kondo14}, see also \cite{Kondo04} for the related works. 

\begin{enumerate}
\item[0.]
The gauge-invariant mass $M$ for the remaining field $\mathscr{X}_\mu(x)$ can be generated 
dynamically through the gauge-invariant vacuum condensation of mass dimension two:%
\footnote{
The condensate $\Phi$ is a gauge-invariant version (which is made possible in our formulation) of the BRST-invariant vacuum condensation of mass dimension-two obtained from the on-shell BRST invariant operator of mass dimension two proposed in \cite{Kondo01}.
} 
\begin{equation}
\Phi:= \left\langle  \mathscr{X}_\rho^A  \mathscr{X}^{\rho A}  \right\rangle
=  \left\langle  2{\rm tr}[ \mathscr{X}_\rho  \mathscr{X}^{\rho }  ] \right\rangle ,
\label{dim-2-vc}
\end{equation} 
which occurs due to the quartic  self-interactions among the gluons represented by the remaining fields in the Yang-Mills theory. 
The dynamical gluonic mass $M$ is obtained  from the minimum of the effective potential $V_{\rm eff}(\Phi)$ of the vacuum condensate $\Phi$, which is also written as $V_{\rm eff}(M)$. 
Another way of understanding the mass term  is also given from the viewpoint of the gluonic Higgs field, which can be elucidated only in our formulation.
See section \ref{section:gluon-mass-term}.

\end{enumerate}
The above ideas enable us to calculate the gauge-invariant dynamical gluonic mass $M$ also at finite temperature in  our reformulation.
In fact, the temperature dependence of the dynamical mass $M(T)$ is obtained from the minimum of the effective potential $V_{\rm eff}(M)$ at finite temperature. 
At the same time, we want to calculate the Polyakov loop average $L$ to discuss the confinement/deconfinement transition.  Therefore, we need to calculate the simultaneous effective potential $V_{\rm eff}(\Phi, L)$ as a function of the two variables $M$ and $L$. 
It is confirmed by comparing \cite{Kondo14} and \cite{EGP11} that the gauge-invariant vacuum condensation of mass dimension two 
$ 
\Phi=\left\langle  \mathscr{X}_\rho^A  \mathscr{X}^{\rho A}  \right\rangle
$
can be related to the well-known gauge-invariant gluon condensation of mass dimension four, i.e., 
$
 \left\langle  \mathscr{F}_{\mu\nu}^A  \mathscr{F}^{\mu\nu A}  \right\rangle
$
responsible for the trace anomaly, which determines the non-perturbative vacuum, see also \cite{CSME00}. 
In this paper, however, we treat the mass $M$ just as a constant without the temperature dependence by restricting to the effective potential $V_{\rm eff}(L)$ of the Polyakov loop average $L$ alone  for simplicity. Hence, $M$ is equal to the value at zero temperature. 
The result of the effective potential $V_{\rm eff}(\Phi, L)$ will be reported in a subsequent work.

The following results are obtained based on an analytical calculation of the effective potential $V_{\rm eff}(L)$ of the Polyakov loop average $L$ alone in the $SU(2)$ and $SU(3)$ Yang-Mills theories at finite temperature $T$ in $D=4$ dimensions 
by including the gauge-invariant and dynamical ``gluonic mass'' $M$.

\begin{enumerate}

\item
There exists a confinement/deconfinement phase transition at a critical temperature $T_d$ in the respective Yang-Mills theory at finite temperature $T$ signaled by the Polyakov loop average $\langle L(\bm{x}) \rangle$, i.e., 
non-vanishing $\langle L(\bm{x}) \rangle \neq 0$ for high temperature $T>T_d$, and vanishing $\langle L(\bm{x}) \rangle = 0$ for low temperature $T<T_d$.
\footnote{
On a lattice, a rigorous proof for the existence of $SU(N)$ and $U(N)$ gauge theory in $d \ge 3$ dimensions was given  by \cite{BS83}.
} 
The $Z(N)$ center symmetry which is spontaneously broken at high temperature restores at low temperature.%
\footnote{
The center symmetry corresponds to the  aperiodic gauge transformation: $U(t+T^{-1},\bm{x})=U(t,\bm{x})g$ ($U,g \in G$) such that the gauge transformed field is periodic:
$\mathscr{A}_\mu^{U}(t+T^{-1},\bm{x}) = \mathscr{A}_\mu^{U}(t,\bm{x})$
(the periodicity of the gauge field is preserved under the gauge transformation
$\mathscr{A}_\mu^{\prime}(t+T^{-1},\bm{x}) = \mathscr{A}_\mu^{\prime}(t,\bm{x})$) \cite{Svetitsky86}. 
Under such an aperiodic gauge transformation, the action is invariant, while the Polyakov loop operator is not invariant: $L(x) \to gL(x)$.
Such a set of elements $g$ constitutes a discrete subgroup of $G$ called the center.  For $G=SU(N)$, $g=z\bm{1}$ must satisfy $gg^\dagger=\bm{1}$ and $\det g=1$, which yields $zz^*=1$ and $z^N=1$. From this observation, we find $Center(SU(N))=Z(N)$.
}

\item
The critical temperature $T_d$ is estimated in the form of the ratio to the dynamical gluonic mass $M$ in the respective Yang-Mills theory:
\begin{align}
 T_d/M =& 0.34 \ \text{for  $SU(2)$} , 
\nonumber\\
T_d/M =& 
 0.36 \ \text{for  $SU(3)$} . 
\end{align}
It should be emphasized that this ratio is gauge-independent. 
To obtain the critical temperature $T_d$, we need to know the value  $M$ of the gluonic mass.%
\footnote{
Our estimate on $T_d$ is indeed a little bit higher than expected at present. But this is based on the value of the mass $M$ obtained at zero temperature $T=0$.  The gluonic mass $M$ should depend on the temperature $T$. The mass $M$ should be determined in a self-consistent way, not just a given parameter.  
Indeed, if the mass $M$ decreases as the temperature increases: $M(T>0)<M(T=0)$, then the initial value reproduces a better result than the naive estimate.  
 Therefore, our approach has the potential to give better numerical estimate on $T_d$ without further improvements.  
The direct measurement of the gluonic mass $M$ on the lattice at finite temperature is under way \cite{KS15}. 
}
The values of the gluonic mass $M$ have been measured on the lattice at zero temperature $T=0$ by 
Shibata et al. \cite{SKKMSI07,Shibata-lattice2007,SKKS13}: 
\begin{align}
 M(T=0) =& 1.1  \ \text{GeV for  $SU(2)$} , 
\nonumber\\
 M(T=0) =& 0.8 \sim 1.0 \ \text{GeV for  $SU(3)$} . 
\end{align}
A naive use of these values of $M$ leads to the estimate on $T_d$:
\begin{align}
 T_d =& 374 \ \text{MeV for  $SU(2)$} , 
\nonumber\\
 T_d =& 288 \sim 360 \ \text{MeV for  $SU(3)$} . 
\end{align}
Incidentally, the numerical simulations on a lattice give the values \cite{LP13}:
\begin{align}
 T_d =& 295 \ \text{MeV for  $SU(2)$} , 
\nonumber\\
 T_d =& 270   \ \text{MeV for  $SU(3)$} , 
\end{align}
while the continuum approach, e.g., the most recent FRG studies give \cite{BGP10,FP13}
\begin{align}
 T_d =& 230 \ \text{MeV for  $SU(2)$} , 
\nonumber\\
 T_d =& 275  \ \text{MeV for  $SU(3)$} . 
\end{align}

\item
The order of the phase transition at $T_d$ is the second  order for $SU(2)$ and (weakly) first order for $SU(3)$ Yang-Mills theory.
This result is shown to be consistent with the standard  argument based on the Landau theory of phase transition using the expansion of the effective potential $V_{\rm eff}(L)$ into the power series of the Polyakov  loop average $L$ as the order parameter.
In particular, the first order transition in the $SU(3)$ Yang-Mills theory is induced by  the cubic term $L^3$ of the Polyakov loop average $L$ in the effective potential $V_{\rm eff}(L)$.

\item
The mechanism for quark confinement or deconfinement at finite temperature is elucidated  without detailed numerical analysis in this framework by taking into account the gluonic mass $M$. 
In high temperature $T \gg M$ the gluonic mass $M$ becomes negligible and all the relevant degrees of freedom behave as massless modes, and the effective potential can be calculated in the perturbation theory 
so that the minimum of the effective potential $V_{\rm eff}(L)$ is given at the non-vanishing Polyakov loop average $L \neq 0$ implying deconfinement \cite{Weiss81,GPY81}.
Whereas in low temperature $T \ll M$ the ``massive'' spin-one gluonic degrees of freedom (i.e., two transverse modes and one longitudinal mode) are surpressed and the remaining unphysical massless degrees of freedom (i.e., a scalar mode, and ghost--antighost modes) become dominant. Consequently,  the signature of the effective potential $V_{\rm eff}(L)$ is reversed so that the minimum of the effective potential is given at the vanishing Polyakov loop average $L=0$ implying confinement.%
\footnote{
This observation is in line with the general arguments given in  \cite{BGP10} in the FRG and agrees with the statement given in \cite{RSTW15,RSTW15b} in the approach \cite{TW11}.
}

\item
The above results are shown using the first approximation based on the  analytical calculations of the ``one-loop type'' (which is different from the one-loop calculation in perturbation theory).  This results of the first approximation  offer an effective starting point for the more systematic analysis of the non-perturbative studies.%
\footnote{
The first approximation does not depend on the gauge coupling constant explicitly, although the coupling constant dependence could be included into the result through the gluonic mass implicitly.
The explicit dependence appears if we calculate the gluonic mass from the simultaneous effective potential $V_{\rm eff}(\Phi, L)$. 
The improvement by the FRG approach depends explicitly on the coupling constant. 
}  
These initial results are regarded as the initial condition in solving the flow equation of the Wetterich type  and they can be improved in a systematic way in the FRG framework according to the prescription given in the previous paper \cite{Kondo10} where   
the crossover between confinement/deconfinement and chiral symmetry breaking/restoration has been discussed from the first principle, i.e., QCD, without explicitly introducing the gluonic mass. 
[
Remember that the first approximate solution of the Wetterich equation is given by the ``one-loop type'' expression with the additional infrared regulator term which plays the role of the mass term in a certain sense.
] 
But, the FRG improvement does not change the above conclusions in an essential manner. 
The above $T_d$ gives a lower bound on the true critical temperature $T_c$, since the flow evolves towards enhancing the confinement, under the assumption that $M$ does not change so much along the flow. 

\item
Remark: 
We must be cautious in treating the thermodynamic observables, which needs the value of the absolute minimum $V_{\rm eff}^{\rm min}= V_{\rm eff}(L_{\rm min})$ of the effective potential $V_{\rm eff}$, i.e., the vacuum energy. We do not need such information to derive the above results which are obtained only from the location $L_{\rm min}$ of $L$ giving the minimum $V_{\rm eff}^{\rm min}$:  
\begin{align}
 V^\prime_{\rm eff}(L_{\rm min}) := \frac{\partial V_{\rm eff}(L)}{\partial L}\Big|_{L=L_{\rm min}}=0 .
\end{align}
The thermodynamic  pressure $P(T)=-V_{\rm eff}^{\rm min}(T)=- V_{\rm eff}(L_{\rm min}(T))$  remains positive in the low-temperature confined phase $L=0$ in the first approximation of our formulation, in sharp contrast to the positivity violation reported in the preceding work at one loop \cite{SR12,RSTW15b}. 
For the entropy density $\mathcal{S}(T) := \frac{dP(T)}{dT}$, we find the positivity violation near the critical temperature and need the improvement of the naive first approximation. 
We discuss the theoretical and physical  reasons for these artifacts in Section \ref{section:pressure}. 


\end{enumerate}

Besides the numerical simulations on the lattice, there are other approaches, see e.g., \cite{TW11,RSTW15,RSTW15b,Fischer09,RH13,SR12,FK13}. 
Among them, especially, the authors of 
\cite{RSTW15,RSTW15b} have introduced a different kind of gluonic mass term  in the gauge-fixed Yang-Mills theory at finite temperature and have investigated the effect of the mass term on confinement/deconfinement phase transition. 
They have found that the phase transition is quite well described by  the one-loop calculations in the perturbation theory, once the gluonic mass is introduced to the Yang-Mills theory. 
Their works are very interesting in its own right, but quite surprising. 
One must answer what is the meaning of the gluonic mass and 
why the one-loop calculation is enough  
 (although they tried to improve the one-loop  result by including the two-loop result \cite{RSTW15b}).
We will give a partial answer to these questions from our point of view. 
It should be remarked that their mass term is somewhat similar to ours at first glance, but its theoretical origin and the content are totally different from ours. 

This paper is organized as follows.

In Section II, we give the reformulation of the $SU(N)$ Yang-Mills theory written in terms of the new variables. 
We show that the gluonic mass term is introduced in the gauge invariant way in the reformulated Yang-Mills theory. 
We discuss also the meaning of the gluonic mass term. 

In Section III, we discuss how the Polyakov loop operator is expressed in terms of the new variable. Then we explain our standpoint to give a prescription of calculating the effective potential of the Polyakov loop average in the reformulated  $SU(N)$ Yang-Mills theory.

In Section IV, we show the existence of confinement/deconfinement phase transition at finite temperature in $SU(2)$ Yang-Mills theory by examining the effective potential of the Polyakov loop average.  This result clarifies the mechanism of the phase transition. 
We show that the critical temperature $T_d$ is estimated as the ratio $T_d/M$ to the dynamical gluonic mass $M$, and that the phase transition is continuous, namely, the second order. 
These are the initial results due to the analytical calculation of the ``one-loop type'' obtained in the first approximation.  
Moreover, we discuss how the initial results obtained in the first approximation are improved by using the flow equation of the Wetterich type in the framework of FRG in a systematic way. 
Finally, we point out the artifacts in treating the thermodynamic quantities, e.g., the pressure and the entropy, and propose the possible resolution in our framework.
 
In Section V, we show the existence of the  confinement/deconfinement phase transition at finite temperature  in $SU(3)$ Yang-Mills theory. 
We show that the critical temperature $T_d$ is estimated as the ratio $T_d/M$ to the dynamical gluonic mass $M$, and that the phase transition is discontinuous, namely, the weakly first order. 

The final section is devoted to conclusions and discussions. 
Some technical materials are collected in the Appendices.


\section{Reformulating Yang-Mills theory using new variables}

\setcounter{equation}{0}

\subsection{Gauge-covariant decomposition of Yang-Mills field}

In this section, we give a brief introduction to the reformulation of the Yang-Mills theory needed in what follows, see e.g., a review \cite{KKSS15} for the details. 
In this paper we consider the $SU(N)$ Yang-Mills theory in the $D$-dimensional space-time.
We use the gauge-covariant decomposition of the $SU(N)$ Yang-Mills field $\mathscr{A}_\mu(x)$ into two pieces $\mathscr{V}_\mu(x)$ and $\mathscr{X}_\mu(x)$ ($\mu=0,1,...,D-1$):
\begin{equation}
 \mathscr{A}_\mu(x) = \mathscr{V}_\mu(x) + \mathscr{X}_\mu(x) \in su(N) := Lie(SU(N)),
 \label{decomp}
\end{equation}
where $\mathscr{G}:= Lie(G)$ denotes the Lie algebra of a Lie group $G$.
When the original Yang-Mills field $\mathscr{A}_\mu(x)$ obeys the ordinary gauge transformations given by
\begin{align}
  \mathscr{A}_\mu(x) & \rightarrow \mathscr{A}^\prime_\mu(x) := U(x) [\mathscr{A}_\mu(x)  + ig_{{}_{\rm YM}}^{-1}  \partial_\mu ] U(x)^{-1}  
 ,
\end{align}
 the first piece $\mathscr{V}_\mu(x)$ called the \textbf{restricted field} and the second piece $\mathscr{X}_\mu(x)$ called the \textbf{remaining field} are required to obey the gauge transformation:
\begin{align}
  \mathscr{V}_\mu(x) & \rightarrow \mathscr{V}^\prime_\mu(x) := U(x) [\mathscr{V}_\mu(x) + ig_{{}_{\rm YM}}^{-1}   \partial_\mu ] U(x)^{-1} ,
\nonumber\\   \mathscr{X}_\mu(x) & \rightarrow \mathscr{X}^  
\prime_\mu(x) :=  U(x) \mathscr{X}_\mu(x) U(x)^{-1}  
.
\end{align}
Therefore, we have the same form of the decomposition after the gauge transformation:
\begin{equation}
 \mathscr{A}^\prime_\mu(x) = \mathscr{V}^\prime_\mu(x) + \mathscr{X}^\prime_\mu(x) \in su(N) .
 \label{decomp2}
\end{equation}

Such a decomposition can be constructed by introducing a Lie algebra valued field $\bm n(x)$ called the \textbf{color direction field} or \textbf{color field} for short which is supposed to obey the gauge transformation in the adjoint representation:
\begin{align}
  \bm n(x) & \rightarrow \bm n^\prime(x) :=U(x) \bm n(x) U(x)^{-1}  \in Lie(G/\tilde H) .
\end{align}
Here $\tilde H$ is a subgroup of $G$ called the \textbf{maximal stability subgroup} \cite{KT00b,KT00,Kondo08}. 
There are a number of options depending on the choice of $\tilde H$. 
For $G=SU(N)$, the maximal stability subgroup $\tilde H$ is equal to the maximal torus subgroup $\tilde H=H:=U(1)^{N-1}$ in the \textbf{maximal option}, and $\tilde H=U(N-1)$ in the \textbf{minimal option}.

In this paper we discuss only the maximal option of the $SU(N)$ Yang-Mills theory and omit other options, see \cite{KSM08,KKSS15}.
The group $G=SU(N)$ has the rank  $r=N-1$.
In the maximal option, it is possible to construct a set of $r$ Lie algebra $\mathscr{G}$-valued fields $\bm{n}_j(x)$  ($j=1, \cdots, r$) by the repeated multiplication of the original color field $\bm{n}(x)$:
\begin{align}
 \bm{n}_j(x) =n_j^A(x) T_A  \in Lie(G/H) \quad  ( j=1, \cdots, r ) 
 ,
\end{align}
where $T_A$ ($A=1,...,{\rm dim}G=N^2-1$) are the generators of $su(N)$: $T_A=\frac{1}{2}\sigma_A$ with $\sigma_A$ being the Pauli matrices for $G=SU(2)$ and $T_A=\frac{1}{2}\lambda_A$ with $\lambda_A$ being the Gell-Mann matrices for $G=SU(3)$. 
The color fields are orthonormal:
\begin{align}
  \bm{n}_j(x) \cdot \bm{n}_k(x)  &:= 2 {\rm tr}(\bm{n}_j(x)\bm{n}_k(x))
  = n_j^A(x) n_k^A(x) 
= \delta_{jk}, \nonumber\\&
 j,k \in \{ 1, 2, \cdots, r \} 
 ,
 \label{C27-orthonormal}
\end{align}
and they mutually commute: 
\begin{align}
  [ \bm{n}_j(x), \bm{n}_k(x)] = 0, \quad j,k \in \{ 1, 2, \cdots, r \} 
 .
\label{C27-com1}
\end{align}
Therefore, all $\bm{n}_j(x)$ have the same gauge transformation as $\bm{n}(x)$:
\begin{align}
   \bm n_j(x)   \rightarrow \bm n_j^\prime(x) & :=  U(x) \bm n_j(x) U(x)^{-1}  
\nonumber\\ & \in   Lie(SU(N)/U(1)^{N-1})  \quad (j=1,...,r) .
\end{align}
For $G=SU(3)$, we introduce the two color fields denoted by $\bm{n}_3$ and $\bm{n}_8$:
\begin{align}
 \bm{n}_3(x) =n_3^A(x) T_A  , \
 \bm{n}_8(x) =n_8^A(x) T_A \in Lie(SU(3)/U(1)^2) 
 .
\end{align}
For $G=SU(3)$, $\bm{n}$ is constructed as a linear combination of $\bm{n}_3$ and $\bm{n}_8$.  A simple choice is $\bm{n}(x)=\bm{n}_3(x)$. Then $\bm{n}_8$ is constructed from $\bm{n}_3$. 
Indeed, the two color fields are related as
\begin{align}
  \bm{n}_3(x)  \bm{n}_3(x)  
=  \frac{1}{6}  \mathbf{1}  + \frac12  \frac1{\sqrt3}\bm{ n}_8(x) 
 .
\end{align}

Such color fields  $\bm{n}_j(x)$ are constructed 
using the adjoint orbit representation from the generators $H_j$ of the Cartan subalgebra $\mathscr{H}$ of $\mathscr{G}$:
\begin{align}
  \bm{n}_j(x) = U^{\dagger}(x) H_j U(x) \in Lie(G/H), \quad j \in \{ 1, 2, \cdots, r \} 
 .
\label{C27-ador}
\end{align}
In fact, the fields $\bm{n}_j(x)$ defined in this way satisfy the ortho-normality condition (\ref{C27-orthonormal}), since
\begin{align}
   \bm{n}_j(x) \cdot \bm{n}_k(x)  
=  2 {\rm tr}( H_j  H_k ) 
=  H_j \cdot H_k  = \delta_{jk}
 .
\end{align}
The commutativity (\ref{C27-com1}) is also satisfied, 
since $H_j$ are the Cartan subalgebra obeying 
\begin{align}
  [ H_j ,H_k ] = 0, \quad j,k \in \{ 1, 2, \cdots, r \} 
 .
\end{align}

Once a set of color  fields $\bm{n}_j(x)$ satisfying the above properties is given, 
  the respective pieces $\mathscr V_\mu(x)$ and $\mathscr X_\mu(x)$ of the decomposition  are uniquely determined by imposing the conditions called the \textbf{defining equation}:

(I) all color fields $\bm{n}_j(x)$ are covariantly constant in the restricted background field $\mathscr{V}_\mu(x)$:
\begin{align}
  0 =  \mathscr{D}_\mu[\mathscr{V}] \bm{n}_j(x) 
=& \partial_\mu \bm{n}_j(x) -  ig [\mathscr{V}_\mu(x), \bm{n}_j(x)]
\nonumber\\&
 (j=1,2, \cdots, r) 
 ,
\label{C27-defVL}
\end{align}

(II)  the remaining field $\mathscr{X}_\mu(x)$  is orthogonal to all $\bm{n}_j(x)$:
\begin{align}
 0 =  \mathscr{X}_\mu(x) \cdot  \bm{n}_j(x) 
 :=& 2{\rm tr}(\mathscr{X}_\mu(x) \bm{n}_j(x) )  = \mathscr{X}_\mu^A(x) n_j^A(x)  
\nonumber\\&
  (j=1,2, \cdots, r) 
\label{C27-defXL}
 . 
\end{align}

The  defining equation (I) follows from 
\\
(I') The single color field $\bm{n}(x)$ is covariantly constant in the background field $\mathscr{V}_\mu(x)$:
\begin{align}
  0 = \mathscr{D}_\mu[\mathscr{V}] \bm{n}(x) 
=\partial_\mu \bm{n}(x) -  ig [\mathscr{V}_\mu(x), \bm{n}(x)] .
\label{C27-defVL-n}
\end{align}
The defining equation (II) is also given by 
\\
(II')  $\mathscr{X}^\mu(x)$  does not have the ${H}$-commutative part $\mathscr{X}^\mu(x)_{{H}}
:= \mathscr{X}^\mu(x) -  [\bm{n}_j(x) , [\bm{n}_j(x) ,    \mathscr{X}^\mu(x) ]]
$:
\begin{align}
 & 0 =  \mathscr{X}^\mu(x)_{{H}} 
  \Longleftrightarrow \mathscr{X}^\mu(x)  =   [\bm{n}_j(x) , [\bm{n}_j(x) , \mathscr{X}^\mu(x) ]]
\label{C27-defXL2a}
 . 
\end{align}
In what follows, the summation over the index $j=1,...,r$ should be understood when it is repeated, unless otherwise stated. 
It is possible to show \cite{KKSS15} the equivalence between (\ref{C27-defXL}) and (\ref{C27-defXL2a}).

By solving the defining equations, the decomposed fields  $\mathscr V_\mu(x)$ and $\mathscr X_\mu(x)$ are  determined uniquely: 
\begin{align}
   \mathscr{X}_\mu(x) =& -ig_{{}_{\rm YM}}^{-1}    [\bm{n}_j(x), \mathscr{D}_\mu[\mathscr{A}]\bm{n}_j(x) ] 
\in Lie(G/{H}) ,
\nonumber\\
   \mathscr{V}_\mu(x) =&   \mathscr{C}_\mu(x) + \mathscr{B}_\mu(x) \in Lie(G) ,
\nonumber\\
  \mathscr{C}_\mu(x) =&     \bm{n}_j(x)  (\bm{n}_j(x) \cdot \mathscr{A}_\mu(x)) 
=   \bm{n}_j(x) \mathscr{C}_\mu^j(x)  \in Lie(H)  ,
\nonumber\\
  \mathscr{B}_\mu(x) =&  
   ig_{{}_{\rm YM}}^{-1}   [\bm{n}_j(x), \partial_\mu  \bm{n}_j(x)] \in Lie(G/{H})
 .
\label{C27-NLCV-maximal}
\end{align}
Thus, once a full set of color fields $\bm{n}_j(x)$ is given, the original gauge field  has the unique decomposition  called the \textbf{Cho-Faddeev-Niemi  decomposition} \cite{Cho80c,FN99a,KSM08}. 
In this stage, the decomposed fields are written in terms of $\bm{n}_j(x)$ and $\mathscr{A}_\mu(x)$.
For $SU(2)$, in particular, we have a single color field $\bm{n}(x)$ and the decomposition reads
\begin{align}
   \mathscr{X}_\mu(x) =& -ig_{{}_{\rm YM}}^{-1}    [\bm{n} (x), \mathscr{D}_\mu[\mathscr{A}]\bm{n} (x) ] 
\in Lie(SU(2)/U(1)) ,
\nonumber\\
   \mathscr{V}_\mu(x) =&   \mathscr{C}_\mu(x) + \mathscr{B}_\mu(x) \in Lie(SU(2)) ,
\nonumber\\
  \mathscr{C}_\mu(x) =&     \bm{n} (x)  (\bm{n} (x) \cdot \mathscr{A}_\mu(x)) 
=   \bm{n} (x) \mathscr{C}_\mu (x)  \in Lie(U(1))  ,
\nonumber\\
  \mathscr{B}_\mu(x) =&  
   ig_{{}_{\rm YM}}^{-1}   [\bm{n} (x), \partial_\mu  \bm{n} (x)] \in Lie(SU(2)/U(1))
 .
\label{NLCV-DU2}
\end{align}
This is the \textbf{Cho-Duan-Ge decomposition} \cite{Cho80,DG79,FN99,Shabanov99,KMS05,KMS06,Kondo06}.

The advantages of the decomposition are as follows.
\\ 
(a) [restricted field dominance] 
The original Wilson loop operator and the Polyakov loop operator are exactly reproduced from $\mathscr{V}_\mu$ alone \cite{Kondo08,KondoIV,Kondo08b}: 
\begin{equation}
 W_C[\mathscr{A}] = W_C[\mathscr{V}] , \quad
 L_{\bm{x}}[\mathscr{A}] = L_{\bm{x}}[\mathscr{V}] ,
 \label{W-dominant}
\end{equation}
(b) [gauge-invariant field strength] 
The gauge-invariant field strength $\mathscr{G}_{\mu\nu}^j$ is obtained from the field strength of the restricted field $\mathscr{F}_{\mu\nu}[\mathscr{V}] := \partial_\mu \mathscr{V}_\nu - \partial_\nu \mathscr{V}_\mu -ig_{{}_{\rm YM}} [ \mathscr{V}_\mu,   \mathscr{V}_\nu ]$ in the $\bm{n}_j$ direction \cite{KKSS15}:
\begin{align}
 \mathscr{G}_{\mu\nu}^j(x) ={\rm tr} \{ \bm{n}_j(x) \mathscr{F}_{\mu\nu}[\mathscr{V}](x) \} .
 \label{G-NF}
\end{align}

\subsection{Reformulation of Yang-Mills theory}

The goal of the reformulation is to change the original field variables $\mathscr{A}_\mu(x)$ into the new field variables.%
\footnote{
The reduction condition is necessary for  the reformulated gauge theory written in terms of the new field variables  based on change of field variables to be  equivalent to the original gauge theory.  It has been shown  \cite{KMS05,KSM08} that the reduction condition is regarded as the gauge-fixing condition of breaking the enlarged gauge symmetry $G \times G/H$ into the original gauge symmetry $G$. Even after imposing the reduction condition, therefore, the original gauge symmetry is retained in the reformulated gauge theory. 
The reduction condition is given by the global minimum condition of a functional defined as the integration over the whole space-time, which is minimized under the gauge transformation of the enlarged gauge symmetry \cite{KMS05,KSM08}.  Therefore, the reduction condition is free from the Gribov copy problem in principle.  In fact, we can transfer this framework to the lattice to perform the full non-perturbative studies \cite{KKMSSI05,IKKMSS06,SKKMSI07,Shibata-lattice2007,SKKS13}.
The reduction condition and the associated ``Faddeev-Popov'' determinant (which was called the Shabanov determinant \cite{Shabanov99}) given in Section II are just the local representations corresponding to the local minimum given in the framework of the local field theory.
For the details, see the original references \cite{KMS05,KMS06,Kondo06,KSM08} or the review \cite{KKSS15}.
The gauge fixing of the original gauge symmetry $G$ is discussed in the next section III. 
}
For this purpose, the color field 
$
\bm{n}(x) \in Lie(G/\tilde{H})
$
must be written in terms of the original $\mathscr{A}_\mu(x) \in Lie(G)$.
This is achieved by solving the  \textbf{reduction condition $\bm\chi(x)=0$} for a given $\mathscr{A}_\mu(x)$.
A choice of the reduction condition is  
\begin{equation}
 {\bm\chi[\mathscr{A},\bm{n}]}
 :=[ \bm{n}_j(x) ,  \mathscr{D}^\mu[\mathscr{A}]\mathscr{D}_\mu[\mathscr{A}]\bm{n}_j(x) ]
   \in Lie(G/\tilde{H})
  .
\label{eq:diff-red}
\end{equation}
Thus, all the new field variables have been written in terms of  the original variables $\mathscr{A}_\mu$.


In the original Yang-Mill theory, the average of $F$ is written (omitting the gauge fixing of $G$) as
\begin{align}
 \langle F[\mathscr{A}]  \rangle_{\rm YM}
  = Z_{{\rm YM}}^{-1} \int \mathcal{D} \mathscr{A}_\mu^A e^{iS_{\rm YM}[\mathscr{A}] }  F[\mathscr{A}] 
 .
\end{align}
In the reformulated Yang-Mills theory, the average of $F$ is rewritten  as
\begin{align}
   \langle F[\mathscr{A}] \rangle_{\rm YM^\prime}
  =& Z_{{\rm YM}^\prime}^{-1} \int \mathcal{D}n^\beta \mathcal{D}\mathscr{C}_\mu^k   \mathcal{D}\mathscr{X}_\mu^b  \tilde{J} \delta(\tilde{\bm\chi})  
   \Delta_{\rm FP}^{\rm red}
  \nonumber\\    & \times 
 e^{i\tilde S_{\rm YM}[\bm n, \mathscr{C},\mathscr{X} ]} F[\bm n^\beta, \mathscr C_\nu^k  ,  \mathscr X_\nu^b]
 .
\end{align}
(See \cite{KMS05,KMS06,Kondo06} for $SU(2)$, \cite{KSM08} for $SU(N)$, which correspond to section 4.4 for $SU(2)$, and section 5.6 for $SU(N)$   in the review \cite{KKSS15}.)
Here
(i) $\tilde{\bm\chi}=0$ is the reduction condition rewritten in terms of the new  variables:
\begin{equation}
\tilde{\bm\chi} 
 :=\tilde{\bm\chi} [\bm n, \mathscr{C},\mathscr{X}]
 :=D^\mu[\mathscr{V}]\mathscr{X}_\mu 
 . 
\end{equation}
This constraint can be incorporated into the Lagrangian  by introducing the Lagrange multiplier field, namely, the Nakanishi-Lautrup field $\mathscr{N}(x)$:
\begin{align}
    \delta(\tilde{\bm\chi}) 
  =&   \int \mathcal{D}\mathscr{N}^A  
 e^{i \int d^Dx \mathscr{L}_{\rm red} } 
 , 
 \\
  \mathscr{L}_{\rm red}   
=&  \mathscr{N}^A (\mathscr{D}_\mu[\mathscr{V}] \mathscr{X}^\mu)^A 
=  2 {\rm tr}[ \mathscr{N}   \mathscr{D}_\mu[\mathscr{V}] \mathscr{X}^\mu  ] .
\end{align}
(ii) $\Delta_{\rm FP}^{\rm red}$ is the ``Faddeev-Popov'' determinant associated with the reduction condition.
 The precise form is obtained by the BRST method as
\begin{equation}
  \Delta_{\rm FP}^{\rm red}[\bm n,\mathscr{C},\mathscr X] 
= \det \{-\mathscr{D}_\mu[\mathscr V-\mathscr X] \mathscr{D}^\mu[\mathscr V+\mathscr  X] \}  .
\end{equation}
(See \cite{KMS05}, section 4.6 and Appendix E for $SU(2)$, and section 5.8 and Appendix H for $SU(N)$ in \cite{KKSS15}.)
The determinant can be incorporated into the Lagrangian by introducing the ``ghost'' $\mathscr{\eta}(x)$ and ``antighost'' field $\mathscr{\bar \eta}(x)$ as 
\begin{align}
& \Delta_{\rm FP}^{\rm red}[\bm n, \mathscr{C},\mathscr X] 
  =   \int \mathcal{D}\mathscr{\eta}^A \mathcal{D}\mathscr{\bar \eta}^A  
 e^{i \int d^Dx \mathscr{L}_{\rm FP} } 
 , 
 \nonumber\\
 & \mathscr{L}_{\rm FP}   
=  i\mathscr{\bar \eta}^A \{- \mathscr{D}_\mu[\mathscr V-\mathscr X] \mathscr{D}^\mu[\mathscr V+\mathscr  X]\}^{AB} \mathscr{\eta}^B   
  .
\end{align}
(iii) $\tilde{J}$  is the Jacobian associated with the change of variables.  By a suitable choice of the basis in the color space, 
\begin{equation}
   \tilde{J} = 1 ,
\end{equation}
irrespective of the choice of the reduction condition.
(See \cite{Kondo06} for $SU(2)$, \cite{KSM08} for $SU(N)$, which correspond to section 4.5 for $SU(2)$, and section 5.7 for $SU(N)$  in the review \cite{KKSS15}.)

Notice that the reformulated Yang-Mills theory retains the gauge symmetry of the gauge group $G=SU(N)$ of the original Yang-Mills theory at any stage even after imposing the reduction condition and introducing the ``Faddeev-Popov'' determinant. 

 \begin{tabular}{l|cl}
    &  original  & $\rightarrow$ reformulated  \\ \hline
  field variables & $\mathscr A_\mu^A$ & $\rightarrow$ $\bm n^\beta, \mathscr C_\nu^k  ,  \mathscr X_\nu^b$ \\ 
  action & $S_{\rm YM}[\mathscr A]$ & $\rightarrow$ $\tilde S_{\rm YM}[\bm n, \mathscr{C},\mathscr{X}]$ \\
  measure & $\mathcal{D}\mathscr{A}_\mu^A$ 
& $\rightarrow$  $\mathcal{D}n^\beta \mathcal{D}\mathscr{C}_\nu^k   \mathcal{D}\mathscr{X}_\nu^b
\tilde{J}  \delta( {\tilde{\bm\chi}}) \Delta_{\rm FP}^{\rm red}$  \\
  operator & $F[\mathscr{A}]$ 
& $\rightarrow$ $F[\bm n^\beta, \mathscr C_\nu^k  ,  \mathscr X_\nu^b]$ \\
 \end{tabular}

Remarkably, this reformulation  allows us to introduce the \textbf{gauge-invariant ``mass term''} for the remaining field $\mathscr{X}_\mu$ which is a piece  obtained by the gauge-covariant decomposition  $\mathscr{A}_\mu=\mathscr{V}_\mu+\mathscr{X}_\mu$ of the original gauge field $\mathscr{A}_\mu$:
\begin{equation}
\mathscr{L}_{\rm m}
=  M^2 {\rm tr}(\mathscr{X}_\mu  \mathscr{X}^{\mu})
=  \frac{1}{2} M^2 \mathscr{X}_\mu^A \mathscr{X}^{\mu A} .
\label{mass1}
\end{equation}
In fact, the numerical simulations on the lattice exhibit the \textbf{dynamical mass generation for the remaining field} at zero temperature, see  \cite{SKKMSI07} for $SU(2)$, \cite{Shibata-lattice2007} for $SU(3)$ maximal option, and \cite{SKKS13} for $SU(3)$ minimal option:
\begin{align}
 & M(T=0) \simeq 1.1 \ \text{GeV for $SU(2)$}, 
\nonumber\\
 & M(T=0) \simeq 0.80 \sim 1.0 \ \text{GeV for $SU(3)$  maximal option} ,
\nonumber\\
 & M(T=0) \simeq 1.1  \sim 1.15 \ \text{GeV for  $SU(3)$ minimal option} .
\end{align}
%

In the reformulated Yang-Mills theory, thus, the average is obtained  as
\begin{align}
   \langle F[\mathscr{A}] \rangle_{\rm YM^\prime}
  =& Z_{{\rm YM}^\prime}^{-1} \int \mathcal{D}n^\beta \mathcal{D}\mathscr{C}_\nu^k   \mathcal{D}\mathscr{X}_\mu^b \mathcal{D}\mathscr{N}^A \mathcal{D}\mathscr{\eta}^A \mathcal{D}\mathscr{\bar \eta}^A  
   \nonumber\\    & \times 
 e^{i\tilde S_{\rm YM}^{\rm tot}[\bm n, \mathscr{C}, \mathscr{X}, \mathscr{N},\mathscr{\eta}, \mathscr{\bar \eta} ]} F[\bm n^\beta, \mathscr C_\nu^k  ,  \mathscr X_\nu^b]
 ,
\end{align}
where
\begin{align}
 \tilde S_{\rm YM}^{\rm tot}[\bm n, \mathscr{C}, \mathscr{X}, \mathscr{N}, \mathscr{\eta}, \mathscr{\bar \eta} ]
  =& \int d^Dx  \mathscr{L}_{\rm YM}^{\rm tot}[\bm n, \mathscr{C},\mathscr{X}, \mathscr{N}, \mathscr{\eta} , \mathscr{\bar \eta} ] ,
\nonumber\\  
 \mathscr{L}_{\rm YM}^{\rm tot}[\bm n, \mathscr{C},\mathscr{X}, \mathscr{N}, \mathscr{\eta} , \mathscr{\bar \eta} ]  
=&  \mathscr{L}_{\rm YM}[\bm n, \mathscr{C},\mathscr{X} ] 
\nonumber\\ &
+ \mathscr{L}_{\rm red}[\bm n, \mathscr{C}, \mathscr{X}, \mathscr{N} ]
\nonumber\\ &
+ \mathscr{L}_{\rm FP}[\bm n, \mathscr{C}, \mathscr{X}, \mathscr{N}, \mathscr{\eta} , \mathscr{\bar \eta}  ]
\nonumber\\ &
+    \mathscr{L}_{\rm m}[ \mathscr{X} ]  .
\end{align}
In the reformulated Yang-Mills theory, the Polyakov loop average is obtained  as
\begin{align}
   & \langle L_{\bm{x}}[\mathscr{A}] \rangle_{\rm YM^\prime}
\nonumber\\
  =& Z_{{\rm YM}^\prime}^{-1} \int \mathcal{D}n^\beta \mathcal{D}\mathscr{C}_\nu^k   \mathcal{D}\mathscr{X}_\mu^b \mathcal{D}\mathscr{N}^A \mathcal{D}\mathscr{\eta}^A \mathcal{D}\mathscr{\bar \eta}^A
\nonumber\\  & \times 
 e^{i\tilde S_{\rm YM}^{\rm tot}[\bm n, \mathscr{C}, \mathscr{X}, \mathscr{N},\mathscr{\eta}, \mathscr{\bar \eta} ]} L_{\bm{x}}[\mathscr{V}]
\nonumber\\
  =& Z_{{\rm YM}^\prime}^{-1} \int \mathcal{D}n^\beta \mathcal{D}\mathscr{C}_\nu^k  L_{\bm{x}}[\mathscr{V}]
\nonumber\\  & \times 
 \int \mathcal{D}\mathscr{X}_\mu^b \mathcal{D}\mathscr{N}^A \mathcal{D}\mathscr{\eta}^A \mathcal{D}\mathscr{\bar \eta}^A e^{i\tilde S_{\rm YM}^{\rm tot}[\bm n, \mathscr{C}, \mathscr{X}, \mathscr{N},\mathscr{\eta}, \mathscr{\bar \eta} ]} 
 .
 \label{formulation-L}
\end{align}

We can show that the Lagrangian density of the $SU(N)$ Yang-Mills theory
\begin{align}
   \mathscr{L}_{\rm YM} = 
   -\frac{1}{2} {\rm tr}(\mathscr{F}_{\mu\nu}[\mathscr{A}]  \mathscr{F}^{\mu\nu}[\mathscr{A}] )
 ,
\end{align}
is decomposed into the form:
\begin{align}
   \mathscr{L}_{\rm YM} 
=& -\frac{1}{4} \mathscr{F}_{\mu\nu}^A[\mathscr{V}] \mathscr{F}^{\mu\nu}{}^A[\mathscr{V}]
-     \frac{1}{2} \mathscr{X}^{\mu A}  W_{\mu\nu}^{AB} \mathscr{X}^{\nu B} 
\nonumber\\&
+ O(\mathscr{X}^3)
 ,
\nonumber\\
 W_{\mu\nu}^{AB}  :=&  - (\mathscr{D}_\rho[\mathscr{V}]\mathscr{D}^\rho[\mathscr{V}])^{AB} g_{\mu\nu} 
+ 2g_{{}_{\rm YM}}f^{ABC} \mathscr{F}_{\mu\nu}^{C}[\mathscr{V}] 
\nonumber\\&
+ \mathscr{D}_\mu^{AC}[\mathscr{V}] \mathscr{D}_\nu^{CB}[\mathscr{V}]
 ,
\label{C27-W2}
\end{align}
where $\mathscr{F}_{\mu\nu}[\mathscr{V}]$ is the non-Abelian field strength of the restricted field $\mathscr{V}_\mu$.
In the Yang-Mills Lagrangian, the terms linear in the remaining field $\mathscr{X}_\mu$ do not appear. 
In what follows, we neglect all terms cubic and quartic  in the remaining field $\mathscr{X}_\mu$. 
In particular, the cubic terms do not appear for the $SU(2)$ group and for the $SU(3)$ group in the minimal option.

\subsection{A novel viewpoint for the gluonic mass term}\label{section:gluon-mass-term}

A novel viewpoint for understanding the gluonic mass term $\mathscr{L}_{\rm m}$ is as follows.
For simplicity, we discuss only the $SU(2)$ case. 
Then the  gauge-invariant mass term (\ref{mass1}) is rewritten in terms of the original variables $\mathscr{A}_\mu$:
\begin{align}
\mathscr{L}_{\rm m}
 =&   M^2 {\rm tr} \{(\mathscr{A}_\mu - \mathscr{V}_\mu)^2  \} 
\nonumber\\
=&  M^2 {\rm tr} \{ (\mathscr{A}_\mu -c_\mu \bm{n}
  -ig_{{}_{\rm YM}}^{-1} [\partial_\mu \bm{n},   \bm{n} ] )^2 \} 
\nonumber\\
=&  g_{{}_{\rm YM}}^{-2}  M^2 {\rm tr} \{ (D_\mu[\mathscr{A}] \bm{n})^2 \} ,
  \label{mass4}
\end{align}
with the understanding that the color field $\bm{n}$ is expressed in terms of the original gauge field $\mathscr{A}_\mu$ by solving the reduction condition. 
Therefore, $\mathscr{V}_\mu$ (or $c_\mu$ and $\bm{n}$) plays the similar role to the \textbf{St\"uckelberg field} to recover the local gauge symmetry.
Note that $c_\mu$, $\bm{n}$ and $\mathscr{X}_\mu$ are treated as independent variables after the change of variables and the mass term is a polynomial in the new variable $\mathscr{X}_\mu$, although they might be   non-linear composite operators of the original variables $\mathscr{A}_\mu$.

We can identify the color field $\bm{n}(x)$ with the \textbf{gluonic Higgs field} $\phi(x)$: 
\begin{align}
\phi(x) = M \bm{n}(x) \in Lie(SU(2)/U(1)) ,
  \label{mass5}
\end{align}
the mass term is regarded as the kinetic term for the non-linear sigma model: 
\begin{align}
\mathscr{L}_{\rm m}^\prime
 =&   g_{{}_{\rm YM}}^{-2}  {\rm tr} \{ (D_\mu[\mathscr{A}] \phi)^2 \} + u \{ 2{\rm tr}(\phi^2) - M^2 \}  
\nonumber\\
=&   \frac{1}{2}g_{{}_{\rm YM}}^{-2}   (D_\mu[\mathscr{A}] \phi) \cdot (D_\mu[\mathscr{A}] \phi) + u ( \phi \cdot \phi - M^2)  ,
  \label{mass6}
\end{align}
where $u(x)$ is the Lagrange multiplier field for incorporating the constraint: 
\begin{align}
\phi(x) \cdot \phi(x) - M^2=0 .
\label{mass7}
\end{align}
Alternatively, the mass term is regarded as the limit $\lambda \to \infty$ of the model:
\begin{align}
\mathscr{L}_{\rm m}^\prime
=    \frac{1}{2}g_{{}_{\rm YM}}^{-2}   (D_\mu[\mathscr{A}] \phi) \cdot (D_\mu[\mathscr{A}] \phi) + \lambda ( \phi \cdot \phi - M^2)^2  .
  \label{mass8}
\end{align}
The dynamical generation of the gluonic mass is equivalent to the gluonic Higgs phase.
Thus, the Yang-Mills theory with the ``mass term'' for the remaining field $\mathscr{X}_\mu$ is identified with the Yang-Mills-Higgs model with the gluonic Higgs field $\phi(x)$ in the Higgs phase

The proposed mass term (\ref{mass1})   for the gluon should be compared with the conventional gauge-invariant mass term of Kunimasa--Goto type \cite{KG67}:
\begin{align}
\mathscr{L}_{\rm KG}
 =&   M^2 {\rm tr} \{(\mathscr{A}_\mu - ig_{{}_{\rm YM}}^{-1}U\partial_\mu U^\dagger)^2 \}
\nonumber\\
=&   M^2 {\rm tr} \{(UD_\mu[\mathscr{A}]U^\dagger)^2 \} ,
 \quad 
 U(x) = e^{-i\chi(x)/v} .
 \label{mass3}
\end{align}
This mass term is non-polynomial in the St\"uckelberg field $\chi(x)$.  This fact makes the field theoretical treatment very difficult.

\subsection{Mechanism of dynamical gluonic mass generation}

It is possible to  argue that there occurs a novel \textbf{vacuum condensation of mass dimension--two} for the field $\mathscr{X}^\mu$, i.e.,%
\footnote{
We adopt the Minkowski metric $g_{\mu\nu}={\rm diag}(1,-1,-1,-1)$.  After the Wick rotation to the Euclidean region, the Minkowski metric tensor $g_{\mu\nu}$ is replaced by 
$
 -\delta_{\mu\nu} ={\rm diag}(-1,-1,-1,-1)
$.  Therefore, we have
$
 -\mathscr{X}_\mu^2 \rightarrow (\mathscr{X}^E_\mu)^2 > 0
$.
}
\begin{equation}
 \left< -\mathscr{X}_\mu^A \mathscr{X}^{\mu A} \right> \ne 0 .
\end{equation}
and that the field $\mathscr{X}^\mu$ acquires the mass dynamically through this condensation.  
This is a gauge-invariant version (which is made possible in our formulation) of the BRST-invariant vacuum condensation of mass dimension two obtained from the on-shell BRST invariant operator of mass dimension two proposed in \cite{Kondo01}.
A naive way to see this is to apply the mean-field like argument or the Hartree--Fock approximation to the four-gluon interaction, i.e., the quartic self-interaction among $\mathscr{X}^\mu$ which leads to 
the gauge-invariant mass term for  $\mathscr{X}_\mu$ gluons and a gauge-invariant gluonic mass $M$:
\begin{align}
  & -\frac{1}{4}(g_{{}_{\rm YM}} \mathscr{X}_\mu \times \mathscr{X}_\nu)^2 
\nonumber\\
=&   -\frac{1}{4}(g_{{}_{\rm YM}} \mathscr{X}_\mu \times \mathscr{X}_\nu) \cdot (g_{{}_{\rm YM}} \mathscr{X}^\mu \times \mathscr{X}^\nu) 
 \nonumber\\
\to & \frac{1}{2}g_{{}_{\rm YM}}^2 \mathscr{X}^A_\mu \left[\left\langle -\mathscr{X}_\rho^C \mathscr{X}^{\rho C} \right\rangle \delta^{AB} - \left\langle -\mathscr{X}^A_\rho \mathscr{X}^{\rho B} \right\rangle \right] \mathscr{X}^{\mu B} 
 \nonumber\\
=&   \frac{1}{2} M^2 \mathscr{X}_\mu \cdot \mathscr{X}^\mu ,
\quad 
 M^2 = \frac23 g_{{}_{\rm YM}}^2 \left\langle -\mathscr{X}_\rho^C \mathscr{X}^{\rho C} \right\rangle .
\label{mass-term}
\end{align} 
The more detailed and systematic analyses have been given in \cite{Kondo06}. 
Consequently, the gauge-invariant mass for the ``off-diagonal'' gluon $\mathscr{X}_\mu$ is generated. 
Then the $\mathscr{X}_\mu$ gluon modes  decouple in the low-energy (or long-distance) region below the mass scale $M$.  
Consequently, the infrared \textbf{``Abelian'' dominance} or \textbf{restricted field dominance} for the large Wilson loop average follows immediately from the fact that the Wilson loop operator is written in terms of $\mathscr{V}_\mu$ alone according to the non-Abelian Stokes theorem for the Wilson loop operator (\ref{W-dominant}) and that the Wilson loop average is entirely estimated by the restricted field alone.

The dynamical gluonic mass has various implications to the nonperturbative aspects of the Yang-Mills theory, e.g., the glueball mass \cite{KOSSM06}, the stability of the vacuum  \cite{Kondo14} and to  the low-energy effective theory for gluon confinement \cite{Kondo11}.  Notice that there is an unsolved problem of a physical unitarity for the massive vector model \cite{Kondo12}.



\section{Effective potential of the Polyakov loop average
}

\setcounter{equation}{0}

\subsection{Polyakov loop operator}

The $SU(N)$ \textbf{Polyakov loop operator} $L_{\bm{x}}[\mathscr{A}]$ is defined by taking the trace of the \textbf{holonomy operator} $P_{\mathscr{A}}(\bm{x})$:
\begin{align}
 & L_{\bm{x}}[\mathscr{A}] :=   {\rm tr}(P_{\mathscr{A}}(\bm{x}))/{\rm tr}(\bm{1}) ,
\nonumber\\
 & P_{\mathscr{A}}(\bm{x}) = \mathscr{P} \exp \left[ ig \int_{0}^{1/T} d\tau \mathscr{A}_0 (\bm{x},\tau)    \right] \in SU(N) ,
\end{align}
where $\mathscr{P}$ is the path ordering. 
The Polyakov loop operator $L_{\bm{x}}[\mathscr{A}]$ is gauge invariant: $L_{\bm{x}}[\mathscr{A}^\prime]=L_{\bm{x}}[\mathscr{A}]$. 
In our reformulation, $\mathscr{A}_0^A$ in $L_{\bm{x}}[\mathscr{A}]$ can be replaced by the restricted field $\mathscr{V}_0^A$ exactly:
\begin{align}
 & L_{\bm{x}}[\mathscr{V}] :=   {\rm tr}(P_{\mathscr{V}}(\bm{x}))/{\rm tr}(\bm{1}) ,
\nonumber\\
 & P_{\mathscr{V}}(\bm{x}) = \mathscr{P} \exp \left[ ig \int_{0}^{1/T} d\tau \mathscr{V}_0 (\bm{x},\tau)      \right] \in SU(N)
.  
\end{align}
The Polyakov loop operator $L_{\bm{x}}[\mathscr{V}]$ is also gauge invariant: $L_{\bm{x}}[\mathscr{V}^\prime]=L_{\bm{x}}[\mathscr{V}]$, and coincides exactly with the original one $L_{\bm{x}}[\mathscr{A}]$ \cite{Kondo08,KondoIV}: 
\begin{align}
  L_{\bm{x}}[\mathscr{A}]  =  L_{\bm{x}}[\mathscr{V}] 
.  
\end{align}

\subsection{Gauge fixing}

The reformulation given in the above is gauge invariant in the sense that the original gauge symmetry for the gauge group $G=SU(N)$ in the Yang-Mills theory is retained at any stage.  Now we proceed to discuss the gauge fixing in the reformulated Yang-Mills theory.

The choice of the gauge does not change the average value of the gauge invariant quantity such as the Polyakov loop average. 
In the actual calculations in this paper, therefore, we take a special gauge: the color field has the uniform direction of the Cartan subalgebra corresponding to the maximal torus subgroup $H=U(1)^{N-1}$, which facilitates the calculation of the Polyakov loop average.
This is equivalent to perform the local gauge transformation to diagonalize the color field. 
Then the color fields $\bm{n}_j(x)$ are no longer the local field variables and the integration over $\bm{n}$ in the path integral formula derived above becomes unnecessary.
For $SU(2)$, the resulting color field is chosen to be
\begin{equation}
  \bm{n}^\prime(x) \equiv \frac{\sigma_3}{2} \Longleftrightarrow  n^\prime{}^{A}(x) = \delta^{A}_{3}   .
\label{n-approx-SU2}
\end{equation}
Then the restricted field is given by
\begin{align}
   \mathscr{V}^\prime_\mu(x) =&   \mathscr{C}^\prime{}_\mu(x) \bm{n}^\prime(x) + ig_{{}_{\rm YM}}^{-1}   [\bm{n}^\prime(x), \partial_\mu  \bm{n}^\prime(x)] 
\nonumber\\
=&  \mathscr{C}^\prime{}_\mu(x) \frac{\sigma_3}{2}   
 .
\end{align}
For $SU(3)$, the color field is taken to be a linear combination of the two diagonal  generators  $H_1$ and $H_2$ belonging to the Cartan subalgebra:
\begin{equation}
  \bm{n}^\prime_3(x) \equiv  \frac{\lambda_3}{2}, \quad
\bm{n}^\prime_8(x) \equiv  \frac{\lambda_8}{2}   \Longleftrightarrow  n^\prime{}_j^A(x) = \delta^{A}_{j} .
\label{n-approx-SU3}
\end{equation}
Then the restricted field is given by
\begin{align}
   \mathscr{V}^\prime_\mu(x) =&   \mathscr{C}^\prime{}_\mu^j(x) \bm{n}^\prime_j(x) + ig_{{}_{\rm YM}}^{-1}   [\bm{n}^\prime_j(x), \partial_\mu  \bm{n}^\prime_j(x)] 
\nonumber\\
=&  \mathscr{C}^\prime{}_\mu^3(x) \frac{\lambda_3}{2}  +  \mathscr{C}^\prime{}_\mu^8(x) \frac{\lambda_8}{2}   
 .
\end{align}

In this way, the restricted field $\mathscr{V}_\mu$ is chosen to be in the Cartan subalgebra.
In what follows, we omit the prime. 
The restricted field is separated into the background part and the quantum fluctuation part:
\begin{align}
   \mathscr{V}_\mu(x) =&  \bar{\mathscr{V}}_\mu(x) + \tilde{\mathscr{V}}_\mu(x)  .
\end{align}
which is realized by separating $\mathscr{C}_\mu^j(x)$ into the the background part and the quantum fluctuation part.
For later convenience, we choose the specific background $\bar {\mathscr{C}}_\mu(x)=g^{-1}T  \varphi \delta_{\mu 0}$ of $\mathscr{C}_\mu(x)$ for $SU(2)$, and $\bar {\mathscr{C}}_\mu^j(x)=g^{-1}T  \varphi_j \delta_{\mu 0}$ of $\mathscr{C}_\mu^j(x)$ for $SU(3)$.
\\
For $SU(2)$,
\begin{align}
& \mathscr{V}_\mu (\bm{x},\tau)= g^{-1}T \varphi \delta_{\mu 0}  \frac{\sigma_3}{2}  
+ \mathscr{\tilde V}_\mu (\bm{x},\tau) ,
 \nonumber\\
& \Longleftrightarrow 
\mathscr{V}_\mu^A(\bm{x},\tau)
=   g^{-1}T \varphi \delta_{\mu 0}  \delta_{A3} 
+ \mathscr{\tilde V}_\mu^A(\bm{x},\tau) 
 ,
 \label{sepa-SU2}
\end{align}
For $SU(3)$, 
\begin{align}
& \mathscr{V}_\mu (\bm{x},\tau)= g^{-1}T  \varphi_3 \delta_{\mu 0} \frac{\lambda_3}{2} +   g^{-1}T \varphi_8  \delta_{\mu 0} \frac{\lambda_8}{2}   
+ \mathscr{\tilde V}_\mu (\bm{x},\tau) ,
 \nonumber\\
& \Longleftrightarrow
\mathscr{V}_\mu^A(\bm{x},\tau)
=  g^{-1}T  \delta_{\mu 0}  ( \varphi_3 \delta_{A3} +   \varphi_8 \delta_{A8})
+ \mathscr{\tilde V}_\mu^A(\bm{x},\tau) 
 ,
 \label{sepa-SU3}
\end{align}
where $\varphi$, $\varphi_3$ and $\varphi_8$ are dimensionless quantities.

We take the approximation in which the quantum fluctuation parts $\mathscr{\tilde V}_\mu^A$ are neglected, $\mathscr{\tilde V}_\mu^A \ll 1$ .  
Then the holonomy operator $P(\bm{x})$ takes the simple form without the path ordering: 
\begin{align}
P =& \exp \left[i \varphi  \frac{\sigma_3}{2}  \right]  \in U(1) \subset SU(2) ,  
\nonumber\\ 
P =& \exp \left[i \varphi_3 \frac{\lambda_3}{2}+i \varphi_8 \frac{\lambda_8}{2} \right]  \in U(1) \times U(1) \subset SU(3), 
\end{align}
where $\sigma_3$ is the diagonal Pauli matrix and  $\lambda_3$, $\lambda_8$ are the diagonal Gell-Mann matrices. 
The $SU(2)$ Polyakov loop operator $L$ becomes a real-valued function of an angle $\varphi$:
\begin{align}
  L(\varphi) :=  \frac12 {\rm tr}(P) 
= \frac12 {\rm tr} \left\{ \exp \left[i \varphi  \frac{\sigma_3}{2}  \right] \right\} 
= \cos \frac{\varphi}{2}   \in \mathbb{R} ,
\end{align}
and the $SU(3)$ Polyakov loop operator  $L$  becomes a complex-valued function of the two angles $\varphi_3$ and $\varphi_8$:
\begin{align}
 & L(\varphi_{3},\varphi_{8}) :=  \frac13 {\rm tr}(P)
\nonumber\\ 
=& \frac13 {\rm tr} \left\{ \exp \left[i \varphi_3 \frac{\lambda_3}{2}+i \varphi_8 \frac{\lambda_8}{2} \right] \right\} 
\nonumber\\
=& \frac13 \left\{ e^{i\frac12 \left(  \varphi_{3} + \frac{1}{\sqrt{3}} \varphi_{8}  \right)}+e^{i\frac12 \left(-\varphi_{3} + \frac{1}{\sqrt{3}} \varphi_{8}   \right)}+e^{i\frac12 \left( - \frac{2}{\sqrt{3}} \varphi_{8}   \right)} \right\}
\nonumber\\
=& \frac13 \left[ e^{ -i  \frac{1}{\sqrt{3}} \varphi_{8}    } + 2 e^{i   \frac{1}{2\sqrt{3}} \varphi_{8} } \cos \left(\frac{\varphi_{3}}{2} \right)  \right] 
\in \mathbb{C} 
.  
\label{SU(3)-PL}
\end{align}

\subsection{Our standpoint}

The standpoint of our approach presented in this paper, the first approximation and its improvements, is completely different from the other work based on the systematic loop calculations in the perturbation theory \cite{RSTW15}.
Although the standpoint of our approach has been explained in the previous work \cite{Kondo10} and this paper is also written in this setting, we repeat it below for avoiding the misunderstanding on our approach.

In this paper we aim at a purely non-perturbative approach in which we look for \textit{the initial approximation which captures the essential features of the problem in question as much as possible at the initial stage}, which is the spirit of the first approximation. 
We do not intend to do the one-loop calculation in the perturbation theory and do not intend to do the systematic loop calculations of higher orders, either.
Our approach is different from \cite{RSTW15} conceptually in this aspect.
We use the terminology ``one-loop type'' to distinguish it from the one-loop in the perturbation theory.

In the first approximation, we take into account only the quadratic terms in the fields to obtain the effective action  (except for the restricted field), which leads to the ``one-loop type'' calculations.  It is well known that the effective action $\Gamma$ obtained from the classical action $S$ by the Legendre transform of the generating functional of the connected Green functions is equal to the classical action $S$ plus the additional part represented by the logarithmic determinant resulting from the Gaussian integrations over the quadratic parts.
Therefore, the action $S_{\rm eff}$ to be calculated by integrating out all  fields in the first approximation in our paper  agrees with the effective action $\Gamma$, up to the special treatment of the restricted field $\mathscr{V}_\mu$ as explained below.

The reason of the special treatment of the restricted field is as follows.
In our formulation, the Polyakov loop operator $L[\mathscr{A}]$ is completely written in terms of the restricted field $\mathscr{V}$, i.e., $L[\mathscr{A}]=L[\mathscr{V}]$. 
If we take the gauge (III.4) or (III.6), then $\mathscr{C}$ is equivalent to $\mathscr{V}$ as indicated in (III.5) or (III.7).
By integrating out all the fields up to the quadratic parts other than the restricted field $\mathscr{V}$, i.e.,  $\mathscr{C}$, we obtain the effective theory written in terms of the  the restricted field $\mathscr{V}$ alone. 
This is along the spirit of the first approximation mentioned in the above. 
Then, we estimate the Polyakov loop average by the minimum of the effective potential obtained from the effective theory.

The resulting effective theory is identified with the low-energy effective theory in the following sense.
We use the results obtained in the first approximation  as the input for performing the FRG approach to improve the first result.  
In this sense, the first approximation is regarded as the initial condition corresponding to the large flow parameter $\kappa$ at which the FRG analysis start.
Or the first approximation can be regarded as a preliminary Ansatz for solving the flow equation of FRG. 
The effective action with the flow parameter $\kappa$ in the FRG approach  means that the high-energy modes above $\kappa$, i.e., $p > \kappa$ are already integrated out to obtain the low-energy effective theory which is valid below $\kappa$, i.e., $p<\kappa$. 
For large $\kappa$ at the initial step, the high-energy mode for the restricted field is not integrated yet or not to be integrated out, since we identify the restricted field with the low-energy modes.  This is the reason why the restricted field is avoided to be integrated in the first or initial approximation. 
While the remaining fields are regarded as the high energy modes to be integrated even for large $\kappa$, since they behave as the massive modes and decouple in the low-energy region. 

Of course, the above setting is just an approximation and cannot be the rigorous treatment and hence this first approximation must and will be improved afterwards by a systematic method. 
In fact, we intend to improve the first result by the non-perturbative FRG  at once (not by the systematic order by order loop expansion). 
This is the standpoint of our approach adopted in this paper. 

\subsection{Effective action and effective potential}

In order to obtain the \textbf{effective potential} $V_{\rm eff}$ written in terms of the restricted field $\mathscr{V}_\mu$ (similarly  for the the Polyakov loop $L$), 
we perform the functional integration  over the field variables other than the restricted field $\mathscr{V}_\mu $ or $\mathscr{C}_\mu$: the remaining field $\mathscr{X}_\mu$ (massive gluon modes), the Nakanishi-Lautrup field $\mathscr{N}$ (massless scalar mode), and the Faddeev-Popov ghost and antighost fields $\mathscr{\eta}, \mathscr{\bar{\eta}}$.
\begin{align}
  &  e^{iS_{\rm eff}[\mathscr{C} ]} 
\nonumber\\
=& 
 \int \mathcal{D}\mathscr{X}_\mu^b \mathcal{D}\mathscr{N}^A \mathcal{D}\mathscr{\eta}^A \mathcal{D}\mathscr{\bar \eta}^A e^{i\tilde S_{\rm YM}^{\rm tot}[\bm n, \mathscr{C}, \mathscr{X}, \mathscr{N},\mathscr{\eta}, \mathscr{\bar \eta} ]} 
 .
\end{align}
Then we obtain the effective action $S_{\rm eff}$: 
\begin{align}
 & S_{\rm eff} 
\nonumber\\
=& \frac{D-1}{2} {\rm Tr} \ln [ - D_{\mu}^{2} [G] + M^2 ] 
+  \frac{D-1}{2} {\rm Tr} \ln [ - \bar{D}_{\mu}^{2} [G] + M^2 ]  
\nonumber\\
 & + \frac{1}{2} {\rm Tr} \ln [ - D_{\mu}^{2} [G]   ] +  \frac{1}{2} {\rm Tr} \ln [ - \bar{D}_{\mu}^{2} [G]   ]  
\nonumber\\
&-   {\rm Tr} \ln [ - D_{\mu}^{2} [G]  ] -    {\rm Tr} \ln [ - \bar{D}_{\mu}^{2} [G]  ]  
\end{align}
where $G_\mu$ is the restricted field variable replaced from $\mathscr{V}_\mu$ in a new basis (see Appendix \ref{Appendix:path-integral}) and
\begin{align}
  D_\mu[G] :=\partial -igG_\mu, \quad
 \bar{D}_\mu[G] :=\partial +igG_\mu .
\end{align}
Here the first term comes from the integration over the remaining field $\mathscr{X}_\mu$, the second term from  the integration over the Nakanishi-Lautrup field $\mathscr{N}$, and third term from  the integration over the FP ghost and antighost fields. 
See Appendix \ref{Appendix:path-integral} for the details of the calculations.
Finally, we obtain
\begin{align}
 & S_{\rm eff} 
\nonumber\\
=&  \frac{D-1}{2} {\rm Tr} \ln [ - D_{\mu}^{2} [G] + M^2 ] 
+  \frac{D-1}{2} {\rm Tr} \ln [ - \bar{D}_{\mu}^{2} [G] + M^2 ] 
\nonumber\\&
- \frac{1}{2} {\rm Tr} \ln [ - D_{\mu}^{2} [G]   ]
- \frac{1}{2} {\rm Tr} \ln [ - \bar{D}_{\mu}^{2} [G]   ]
 .
\end{align}
Here we have taken into account only the terms quadratic in the fields, in addition to the previous approximation where the quantum fluctuation parts $\mathscr{\tilde V}_\mu^A$ are neglected.  
We call the approximations taken up to this stage the \textbf{first approximation}. 
The corrections to the first approximation are incorporated by using the FRG afterwards where the explicit dependence on the gauge coupling constant enters in the results.

In a new basis the restricted field variable $G_\mu(x)$ is separated into the background field $\underline G_\mu(x)$ and the quantum fluctuation part $\tilde G_\mu(x)$, i.e., $G_\mu(x)=\underline G_\mu(x)+\tilde G_\mu(x)$.
This separation corresponds to (\ref{sepa-SU2}) and (\ref{sepa-SU3}). 
Therefore, we have the specific uniform (i.e., $x$-independent) background $\underline G_\mu(x)=G_0 \delta_{\mu 0}$ for the restricted field variable in a new basis:
\begin{align}
  G_\mu(x) = G_0 \delta_{\mu 0} + \tilde G_\mu(x) .
\end{align}
Then, the covariant Laplacian $-D_{\mu}^{2} [G]$ or $- \bar{D}_{\mu}^{2} [G]$  is given 
\begin{align}
 & -(\partial_\rho \mp igG_\rho(x))^2
\nonumber\\
=&   - \partial_\rho^2  \pm 2igG_\rho(x) \partial_\rho + g^2 G_\rho(x)^2 \pm ig \partial_\rho G_\rho(x)    
\nonumber\\
=& - \partial_\ell^2 - \partial_0^2  \pm 2igG_0 \partial_0   + g^2 G_0^2  
+ [\text{$\tilde G(x)$-dependent terms}] 
\nonumber\\
=& - \partial_\ell^2 + (i \partial_0 \pm  gG_0)^2  
+ [\text{$\tilde G(x)$-dependent terms}] 
 .
\end{align}
Therefore, the covariant Laplacian has the momentum representation at finite temperature: 
\begin{align}
   -(\partial_\rho \mp igG_\rho)^2
\to & 
 \bm{p}^2 + (\omega_n \pm  gG_0)^2  
+ [\text{$\tilde G$-dependent terms}] 
 .
\end{align}
Note that $G_0$ is given by
\begin{align}
\text{$SU(2)$:} \quad 
G_0 =&  g^{-1}T \alpha \varphi, \ \alpha = \pm 1 ,
 \nonumber\\
\text{$SU(3)$:} \quad 
G_0 =&  \pm g^{-1}T  \vec{\alpha}^{(i)} \cdot \vec{\varphi} , \quad \vec{\varphi} :=(\varphi_3, \varphi_8  )  
 ,
\end{align}
where $\vec{\alpha}^{(i)}$ are the positive root vectors of $SU(3)$:
\begin{align}
  \vec{\alpha}^{(1)} = (1,0), \quad 
  \vec{\alpha}^{(2)} = \left(\frac{1}{2},\frac{\sqrt{3}}{2} \right), \quad 
  \vec{\alpha}^{(3)} = \left(\frac{-1}{2},\frac{\sqrt{3}}{2} \right) . 
\end{align}


\section{$SU(2)$ Yang-Mills theory}

\setcounter{equation}{0}

\subsection{Existence of $SU(2)$ confinement/deconfinement transition}

Symmetries of the  $SU(2)$  Polyakov loop average $L$ are as follows: 
\\
i) periodicity of $4\pi$ in  $\varphi$:
\begin{align}
  L(\varphi) 
 = L(\varphi+4\pi) 
 ,
\end{align}
ii) reflection invariance: 
\begin{align}
  L(\varphi) 
 = L(-\varphi) 
 .
\end{align}
The Polyakov loop operator is written in a simple form:
\begin{align}
 L =  \cos \frac{\varphi}{2} \in (-1,1] , \quad \varphi \in [0,2\pi ).
\end{align}

According to (\ref{formulation-L}), thus, the Polyakov loop average can be calculated at the absolute minimum of the effective potential. 
The vanishing Polyakov loop average  $L=0$, i.e., \textbf{confinement} is realized if the  effective potential $V(\varphi)$ has the minimum at $\varphi=\pi$.
Whereas non-vanishing Polyakov loop average  $L \neq 0$, i.e., \textbf{deconfinement} is realized if the  effective potential $V(\varphi)$ has the minimum at $\varphi \neq \pi$.
\begin{align}
 & L  =  0  \Longleftrightarrow \varphi = \pi \quad (\text{confinement: $Z(2)$ symmetric}),
\nonumber\\ 
 & L  \neq 0  \Longleftrightarrow \varphi \neq \pi \quad (\text{deconfinement: $Z(2)$ breaking}).
\end{align}

Symmetries of the  $SU(2)$  effective potential $V(\varphi)$ are as follows: 
\\
i) periodicity of $2\pi$ in  $\varphi$:
\begin{align}
  V_{\rm eff}(\varphi) 
 = V_{\rm eff}(\varphi+2\pi) 
 ,
\end{align}
ii) reflection invariance: 
\begin{align}
  V_{\rm eff}(\varphi) 
 = V_{\rm eff}(-\varphi) 
 .
\end{align}
These are the result of $Z(2)$ center symmetry: 
\begin{align}
  V_{\rm eff}(\varphi) 
 = V_{\rm eff}(2\pi-\varphi) 
 ,
\end{align}
In fact, these properties are satisfied by the explicit form of the effective potential (\ref{SU2-V-eff}) obtained below. 

The effective action $S_{\rm eff}$ reduces to the effective potential $V_{\rm eff}(\varphi)$ written in terms of the background part $\varphi$, i.e., 
\begin{align}
S_{\rm eff}= V_{\rm eff}(\varphi) T^{-1}\int d^{D-1}x
\end{align}
 by neglecting the quantum fluctuation part $\tilde G$, since the background part is $x$-independent.  
In this approximation, thus, the effective potential has the momentum representation:
\begin{align}
 &V_{\rm eff}(\varphi) 
\nonumber\\
&= 
 \frac{D-1}{2} T \sum_{n \in \mathbb{Z}, \pm}  \int \frac{d^{D-1}p}{(2\pi)^{D-1}}   \ln [(\omega_n \pm T\varphi)^2+\bm{p}^2 + M^2 ] 
\nonumber\\
&-  \frac{1}{2} T \sum_{n \in \mathbb{Z}, \pm}  \int \frac{d^{D-1}p}{(2\pi)^{D-1}}   \ln [(\omega_n \pm T\varphi)^2+\bm{p}^2  ] ,  
\nonumber\\   
& (\omega_n := 2\pi Tn )
 .
 \label{SU2-V-eff}
\end{align}

\begin{figure}[ptb]
\begin{center}
\includegraphics[scale=0.60]{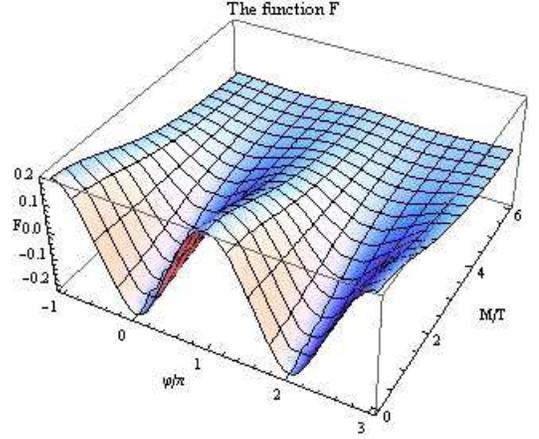}
\end{center}
\vskip -0.3cm
\caption{
The plot of 
$
F_{\hat{M}}(\varphi)
= \int \frac{d^{D-1}\hat{p}}{(2\pi)^{D-1}} \ln [1+e^{-2 \sqrt{\hat{\bm{p}}^2 +\hat{M}^2 }} - 2 e^{-  \sqrt{\hat{\bm{p}}^2 +\hat{M}^2}} \cos ( \varphi)]
$ 
as a function of the angle $\varphi$ for various values of $\hat{M}:=M/T \ge 0$ at $D=4$.
}
\label{fig:function-F}
\end{figure}

The Matsubara sum (summing over the Matsubara frequencies $\omega_n=2\pi Tn$) can be done to obtain the closed form: 
\begin{align}
  & \sum_{n \in \mathbb{Z}}    \ln [(\omega_n +T\theta)^2+\bm{p}^2 + M^2 ] 
\nonumber\\
=&    \ln [1+e^{-2 E_p/T} - 2 e^{- E_p/T} \cos  \theta ] ,
\end{align}
where we have defined
\begin{align}
  E_p := \sqrt{\bm{p}^2+M^2}  .
\end{align}
Then we have 
\begin{align}
 T \sum_{n \in \mathbb{Z}}  \int \frac{d^{D-1}p}{(2\pi)^{D-1}}   \ln [(\omega_n +T\varphi)^2+\bm{p}^2 + M^2 ] 
= T^D F_{\hat{M}}(\varphi) 
 .
\end{align}
Here we have introduced the dimensionless variables $\hat{\bm{p}}$ and $\hat{M}$ normalized by the temperature $T$ as
\begin{align}
\hat{\bm{p}}:=\bm{p}/T, \quad \hat{M} := M/T ,
\end{align}
to define the dimensionless function $F_{\hat{M}}(\varphi)$ of the angle $\varphi$ with a parameter $\hat{M}$  by
\begin{align}
  F_{\hat{M}}(\varphi)
=& \int \frac{d^{D-1}\hat{p}}{(2\pi)^{D-1}} f_{\hat{M}}(\hat{p}^2, \varphi)
 ,
\nonumber\\
  f_{\hat{M}}(\hat{p}^2, \varphi)
:=&  
\ln [1+e^{-2 \sqrt{\hat{\bm{p}}^2 +\hat{M}^2 }} - 2 e^{-  \sqrt{\hat{\bm{p}}^2 +\hat{M}^2}} \cos  \varphi ]  
 .
 \label{F-def1}
\end{align}
In Fig.~\ref{fig:function-F}, 
we have given the plot of   $F_{\hat{M}}(\varphi)$ as a function of the angle $\varphi$ for various values of $\hat{M} \ge 0$ at $D=4$.
We see that $F_{\hat{M}}(\varphi)$ is exponentially suppressed $F_{\hat{M}}(\varphi) \ll 1$ for large $\hat{M}$ and vanishes in the limit $\hat{M} \to \infty$ uniformly in $\varphi$, i.e., irrespective of the value of $\varphi$.
Here we have used
\begin{align}
  F_{\hat{M}}(\varphi)
=&  C_D \int_{0}^{\infty} d \hat{p} \ \hat{p}^{D-2} f_{\hat{M}}(\hat{p}^2, \varphi)  
 , 
\nonumber\\
 C_D :=& \frac{1}{2^{D-2}\pi^{\frac{D-1}{2}} \Gamma(\frac{D-1}{2})}
 .
 \label{F-def2}
\end{align}

\begin{figure}[ptb]
\begin{center}
\includegraphics[scale=0.65]{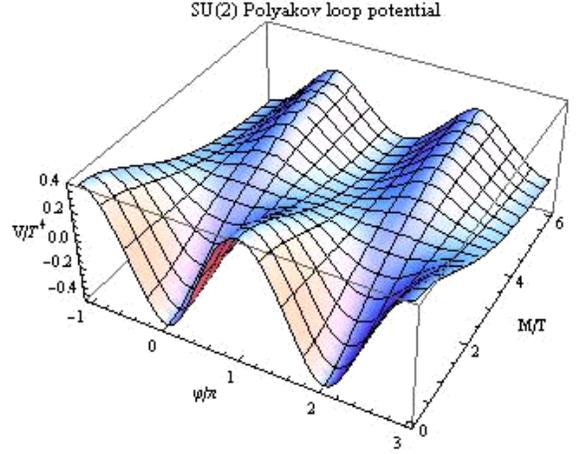}
\end{center}
\vskip -0.3cm
\caption{
The $D=4$ effective potential $\hat{V}$ of the $SU(2)$ Polyakov loop as a function of the angle $\varphi$ for various values of $\hat{M}:=M/T \ge 0$.
}
\label{fig:V-SU2-b}
\end{figure}

Thus, we obtain the dimensionless effective potential of the Polyakov loop average normalized by the temperature as 
\begin{align}
  \hat{V}_{\rm eff}(\varphi) 
:= V_{\rm eff}(\varphi)/T^D = (D-1) F_{\hat{M}}(\varphi) - F_{0}(\varphi) 
 .
\end{align}
In Fig.~\ref{fig:V-SU2-b}, we have given the plot of the  effective potential $\hat{V}_{\rm eff}(\varphi)$ of the $SU(2)$ Polyakov loop as a function of the angle $\varphi$ for various values of $\hat{M}:=M/T \ge 0$ at $D=4$.

\begin{figure}[ptb]
\begin{center}
\includegraphics[scale=0.65]{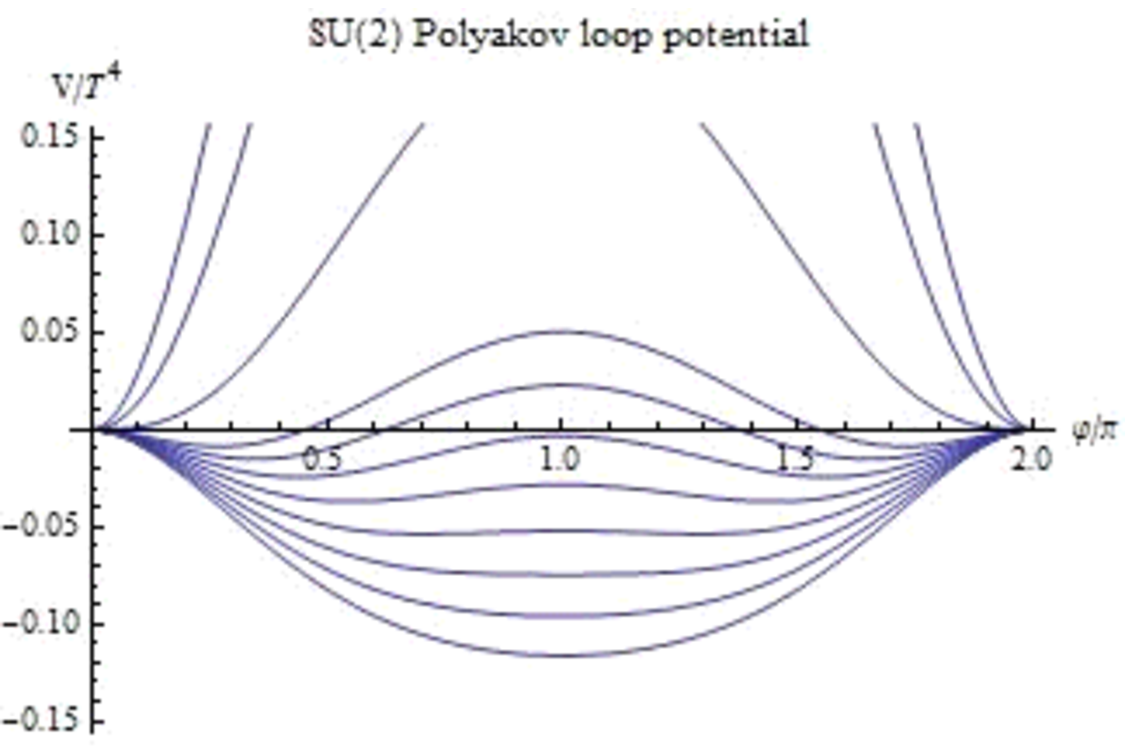}
\includegraphics[scale=0.65]{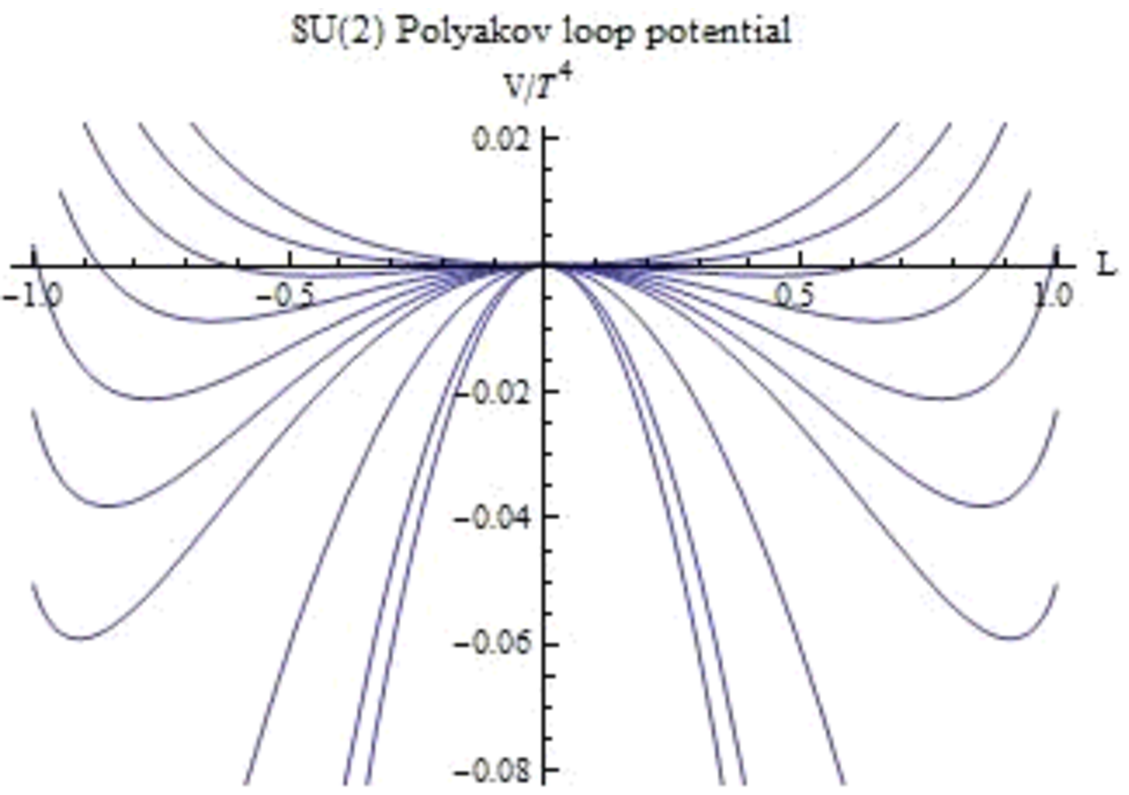}
\end{center}
\vskip -0.3cm
\caption{
The  $D=4$ effective potential $\hat{V}$ of the $SU(2)$ Polyakov loop for   $\hat{M}:=M/T=0.0, 1.0, 2.0, 2.5, 2.6, 2.7, 2.8, 2.9, 3.0, 3.1$, 
(Left) as a function of the angle $\varphi/\pi \in [0, 2)$, 
(Right) as a function of the Polyakov loop average $L = \cos \frac{\varphi}{2} \in (-1,1]$.
}
\label{fig:V-SU2-all}
\end{figure}

At sufficiently high temperature, $\hat{M}=M/T \ll 1$, the gluonic mass $M$ is negligible and the effective potential given by
\begin{align}
  \hat{V}_{\rm eff}^{\rm High}(\varphi) \simeq (D-1) F_{0}(\varphi) -  F_{0}(\varphi) = (D-2) F_{0}(\varphi)  
 .
\end{align}
The high-temperature effective potential $\hat{V}_{\rm eff}^{\rm High}(\varphi)$ has the $Z(2)$ breaking minima at $\varphi=0, 2\pi$, and the extremum at $\varphi=\pi$ is a maximum.
For $D=4$, this effective potential reproduces the well-known Weiss potential \cite{Weiss81,GPY81}. 
\begin{align}
 V_W(\varphi)
=&    T^4 \left[ - \frac{1}{6}  (\varphi-\pi)^2 +  \frac{1}{12\pi^2}   (\varphi-\pi)^4   +   \frac{\pi^2}{12} \right]
\nonumber\\
 ({\rm mod} \ 2\pi)
 .
\end{align}
See the $M/T=0$ part of Fig.~\ref{fig:V-SU2-b} or Fig.~\ref{fig:V-SU2-all}.

At sufficiently low temperature, $\hat{M}=M/T \gg 1$, on the other hand, $F_{\hat{M}}(\varphi)$ is exponentially  suppressed $F_{\hat{M}}(\varphi) \ll 1$ and the effective potential reduces to 
\begin{align}
  \hat{V}_{\rm eff}^{\rm Low}(\varphi) \simeq  -  F_{0}(\varphi)
 .
\end{align}
The low-temperature effective potential $\hat{V}_{\rm eff}^{\rm Low}(\varphi)$ has the $Z(2)$ symmetric minimum at $\varphi=\pi$.
In fact, the effective potential in the sufficiently low temperature is reversed to the Weiss potential at sufficiently high temperature (See the region $M/T>3$  of Fig.~\ref{fig:V-SU2-b} or Fig.~\ref{fig:V-SU2-all}): 
\begin{align}
  \hat{V}_{\rm eff}^{\rm Low}(\varphi) \simeq 
 - (D-2)^{-1} \hat{V}_{\rm eff}^{\rm High}(\varphi)
 .
\end{align}
This indicates the existence of the phase transition from the high-temperature deconfined phase to the low-temperature confined phase.

The physical interpretation of this phenomenon is as follows. 
At sufficiently high temperature $T \gg T_d$, the gluonic mass $M$ is negligible and hence gluons and ghosts contribute equally to the effective potential. Both gluons and ghosts are equally responsible for the dynamics in the high-temperature deconfined phase. 
At sufficiently low temperature $T \ll T_d$, the massive gluons with the mass $M$ are not excited by the thermal fluctuation of order $T <M$ and do not contribute to the effective potential.  The massless scalar mode, ghosts and antighosts give the dominant contribution to the effective potential at low-temperature confined phase. 
In this sense, we can say that the confinement mechanism at finite temperature is the \textbf{scalar-ghost dominance}  or unphysical mode dominance.

This result is reliable in the region where the background part of the restricted field for calculating the Polyakov loop average gives the dominant contribution to the phase transition compared with the quantum fluctuation part of the restricted field.
Therefore, it is applicable to the vicinity of the critical temperature $T_d$.
The above estimate on the critical temperature is meaningful as far as the gluonic mass $M$ does not discontinuously change across the transition point, even if it changes as the temperature varies. 
The absence of the discontinuous change of $M$, namely, continuous change of $M$ across the transition point will be shown in an analytical way as well as in the numerical way in the subsequent paper \cite{KS15}.


\subsection{$SU(2)$  Transition temperature and order of the transition}


We proceed to  estimate the critical temperature and determine the order of the phase transition. 
This is achieved by the detailed study of the function $F_{\hat{M}}(\varphi)$.
The integrand $f_{\hat{M}}(\hat{p}^2, \varphi)$ of $F_{\hat{M}}(\varphi)$ is expanded into the power series in the angle variable $\varphi$ about $\varphi=\pi$ (at which $L=0$):
\begin{align}
  f_{\hat{M}}(\hat{p}^2, \varphi)
=& a_{\hat{M}}(\hat{p}) + b_{\hat{M}}(\hat{p}) (\varphi - \pi)^2 + c_{\hat{M}}(\hat{p}) (\varphi - \pi)^4 
\nonumber\\  &
+ O((\varphi - \pi)^6) 
 ,
\end{align}
with the coefficients:
\begin{align}
 a_{\hat{M}}(\hat{p}) =& 2 \ln (1+e^{- \sqrt{\hat{\bm{p}}^2 +\hat{M}^2 }}) > 0 ,
\nonumber\\
 b_{\hat{M}}(\hat{p}) =& - \frac{e^{- \sqrt{\hat{\bm{p}}^2 +\hat{M}^2 }}}{(1+e^{- \sqrt{\hat{\bm{p}}^2 +\hat{M}^2 }})^2} < 0 ,
\nonumber\\
 c_{\hat{M}}(\hat{p}) =&  \frac{e^{- \sqrt{\hat{\bm{p}}^2 +\hat{M}^2 }} (1 -4 e^{- \sqrt{\hat{\bm{p}}^2 +\hat{M}^2 }}+e^{-2 \sqrt{\hat{\bm{p}}^2 +\hat{M}^2 }}) }{12(1+e^{- \sqrt{\hat{\bm{p}}^2 +\hat{M}^2 }})^4}  .
\end{align}
Then the effective potential is expanded into the power series in the angle variable $\varphi$ around $\varphi=\pi$:
\begin{align}
 \hat{V}_0(\varphi; \hat{M}) 
:=& V_{\rm eff,0}(\varphi)/T^D 
=  
(D-1) F_{\hat{M}}(\varphi) - F_{0}(\varphi)
\nonumber\\
=& A_{0,\hat{M}} + \frac{A_{2,\hat{M}}}{2!} (\varphi - \pi)^2 + \frac{A_{4,\hat{M}}}{4!} (\varphi - \pi)^4 
\nonumber\\ &
+ O((\varphi - \pi)^6)
 ,
\end{align}
where the coefficients are explicitly given as
\begin{align}
    A_{0,\hat{M}}
=&   C_D \int_{0}^{\infty} d \hat{p} \ \hat{p}^{D-2} [(D-1) a_{\hat{M}}(\hat{p}) - a_{0}(\hat{p})]
\nonumber\\
   \frac{1}{2!} A_{2,\hat{M}}
=&  
 C_D \int_{0}^{\infty} d \hat{p} \ \hat{p}^{D-2} [(D-1) b_{\hat{M}}(\hat{p}) - b_{0}(\hat{p})]
 ,
\nonumber\\
   \frac{1}{4!}  A_{4,\hat{M}}
=&  
  C_D \int_{0}^{\infty} d \hat{p} \ \hat{p}^{D-2} [(D-1) c_{\hat{M}}(\hat{p}) - c_{0}(\hat{p})]
 .
\end{align}

\begin{figure}[ptb]
\begin{center}
\includegraphics[scale=0.45]{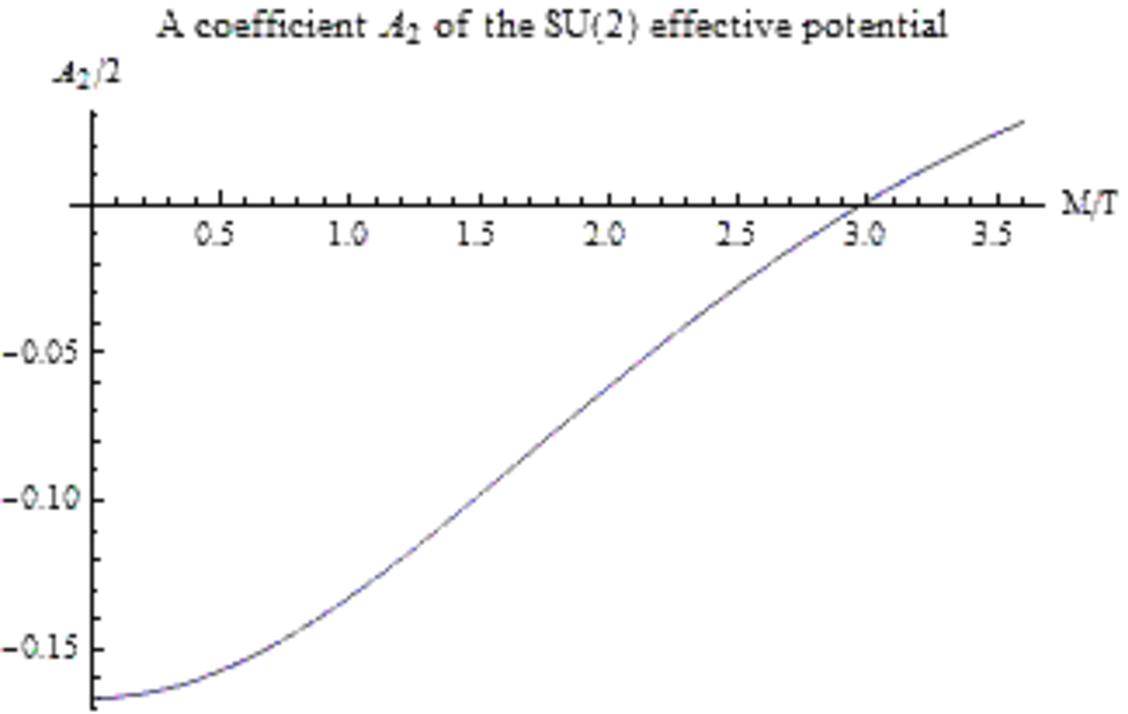}
\includegraphics[scale=0.45]{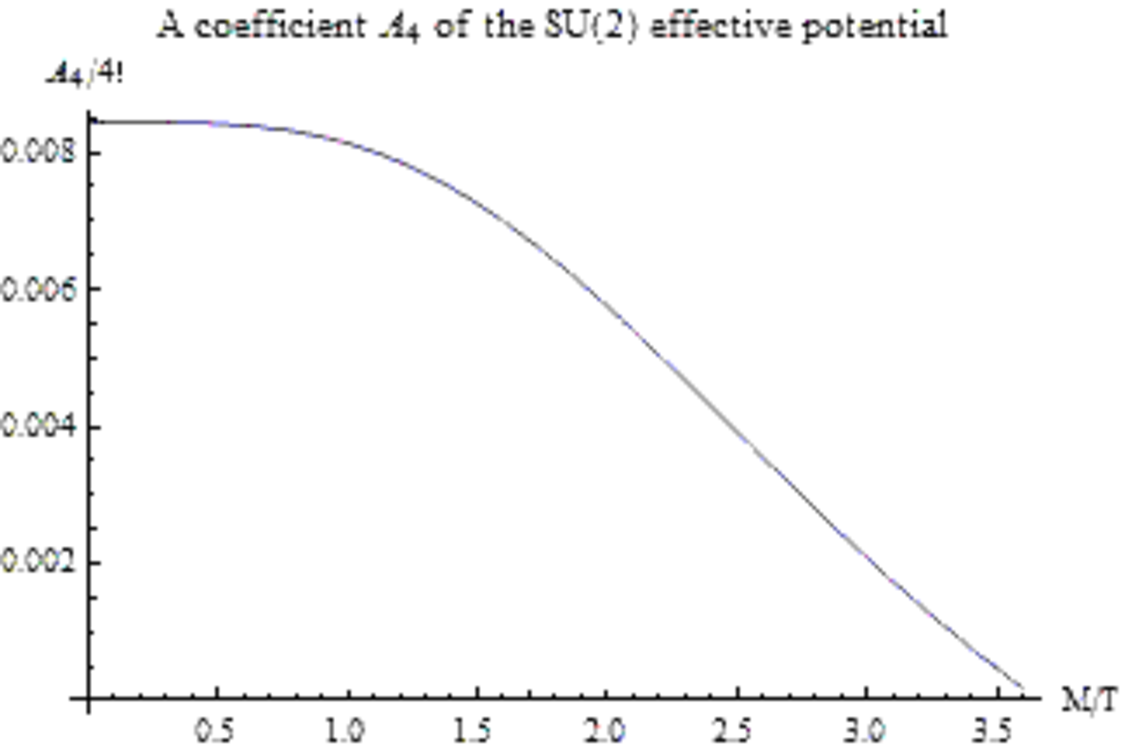}
\end{center}
\vskip -0.3cm
\caption{
The coefficients $A_{2,\hat{M}}$ and $A_{4,\hat{M}}$ of the $SU(2)$ Polyakov loop effective potential $\hat{V}_0(\varphi; \hat{M})$ as a function of $\hat{M}:=M/T$ at $D=4$. 
}
\label{fig:V-SU2-A2A4}
\end{figure}

In the limit $\hat{M} \to 0$, especially, the coefficient $A_{2,0}$ is given by
\begin{align}
 \frac{1}{2!} A_{2,0}
=&  C_D (D-2) \int_{0}^{\infty} d \hat{p} \ \hat{p}^{D-2}  b_{0}(\hat{p}) 
\nonumber\\
=& -  C_D (D-2) \int_{0}^{\infty} d \hat{p} \ \hat{p}^{D-2} \frac{e^{- \hat{p}}}{(1+e^{- \hat{p}})^2}   
 < 0
 ,
\end{align}
which recovers the previous result for $D=4$ with $C_4=\frac{1}{2\pi^2}$    \cite{Kondo10}:
\begin{align}
 \frac{1}{2!}  A_{2,0}
=  -  \frac{1}{\pi^2} \int_{0}^{\infty} d \hat{p} \ \hat{p}^{2} \frac{e^{- \hat{p}}}{(1+e^{- \hat{p}})^2}
= - \frac{1}{6}  
 < 0
 .
\end{align}
Similarly, in the limit $\hat{M} \to 0$, the coefficient $A_{4,0}$ is given by
\begin{align}
 \frac{1}{4!} A_{4,0}
=&  C_D (D-2) \int_{0}^{\infty} d \hat{p} \ \hat{p}^{D-2}  c_{0}(\hat{p}) 
\nonumber\\
=&  C_D (D-2) \int_{0}^{\infty} d \hat{p} \ \hat{p}^{D-2}    \frac{e^{- \hat{p}} (1 -4 e^{- \hat{p}}+e^{-2 \hat{p}}) }{12(1+e^{- \hat{p}})^4}  
 ,
\end{align}
which recovers the previous result for $D=4$  \cite{Kondo10}:
\begin{align}
 \frac{1}{4!} A_{4,0}
=    \frac{1}{\pi^2} \int_{0}^{\infty} d \hat{p} \ \hat{p}^{2}  \frac{e^{- \hat{p}} (1 -4 e^{- \hat{p}}+e^{-2 \hat{p}}) }{12(1+e^{- \hat{p}})^4}  
  =  \frac{1}{12\pi^2}  
 > 0
 .
\end{align}
Fig.~\ref{fig:V-SU2-A2A4} is the plot of the coefficients $A_{2,\hat{M}}/2!$ and $A_{4,\hat{M}}/4!$ of the $SU(2)$ Polyakov-loop effective potential $\hat{V}_0(\varphi; \hat{M})$ as a function of $\hat{M}:=M/T$ at $D=4$. 
We observe that $A_{2,\hat{M}}<0$ for $\hat{M} \in [0,2.9]$ and $A_{2,\hat{M}}>0$ for $\hat{M} >2.9$, while $A_{4,\hat{M}}>0$ for $\hat{M} \in [0,3.6]$.

Therefore, the phase transition from deconfinement to confinement occurs at the temperature $T_d$ at which the coefficient $A_{2,\hat{M}}$ changes its signature  from negative to positive, namely, becomes zero: 
\begin{align}
 & A_{2,\hat{M}} =  0 
\nonumber\\  
\to & \int_{0}^{\infty} d \hat{p} \ \hat{p}^{D-2} [(D-1) b_{\hat{M}}(\hat{p}) - b_{0}(\hat{p})] = 0
 .
\end{align}
This condition determines the critical value for the ratio $\hat{M}_c:=M(T_d)/T_d$ between the gluonic mass $M(T_d)$ and the transition temperature $T_d$.
For $D=4$, we find the critical value:
\begin{align}
   \frac{M(T_d)}{T_d} = 2.9724  \Longleftrightarrow \frac{T_d}{M(T_d)} = 0.33643 
 ,
\end{align}
where $M$ may depend on temperature. 
For instance, 
\begin{align}
   M(T_d) = 1.0 {\rm GeV} \leftrightarrow  T_d =  340 {\rm MeV} 
 .
\end{align}
See Introduction for more information. 


\begin{figure}[ptb]
\begin{center}
\includegraphics[scale=0.45]{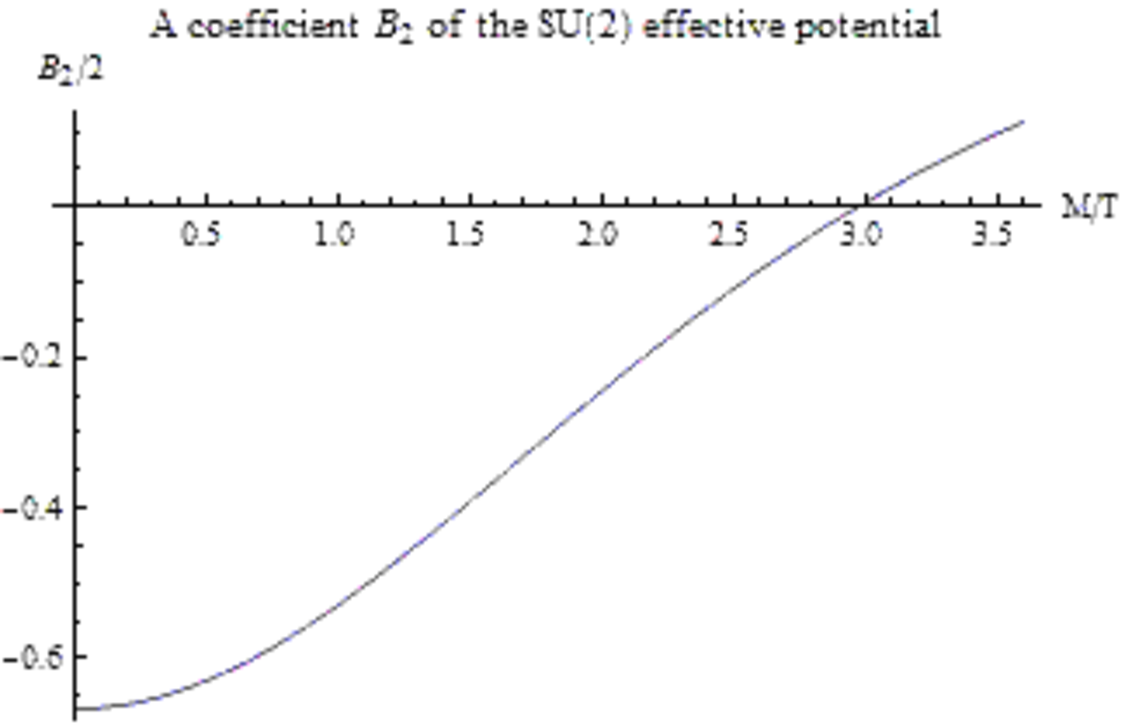}
\includegraphics[scale=0.45]{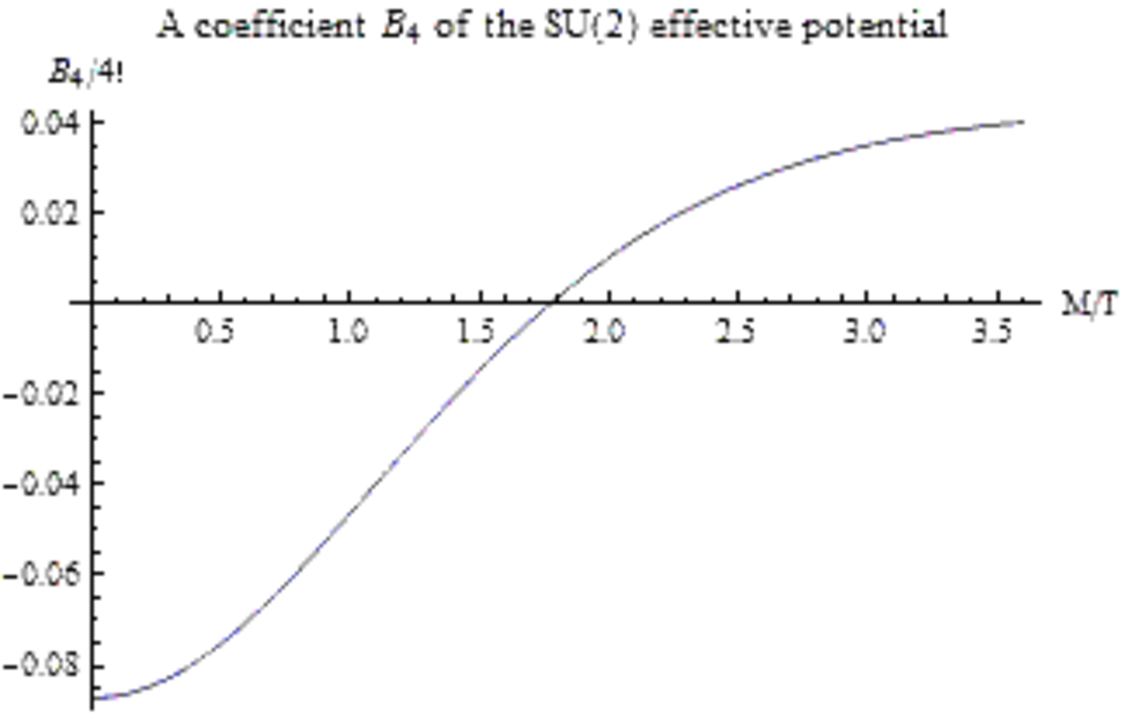}
\end{center}
\vskip -0.3cm
\caption{
The coefficients $B_{2,\hat{M}}$ and $B_{4,\hat{M}}$ of the $SU(2)$ Polyakov loop effective potential $\hat{V}_0(L; \hat{M})$ as a function of $\hat{M}:=M/T$ at $D=4$. 
}
\label{fig:V-SU2-B2B4}
\end{figure}

The above results are rephrased in terms of the Polyakov loop average $L$ directly. 
The angle $\varphi$ and the $SU(2)$ Polyakov loop operator in the first approximation is related as
\begin{align}
   L = \cos \frac{\varphi}{2} \to \cos \varphi = 2L^2 -1 .
\end{align}
Then the effective potential is rewritten in terms of the Polyakov loop average $L$ explicitly.  
In fact, we have the expression for $f_{\hat{M}}(\hat{p}^2, \varphi)$ in terms of $L$:
\begin{align}
  f_{\hat{M}}(\hat{p}^2, \varphi)
:=&  
\ln [1+e^{-2 \sqrt{\hat{\bm{p}}^2 +\hat{M}^2 }} - 2 e^{-  \sqrt{\hat{\bm{p}}^2 +\hat{M}^2}} (2L^2-1)]  
 .
\end{align}
In the similar way to the above, the integrand is expanded into the power series in $L$ about $L=0$:
\begin{align}
  f_{\hat{M}}(\hat{p}^2, \varphi)
=& \tilde a_{\hat{M}}(\hat{p}) + \tilde b_{\hat{M}}(\hat{p}) L^2 + \tilde c_{\hat{M}}(\hat{p}) L^4 + O(L^6) 
 ,
\end{align}
with the coefficients:
\begin{align}
 \tilde a_{\hat{M}}(\hat{p}) =& 2 \ln (1+e^{- \sqrt{\hat{\bm{p}}^2 +\hat{M}^2 }}) > 0 ,
\nonumber\\
 \tilde b_{\hat{M}}(\hat{p}) =& - \frac{4e^{- \sqrt{\hat{\bm{p}}^2 +\hat{M}^2 }}}{(1+e^{- \sqrt{\hat{\bm{p}}^2 +\hat{M}^2 }})^2} < 0 ,
\nonumber\\
 \tilde c_{\hat{M}}(\hat{p}) =& - \frac{8e^{- 2\sqrt{\hat{\bm{p}}^2 +\hat{M}^2 }}  }{(1+e^{- \sqrt{\hat{\bm{p}}^2 +\hat{M}^2 }})^4}  .
\end{align}
The effective potential is  expanded into a power series in $L$ (around $L=0$):
\begin{align}
 \hat{V}_0(L; \hat{M}) 
=& B_{0,\hat{M}} + \frac{B_{2,\hat{M}}}{2!} L^2 + \frac{B_{4,\hat{M}}}{4!} L^4 + O(L^6)
 ,
\end{align}
where the coefficients are given by
\begin{align}
    B_{0,\hat{M}}
=&   C_D \int_{0}^{\infty} d \hat{p} \ \hat{p}^{D-2} [(D-1) \tilde a_{\hat{M}}(\hat{p}) - \tilde a_{0}(\hat{p})]
\nonumber\\
 \frac{1}{2!} B_{2,\hat{M}}
=& 
 C_D \int_{0}^{\infty} d \hat{p} \ \hat{p}^{D-2} [(D-1) \tilde b_{\hat{M}}(\hat{p}) - \tilde b_{0}(\hat{p})]
 ,
\nonumber\\
 \frac{1}{4!} B_{4,\hat{M}}
=& 
  C_D \int_{0}^{\infty} d \hat{p} \ \hat{p}^{D-2} [(D-1) \tilde c_{\hat{M}}(\hat{p}) - \tilde c_{0}(\hat{p})]
 .
\end{align}
Here the fact that only the even powers of $L$ appear in the effective potential reflects the center $Z(2)$ symmetry:
$\hat{V}_0(zL; \hat{M})=\hat{V}_0(L; \hat{M})$
where $z$ satisfies $z^2=1$.
Fig.~\ref{fig:V-SU2-B2B4} is the plot of the coefficients $B_{2,\hat{M}}$ and $B_{4,\hat{M}}$ of the $SU(2)$ Polyakov-loop effective potential $\hat{V}_0(L; \hat{M})$ as a function of $\hat{M}:=M/T$ at $D=4$.

The coefficient $B_{2,\hat{M}}$ has the same behavior as the coefficient  $A_{2,\hat{M}}$. Both coefficients are negative for $0 \le \hat{M}<2.9$ and positive for $\hat{M}>2.9$. 
The coefficient $A_{4,\hat{M}}$ is positive for $0 \le \hat{M}<3.6$. However, the coefficient $B_{4,\hat{M}}$ is negative for $0 \le \hat{M}<1.8$, although it is positive for $\hat{M}>1.8$.  
The angle variable can be used for any temperature above the critical temperature $T_d$. 
Therefore,  it is better to use the angle variable $\varphi$ than the Polyakov loop average itself $L$ at least in this approximation. 
Fortunately, however, the critical value $\hat{M}_c=2.9$ of $A_{2,\hat{M}}=\frac14 B_{2,\hat{M}}$ is contained in the positive region which is common to both $A_{4,\hat{M}}$ and $B_{4,\hat{M}}$.
This allows one to study the vicinity of the critical temperature using the effective potential $\hat{V}_0(L; \hat{M})$ in terms of $L$ even in this approximation.

\subsection{$SU(2)$ Functional renormalization group}

The result of ``one-loop type''  for the effective potential given in the above can be improved using the FRG.  
The flow equation called the \textbf{Wetterich equation} \cite{Wetterich93} is given for the so-called \textbf{effective average action} $\Gamma_k$ which depends on the \textbf{renormalization-group (RG) scale} $k$:
\begin{align}
\partial_t \Gamma_k[\Phi] 
=& \frac12 {\rm STr} \left\{ \left[ \frac{\overrightarrow{\delta}}{\delta \Phi^\dagger} \Gamma_k[\Phi] \frac{\overleftarrow{\delta}}{\delta \Phi} + R_{\Phi,k} \right]^{-1} \cdot \partial_t R_{\Phi,k} \right\}
 ,
\end{align}
where $t$ is the RG time 
$
  t := \ln \frac{k}{\Lambda}
$, 
$
 \partial_t := \frac{\partial}{\partial t} = k \frac{d}{dk} 
$
for some reference scale (UV cutoff) $\Lambda$ 
and $R_{\Phi,k}$ is the \textbf{regulator function} for the field $\Phi$. 
Here ${\rm STr}$ denotes the super-trace introduced to include both  commutative field (gluon) and anticommutative field (quark, ghost). 
The physical result, i.e., the true effective action  is obtained finally  in the limit $k \downarrow 0$, after integrating over the whole range of momentum, which is in accord with the original idea of Wilsonian renormalization group. 
See \cite{FRG} for  reviews of the functional renormalization group. 

Although the choice of the infrared cutoff function $R_k$ is not unique, it is somewhat similar to the mass term with the mass proportional to $k$. 
In the case of massless fields, the control of the flow in solving the differential equation with respect to $k$ is rather subtle, since the mass term originated from the infrared cutoff function $R_k$ disappears eventually in the limit $k \downarrow 0$.
This is indeed the case of the ordinary Yang-Mills theory, i.e., gluodynamics where gluons and ghosts are both massless.

The Wetterich equation tells us that the ``one-loop type'' expression is the first approximation to the solution of the flow equation, if the infrared cutoff function $R_k$ depending on the flow parameter $k$ is included in the relevant ``one-loop type'' expression. 
If we choose the specific gauge in which the color field is uniform, i.e., (\ref{n-approx-SU2}) or (\ref{n-approx-SU3}), then the fields $\Phi$ relevant in the flow equation are the restricted field $\mathscr{V}_\mu(x)$, the remaining field $\mathscr{X}_\mu(x)$, the Nakanishi-Lautrup field $\mathscr{N}$, and the FP ghost and antighost $\mathscr{\eta}(x)$, $\bar{\eta}(x)$, i.e., $\Phi^\dagger = (\mathscr{V}_\mu,\mathscr{X}_\mu,\mathscr{N},\eta,\bar{\eta})$. 
In this section, we use the Euclidean formulation. 
Within the approximation adopted in this paper, the flow equation is written in the form:
\begin{align}
 \partial_t \Gamma_k
 =& \frac12 {\rm Tr} \left\{ \left[ \frac{\overrightarrow{\delta}}{\delta \mathscr{V}^\dagger} \Gamma_k \frac{\overleftarrow{\delta}}{\delta \mathscr{V}} + R_{k} \right]^{-1} \cdot \partial_t R_{k} \right\}
 + \partial_t V_{T,M,k}
  ,
\end{align}
where $V_{T,M,k}$ is the ``one-loop type'' part:
\begin{align}
 V_{T,M,k}
 =&  \frac{D-1}{2} {\rm Tr}  \{  \ln [G^{AB}+M^2 \delta^{AB}+\delta^{AB}R_{k} ] \}  
\nonumber\\&
 +  \frac12 {\rm Tr}  \{ \ln [G^{AB}+\delta^{AB} R_{k} ] \} 
\nonumber\\&
 -   {\rm Tr}  \{ \ln [G^{AB}+\delta^{AB}R_{k} ] \} ,
\nonumber\\ 
 G^{AB} :=& (\mathscr{D}_\rho[\mathscr{V}] \mathscr{D}^\rho[\mathscr{V}])^{AB}
  .
\end{align}
Here, the first, second, and third contributions in $V_{T,M,k}$ comes  respectively from the integration over the remaining field $\mathscr{X}$, the Nakanishi-Lautrup field $\mathscr{N}$,  and  the ghost-antighost fields $\eta, \bar\eta$ \cite{KMS05,KKSS15}.
We have used the same regulator function $R_k$ for gluons and ghosts up to the difference due to the tensor structure of gluons. 
By taking the trace over the Lorentz indices, 
$V_{T,M,k}$ is written as
\begin{align}
V_{T,M,k} 
 :=& \frac{D-1}{2} {\rm Tr}  \{  \ln [G^{AB}+M^2 \delta^{AB}+\delta^{AB}R_{k} ] \} 
\nonumber\\&
 -  \frac12 {\rm Tr}  \{ \ln [G^{AB}+\delta^{AB} R_{k} ] \} 
 .
\end{align}

We neglect back-reactions of the temporal component $\mathscr{V}_{0}$ on the other spatial components $\mathscr{V}_{j}$. 
Assuming an expansion around $\mathscr{V}_{j}=0$, $\Gamma_{k}^{(2)}:=\frac{\overrightarrow{\delta}}{\delta \mathscr{V}^\dagger} \Gamma_k \frac{\overleftarrow{\delta}}{\delta \mathscr{V}}$ is block-diagonal like the regulators. 
Then the flow equation can be decomposed into a sum of two contributions 
under our approximation:
\begin{align}
 \partial_t \Gamma_k
 =  \frac12 {\rm Tr} \left[ \left( \frac{1}{\Gamma_k^{(2)}+R_{k}} \right)_{\mu\nu} \cdot \partial_t R_{k,\mu\nu} \right]
 + \partial_t V_{T,M,k}
  ,
  \label{flow-eq3}
\end{align}
where the gluon regulator $R_{k,\mu\nu}$ is the $(\mu,\nu)$ component of the block-diagonal matrix $R_{k}$ in field space:
\begin{align}
  R_{k,00} =& R_{0,k} = Z_{0,k} R_{{\rm opt},k}(\bm{p}^2), \
\nonumber\\
  R_{k,0j} =& 0 = R_{k,j0} ,
\nonumber\\
  R_{k,j\ell} =& T_{j\ell}(\bm{p}) R_{T,k} 
= T_{j\ell}(\bm{p}) Z_{j,k}  R_{{\rm opt}, k_T}(\bm{p}^2) ,
\end{align}
where $T_{j\ell} := \delta_{j\ell} - \frac{p_jp_\ell}{p_m^2}$ is the transverse projection operator and 
$R_{\rm opt,k}(\bm{p}^2)$ is the ($D-1$ dimensional) optimized choice \cite{Litim00}:%
\footnote{
In our treatment, the difference between the three-dimensional RG scale $k_T$ and the four-dimensional one $k$ is neglected by equating two scales $k_T=k$ just for simplifying the analysis, since it is enough for obtaining a qualitative understanding for the transition. This is not be the case for obtaining quantitative results, see Appendix C of \cite{MP08} for the precise treatment on this issue. 
}
\begin{equation}
 R_{{\rm opt},k}(\bm{p}^2) = (k^2-\bm{p}^2) \theta(k^2-\bm{p}^2) .
\end{equation} 
The first term in the right-hand side of (\ref{flow-eq3}) encodes the quantum fluctuations of $\mathscr{V}_0$, while the second one encodes those of the other components of the gauge field and ghosts.  
In the present truncation, the second term is a total derivative with respect to $t$, and does not receive contributions from the first term.  Therefore, we can evaluate the flow of the second contribution, and use its output $V_{T,M,k}(\mathscr{V}_0)$ as an input for the remaining flow. 
\begin{align}
 \partial_t \Gamma_k
 =&  \frac12 T^{-1} \int \frac{d^{D-1}p}{(2\pi)^{D-1}} \left[ \left( \frac{1}{\Gamma_k^{(2)}
+R_k} \right)_{00} \partial_t R_{0,k} \right]
\nonumber\\&
 +   \partial_t V_{T,M,k} 
  ,
\end{align}

In the present case, we suppose that the remaining gluons $\mathscr{X}_\mu$ exhibit the massive behavior. Therefore, if $k$ decreases and becomes smaller than the gluonic mass $M$, the gluonic mass $M$ becomes dominant and persists in the limit $k \downarrow 0$, even if $M$ can depend on $k$.  In other words, the existence of the gluonic mass $M$ guarantees a stable renormalization group flow under good control. 
Yet, the ghost is still massless and the control of the infrared cutoff effect must be taken into account with great case. 
In this way, the simple ``one-loop type'' calculation based on the massive gluon picture can give rather better results than those expected from the naive  one-loop calculations in perturbation theory. 
However, this is totally different from the perturbation theory.
Indeed, the gluonic mass must be generated in the dynamical way, which is obviously a nonperturbative result. 
Thus, the justification of the one-loop calculation as a good approximation in the present case comes from the observation that the one-loop form in the presence of the infra cutoff function $R_k$ is a first approximation to the solution of the exact Wetterich equation, provided the gluonic mass exists.

In the presence of the infrared cutoff function $R_k$, the approximate solution of the Wetterich equation is given by the ``one-loop type'':
\begin{align}
 & V_{T,M,k}(\varphi) 
\nonumber\\
&= 
  \frac{D-1}{2} T \sum_{n \in \mathbb{Z}}  \int \frac{d^{D-1}p}{(2\pi)^{D-1}}   \ln [(\omega_n \pm T\varphi)^2 \nonumber\\
&+\bm{p}^2 + M^2 + R_k(p)] 
\nonumber\\
&-  \frac{1}{2} T \sum_{n \in \mathbb{Z}}  \int \frac{d^{D-1}p}{(2\pi)^{D-1}}   \ln [(\omega_n \pm T\varphi)^2+\bm{p}^2 + R_k(p) ] 
 ,
\end{align}
where we have introduced the same infrared cutoff function  $R_k$ for all modes. 
In what follows, we assume that the flow parameter $k$ dependence of the gluonic mass $M$ is negligible.
The $V_{T,M,k}(\varphi)$ is obtained by replacing the massless remaining gluon $X$ by the massive remaining gluon $X$ where the counting of the independent degrees of freedom for the massive vector field is different from the massless gauge field.
The $V_{T,M,k}$ is an improvement of the expression $V_{T,k}$  given for $D=4$ as eq.(52)  in the previous work  \cite{Kondo10}, which is recovered in the limit $M \to 0$.

Performing the Matsubara sum, we obtain the expression:
\begin{align}
& V_{T,M,k}(\varphi) 
\nonumber\\
&= 
  \frac{D-1}{2} T \int \frac{d^{D-1}p}{(2\pi)^{D-1}}   \ln [1+e^{-2 \beta \sqrt{\bm{p}^2 +M^2 +R_k(p)}} 
\nonumber\\ &
- 2 e^{- \beta \sqrt{\bm{p}^2 +M^2 +R_k(p) }} \cos \varphi ]
\nonumber\\
&-  \frac{1}{2} T \int \frac{d^{D-1}p}{(2\pi)^{D-1}}  \ln [1+e^{-2 \beta \sqrt{\bm{p}^2  +R_k(p)}}
\nonumber\\ &
- 2 e^{- \beta \sqrt{\bm{p}^2 +R_k(p) }} \cos \varphi ]
 .
\end{align}
In order to perform the momentum integration, we must specify the infrared cutoff function $R_k$. 
By choosing the infrared cutoff function of the optimal type, $R_k(p)=(k^2-p^2)\theta(k^2-p^2)$, we obtain
\begin{align}
 & V_{T,M,k}(\varphi) 
\nonumber\\
&= 
   \frac{D-1}{2} C_D T \int_{0}^{k} dp  \ p^{D-2}
  \ln [1+e^{-2 \beta \sqrt{k^2 +M^2 }} 
\nonumber\\ &
- 2 e^{- \beta \sqrt{k^2 +M^2  }} \cos \varphi] 
\nonumber\\
 &+  \frac{D-1}{2} C_D T \int_{k}^{\infty} dp  \ p^{D-2}   \ln [1+e^{-2 \beta \sqrt{p^2 +M^2}} 
\nonumber\\ &
- 2 e^{- \beta \sqrt{p^2 +M^2  }} \cos \varphi]
\nonumber\\
&-  \frac{1}{2} C_D T  \int_{0}^{k} dp  \ p^{D-2} 
\ln [1+e^{-2 \beta k}- 2 e^{- \beta k} \cos \varphi]
\nonumber\\
&-  \frac{1}{2} C_D T \int_{k}^{\infty} dp  \ p^{D-2}  \ln [1+e^{-2 \beta p} - 2 e^{- \beta p} \cos \varphi]
 .
\end{align}
Therefore, the $k$ dependent part is separated as 
\begin{align}
&  V_{T,k}(\varphi; M) 
\nonumber\\
&= V_{T,0}(\varphi; M) 
 +  \frac{D-1}{2} C_D T \int_{0}^{k} dp  \ p^{D-2}  
\nonumber\\ & \times
\{  \ln [1+e^{-2 \beta \sqrt{k^2 +M^2 }} - 2 e^{- \beta \sqrt{k^2 +M^2  }} \cos \varphi]
\nonumber\\
 &- \ln [1+e^{-2 \beta \sqrt{p^2 +M^2}} - 2 e^{- \beta \sqrt{p^2 +M^2  }} \cos \varphi] \}
\nonumber\\
&- \frac{1}{2} C_D T \int_{0}^{k} dp  \ p^{D-2}  \{  \ln [1+e^{-2 \beta k}- 2 e^{- \beta k} \cos \varphi]
\nonumber\\
&- \ln [1+e^{-2 \beta p}- 2 e^{- \beta p} \cos \varphi] \}
 . 
\end{align}
By introducing the dimensionless RG scale $\hat{k}$ and the dimensionless momentum $\hat{p}$ normalized by the temperature $T$:
\begin{equation}
 \hat{k} := k/T  , \quad \hat{p}:=p/T , 
\label{rescaling}
\end{equation}
this is rewritten for the dimensionless reduced potential defined by 
\begin{align}
\hat{V}_{\hat{k}}(\varphi; \hat{M}) 
:=  V_{T,k}(\varphi; M)/T^D ,
\end{align}
as 
\begin{align}
   \hat{V}_{\hat{k}}(\varphi; \hat{M}) 
 =& \hat{V}_{0}(\varphi; \hat{M}) 
 +  \frac{D-1}{2} C_D \int_{0}^{\hat{k}} d\hat{p}  \ \hat{p}^{D-2}   
\nonumber\\ & \times
\{  \ln [1+e^{-2 \sqrt{\hat{k}^2 +\hat{M}^2 }} - 2 e^{-  \sqrt{\hat{k}^2 +\hat{M}^2  }} \cos \varphi]
\nonumber\\
 &- \ln [1+e^{-2 \sqrt{\hat{p}^2 +\hat{M}^2}} - 2 e^{- \sqrt{\hat{p}^2 +\hat{M}^2  }} \cos \varphi ] \}
\nonumber\\
&- \frac{1}{2} C_D \int_{0}^{\hat{k}} d\hat{p}  \ \hat{p}^{D-2} \{  \ln [1+e^{-2 \hat{k}}- 2 e^{- \hat{k}} \cos \varphi ]
\nonumber\\
&- \ln [1+e^{-2 \hat{p}}- 2 e^{- \hat{p}} \cos \varphi ] \}
 . 
\end{align}

In what follows, we estimate contribution of the $k$-dependent terms to the deconfinement/confinement transition.
We expand the $k$-dependent part of $\hat{V}_{\hat{k}}(\varphi; \hat{M})$ into the power series in $\tilde\varphi:=\varphi - \pi$ around $\varphi=\pi$, using 
$\cos \varphi=\cos (\pi+\tilde \varphi)=-1+\frac12 \tilde\varphi^2-\frac{1}{24}\tilde\varphi^4+O(\tilde\varphi^4)$:
\begin{align}
   \hat{V}_{\hat{k}}(\varphi; \hat{M}) 
=& A_{0,k,\hat{M}} + \frac{A_{2,k,\hat{M}}}{2!} \tilde\varphi^2 + \frac{A_{4,k,\hat{M}}}{4!} \tilde\varphi^4 + O(\tilde\varphi^6)
 ,
\end{align}
where the coefficients $A_{2n,k,\hat{M}}$ are given by
\begin{align}
 & A_{0,k,\hat{M}}
\nonumber\\
=& A_{0,\hat{M}} + C_D \int_{0}^{\hat{k}} d\hat{p}  \ \hat{p}^{D-2}
\nonumber\\
& \times \Biggr[ (D-1) \left\{ \ln (1+e^{-\sqrt{\hat{k}^2 +\hat{M}^2 }}) - \ln (1+e^{-\sqrt{\hat{p}^2 +\hat{M}^2 }}) \right\}
\nonumber\\
&  - \left\{ \ln(1+e^{- \hat{k} }) - \ln(1+e^{- \hat{p} }) \right\} 
\Biggr]
 ,
\label{A0kM}
\end{align}
\begin{align}
 & A_{2,k,\hat{M}}
\nonumber\\
=& A_{2,\hat{M}} + 2! C_D \int_{0}^{\hat{k}} d\hat{p}  \ \hat{p}^{D-2} 
\nonumber\\
&
\times \Biggr[ \frac{D-1}{2} 
 \left\{ \frac{e^{-\sqrt{\hat{p}^2 +\hat{M}^2 }}}{(1+e^{-\sqrt{\hat{p}^2 +\hat{M}^2 }})^2} - \frac{e^{-\sqrt{\hat{k}^2 +\hat{M}^2 }}}{(1+e^{-\sqrt{\hat{k}^2 +\hat{M}^2 }})^2}  \right\} 
\nonumber\\
&- \frac{1}{2} \left\{ \frac{e^{- \hat{p} }}{(1+e^{- \hat{p} })^2} - \frac{e^{- \hat{k}    }}{(1+e^{- \hat{k}} )^2}  \right\} 
\Biggr]
 ,
\label{A2kM}
\end{align}
\begin{align}
 & A_{4,k,\hat{M}}
\nonumber\\
=& A_{4,\hat{M}} + 4! C_D \int_{0}^{\hat{k}} d\hat{p}  \ \hat{p}^{D-2} 
\nonumber\\ &
\times \Biggr[ \frac{D-1}{2} 
 \Biggr\{ \frac{e^{- \sqrt{\hat{p}^2 +\hat{M}^2 }} [1 -4 e^{- \sqrt{\hat{p}^2 +\hat{M}^2 }}+e^{-2 \sqrt{\hat{p}^2 +\hat{M}^2 }}] }{12[1+e^{- \sqrt{\hat{p}^2 +\hat{M}^2 }}]^4} \nonumber\\ &
- (\hat{p} \to \hat{k}) \Biggr\} 
\nonumber\\
&- \frac{1}{2} \left\{ \frac{e^{- \hat{p}} [1 -4 e^{- \hat{p}}+e^{-2 \hat{p}}] }{12[1+e^{- \hat{p}}]^4}  - (\hat{p} \to \hat{k})  \right\} 
\Biggr]
 .
\label{A4kM}
\end{align}
Here $(\hat{p} \to \hat{k})$ means that $\hat{p}$ in the preceding term is replaced by $\hat{k}$. 
In the limit $M \to 0$, these results reduce to those in Appendix B in the previous paper \cite{Kondo10}.

Suppose that the reduced effective potential $\hat{V}_{{\rm eff},k}^{\rm glue}$ is decomposed into two pieces:
\begin{equation}
 \hat{V}_{{\rm eff},k}^{\rm glue} 
:= V_{{\rm eff},k}^{\rm glue}/T^D 
 =   \hat{V}_{\hat{k}} +  \Delta \hat{V}_{\hat{k}} 
  ,
  \label{Veff}
\end{equation}
where the first part $\hat V_{\hat k}$ is the ($k$-dependent) ``perturbative part''  obtained essentially by the ``one-loop type'' calculation with the infrared regulator function $R_k$ being included,  
while the second part $\Delta \hat V_{\hat k}$ represents the ``non-perturbative part'' which is initially zero $\Delta \hat V_{\hat k}|_{k=\Lambda}=0$ and is generated in the evolution of the renormalization group.  
The non-perturbative part $\Delta \hat V_{\hat k}$ is obtained only by solving the flow equation in a numerical way and its analytical form is not available at present.

We expand $\hat{V}_{\hat{k}}(\varphi; \hat{M})$ in the power series in $\tilde\varphi=\varphi-\pi$: 
\begin{align}
 \hat{V}_{\hat{k}}(\varphi; \hat{M}) 
=& A_{0,k,\hat{M}} + \frac{A_{2,k,\hat{M}}}{2!} \tilde\varphi^2 + \frac{A_{4,k,\hat{M}}}{4!} \tilde\varphi^4 + O(\tilde\varphi^6)
 ,
  \label{VTcoeff}
\end{align}
where coefficients  $A_{2n,k,\hat{M}}$ can be drawn as functions of $k$ and $\hat{M}$.
 Suppose that $\Delta \hat V_{\hat{k}}(\varphi; \hat{M})$ is also of the form:
\begin{equation}
       \Delta \hat{V}_{\hat{k}}(\varphi; \hat{M})     
=    a_{0,k,\hat{M}} +  \frac{a_{2,k,\hat{M}}}{2} \tilde \varphi^2   + \frac{a_{4,k,\hat{M}}}{4!} \tilde \varphi^4 + O(\tilde \varphi^6)
 .
 \label{V_T}
\end{equation}
Then the effective potential has the expansion:
\begin{align}
 \hat{V}_{{\rm eff},k}^{\rm glue}(\varphi; \hat{M}) 
=& C_{0,k,\hat{M}} + \frac{C_{2,k,\hat{M}}}{2!} \tilde\varphi^2 + \frac{C_{4,k,\hat{M}}}{4!} \tilde\varphi^4 + O(\tilde\varphi^6)
 ,
  \label{Vehh2}
\end{align}
where 
$
 C_{n,k,\hat{M}} = A_{n,k,\hat{M}} + a_{n,k,\hat{M}}  
$.

After integrating over the fields other than the restricted field $\mathscr{V}_0$, we are lead to  the effective action of $\mathscr{V}_0$: 
\begin{align}
 & \Gamma_{k}[\mathscr{V}_0] 
=    \frac{1}{T} \int d^{D-1}x \left\{ \frac{1}{2}Z_{0,k} \partial_j \mathscr{V}_0(\bm{x})   \partial_j  \mathscr{V}_0(\bm{x})    + V_{{\rm eff},k}^{\rm glue}[\mathscr{V}_0] \right\}  ,
 \nonumber\\ 
 &  V_{{\rm eff},k}^{\rm glue}[\mathscr{V}_0] =  V_{T,k}[\mathscr{V}_0] + \Delta V_k[\mathscr{V}_0]   .
  \label{V_g}
\end{align}
Then the flow equation is reformulated for $\Delta V_k$ with the external input $V_{T,k}$:
\begin{equation}
   \partial_{t}  (\Delta V_k[\mathscr{V}_0])
 = \frac12   \int \frac{d^{D-1}p}{(2\pi)^3} \left[ \left( \frac{1}{\Gamma_k^{(2)}+R_{k}} \right)_{00} \partial_t R_{0,k} \right]
 ,
\end{equation}
where
\begin{equation}
\Gamma_{k}^{(2)}[\mathscr{V}_0] 
=   T^{-1} \left\{  Z_{0,k}   \bm{p}^2       + \partial_{\mathscr{V}_0}^2 V_{k}[\mathscr{V}_0] \right\}  
  .
\end{equation}

Using the specific infrared cutoff function: 
$
  R_{0,k}
= Z_{0,k} (k^2-\bm{p}^2)\theta(k^2-\bm{p}^2) 
$, 
which yields
\begin{align}
  \partial_t R_{0,k}
=  \left[ \partial_{t} Z_{0,k}   (k^2-\bm{p}^2)  + 2 Z_{0,k}  k^2 \right] \theta(k^2-\bm{p}^2)  
 ,
\end{align}
we can perform the momentum integration analytically: 
\begin{align}
 & \beta    \partial_{t}  (\Delta V_k[\mathscr{V}_0])    
\nonumber\\
=&  \frac23  \frac{1}{(2\pi)^2} \frac{  (1+\frac15 \eta_k )  k^5 }{     Z_k^{-1} g^2 \beta^2 \partial_{\varphi}^2 (V_{T,k}[\mathscr{V}_0] + \Delta V_k[\mathscr{V}_0] )  + k^2   }
 ,
\end{align}
where $\alpha_k$ is the running gauge coupling constant defined by   
\begin{equation} 
 g_k^2 := Z_{0,k}^{-1} g^2, \quad \alpha_k := \frac{g_k^2}{4\pi} =  Z_{0,k}^{-1} \frac{g^2}{4\pi}
  ,
\end{equation}
and $\eta_k$ is the anomalous dimension  defined by
\begin{equation}
 \eta_k := \partial_{t} \ln Z_{0,k} = - \partial_{t} \ln \alpha_k .
\end{equation}
The flow equation is simplified for the dimensionless effective potential $\hat{V}_{\hat k}$ and the dimensionless RG scale $\hat{k}$ as
\begin{equation}
     \partial_{\hat k}  \Delta \hat V_{\hat k}[\mathscr{V}_0]    
=    \frac{1}{6\pi^2} \frac{  (1+ \frac15 \eta_k )  \hat k^2 }{   1+\frac{4\pi\alpha_k}{\hat k^2} \partial_{\varphi}^2 (\hat V_{T,\hat k}[\mathscr{V}_0] + \Delta \hat V_{\hat k}[\mathscr{V}_0] )     }
 ,
 \label{flow-eq}
\end{equation}
where all scales are measured in units of temperature. 
The input for the flow equation is just a running gauge coupling constant $\alpha_k$, apart from the gluonic mass $M$ which is assumed to be independent of $k$. 

It is shown \cite{Kondo10} that the flow equation   for the effective potential $V_{{\rm eff},k}^{\rm glue}$ is reduced to a set of coupled flow equations for the coefficients in the effective potential (\ref{Veff}) combined with (\ref{VTcoeff}) and (\ref{V_T}):
\begin{subequations}
\begin{align}
    \partial_{\hat k}  a_{0,k,\hat{M}}     
=&  + \frac{1+\frac15 \eta_k }{6\pi^2} \frac{ \hat k^2 }{1+\frac{4\pi\alpha_k}{\hat k^2}   (A_{2,k,\hat{M}}+a_{2,k,\hat{M}})} 
 ,
\label{flow-coefficient-FRGa}
\\    \partial_{\hat k}  a_{2,k,\hat{M}}     
=&  -  \frac{1+\frac15 \eta_k    }{6\pi^2} \frac{  4\pi\alpha_k  (A_{4,k,\hat{M}}+a_{4,k,\hat{M}})   }{[1+\frac{4\pi\alpha_k}{\hat k^2}   (A_{2,k,\hat{M}}+a_{2,k,\hat{M}})]^2} 
 ,
\label{flow-coefficient-FRGb}
\\
\partial_{\hat k}  a_{4,k,\hat{M}}    
=& +   \frac{1+\frac15 \eta_k }{6\pi^2} \frac{6 [ 4\pi\alpha_k (A_{4,k,\hat{M}}+a_{4,k,\hat{M}})]^2 }{[1+\frac{4\pi\alpha_k}{\hat k^2}   (A_{2,k,\hat{M}}+a_{2,k,\hat{M}})]^3}
 ,
\label{flow-coefficient-FRG}
\nonumber\\
\vdots  
\end{align}
\end{subequations}

These equations are coupled first-order ordinary but nonlinear differential equations for coefficients $a_{n,k,\hat{M}}$, derived in Appendix~C of \cite{Kondo10}.
We see  that this form (\ref{V_T}) is justified as a solution of the flow equation. In fact, it is easy to see  that  
$\partial_{\hat k} a_{1,k,\hat{M}} =0$ and
$\partial_{\hat k} a_{3,k,\hat{M}}=0$ are guaranteed from the flow equation, if the effective potential has no odd power terms  at arbitrary $k$.
Therefore, if an initial condition, $a_{1,k,\hat{M}} =0=a_{3,k,\hat{M}}$ at $k=\Lambda$ is imposed, then   
$a_{1,k,\hat{M}} \equiv 0$ and $a_{3,k,\hat{M}} \equiv 0$ 
are maintained for any $k \in [0, \Lambda]$ by solving the flow equation.
For the given running gauge coupling constant $\alpha_k$ and the gluonic mass $M$, these equations can be in principle solved, since the coefficients $A_{n,k,\hat{M}}$ are given explicitly in (\ref{A2kM}) and (\ref{A4kM}) etc..   
In practice, however,  one must truncate the infinite series of differential equations (\ref{flow-coefficient-FRG}) up to some finite order to obtain manageable set of equations, even if we perform numerical calculations.

We can understand qualitatively why a 2nd order phase transition from the deconfinement phase to the confinement phase can occur by lowering the temperature.
At a certain temperature $T$, the flow starts from the ``one-loop type'' result without non-perturbative part at $k=\Lambda \gg 1$.
Therefore, $a_{n,k,\hat{M}}=0$ at $k=\Lambda \gg 1$ and hence
$C_{2,k,\hat{M}}=A_{2,k,\hat{M}}$ at $k=\Lambda \gg 1$.  Moreover, we assume $C_{4,k,\hat{M}}=A_{4,k,\hat{M}}+a_{4,k,\hat{M}}>0$ for $0 \le k \le \Lambda$, as a necessary condition for realizing a 2nd order transition. Otherwise, we must consider the higher-order terms, e.g. $O(\varphi^6)$.%
\footnote{
 This assumption is assured to be true by  numerical calculations of the full effective potential \cite{MP08,BGP10}, as reproduced in the previous section.    
}
This assumption allows us to analyze just one differential equation for obtaining qualitative understanding.
Then the right-hand side of (\ref{flow-coefficient-FRGb}) is negative for any $k \in [0,\Lambda]$, since the running coupling constant $\alpha_k$ is positive and $1+\frac15 \eta_k$ is positive, see Fig.6 
and Fig.7 
 of \cite{Kondo10}. 
Consequently, $a_{2,k}$ started at zero becomes positive $a_{2,k}>0$ just below $\Lambda$ and increases (monotonically) as $k$ decreases.
See Fig.8  
  of \cite{Kondo10} for the massless case $\hat{M}=0$.
  Thus, the flow always moves in the direction enhancing confinement. 
If we consider the low temperature $T<T_d$ where  $A_{2,k,\hat{M}}>0$, then the final result is always 
  $C_{2,k,\hat{M}}>0$ at $k=0$.
Even if we start from $A_{2,k,\hat{M}}<0$ at temperature $T$ slightly above $T_d$, $T>T_d$ where $C_{2,k,\hat{M}}=A_{2,k,\hat{M}}+a_{2,k,\hat{M}}<0$ at $k=\Lambda \gg 1$, it may happen that  $C_{2,k,\hat{M}}=A_{2,k,\hat{M}}+a_{2,k,\hat{M}}>0$ at $k=0$.
But, the FRG improvement does not change the above conclusions in an essential manner. 
Thus, we conclude without the detailed numerical calculations that the above $T_d$ gives a lower bound on the true critical temperature $T_c$, since the flow evolves towards enhancing the confinement, under the assumption that $M$ does not change so much along the flow. 

Finally, we mention the pressure in the FRG. The pressure $P_{k}(T)$ at the flow parameter $k$ is 
defined in the low-temperature confined phase by $P_{k}=-\hat{V}_{{\rm eff},k}^{\rm glue}(\varphi=\varphi_{\rm min}=\pi) = - C_{0,k,\hat{M}} = - A_{0,k,\hat{M}}-a_{0,k,\hat{M}}$.
The initial condition is  $a_{0,k,\hat{M}}=0$ at $k=\Lambda \gg 1$ or 
$C_{0,k,\hat{M}}=A_{0,k,\hat{M}}$ at $k=\Lambda \gg 1$. 
Then, the flow of the pressure $P_{k}$ is determined from the behavior of $a_{0,k,\hat{M}}$ governed by (\ref{flow-coefficient-FRGa}).
In the low-temperature confined phase, the right-hand side of (\ref{flow-coefficient-FRGa}) is positive, since $C_{2,k,\hat{M}}=A_{2,k,\hat{M}}+a_{2,k,\hat{M}}>0$.
This yields the positivity of the derivative, $\partial_{\hat k} a_{0,k,\hat{M}}>0$, that is to say, $a_{0,k,\hat{M}}$ is monotonically decreasing in decreasing $k$. Therefore, $P_{k}$ is monotonically increasing in decreasing $k$ due to the FRG improvement and finally reaches the largest value for the true pressure $P$ at $k=0$. 
Therefore, the true pressure can be positive, even if the initial pressure is negative in the initial approximation. 
See Fig.~\ref{fig:P-SU2-a}.
The $P_{k}(T)$ increases more rapidly at the temperature $T$ closer to the critical temperature $T_d$ than that at the low temperature $T \ll T_d$, since $C_{2,k,\hat{M}}(T)=A_{2,k,\hat{M}}(T)+a_{2,k,\hat{M}}(T)$ becomes smaller as the temperature is closer to the critical temperature, leading to the larger value of the derivative $\partial_{\hat k}  a_{0,k,\hat{M}}$ according to (\ref{flow-coefficient-FRGa}). 
In this way, the FRG  will improve the positivity violation of the entropy in the first approximation near the critical temperature in the low-temperature confined phase. 
See Fig.~\ref{fig:P-SU2-b}.
This tendency agrees with the  two-loop improvement of the one-loop result \cite{RSTW15b} and 
and is consistent with the other FRG analysis \cite{FP13}.

\subsection{$SU(2)$ Pressure and entropy}\label{section:pressure}

Our effective potential is given by
\begin{align}
&  V^{\rm K}_{\rm eff}(\varphi) 
\nonumber\\
=&   (D-1)  T C_D\int_{0}^{\infty} dp \ p^{D-2} \ln (1+e^{-2\sqrt{p^2+M^2}/T} 
\nonumber\\ &
- 2 e^{-\sqrt{p^2+M^2}/T} \cos \varphi ) 
\nonumber\\ &
 -  T C_D\int_{0}^{\infty} dp \ p^{D-2} \ln (1+e^{-2p/T} - 2 e^{-p/T} \cos \varphi ) .
\end{align}
This should be compared with  the other  effective potential $V^{\rm RSTW}_{\rm eff}(\varphi)$ obtained in \cite{RSTW15}. 
One finds that the effective potential $ V^{\rm RSTW}_{\rm eff}(\varphi)$ contains  two extra terms: In fact, the difference is given by
\begin{align}
 & V^{\rm RSTW}_{\rm eff}(\varphi) - V^{\rm K}_{\rm eff}(\varphi) 
\nonumber\\ 
=& (D-1) T C_D\int_{0}^{\infty} dp \ p^{D-2} \ln (1-e^{-\sqrt{p^2+M^2}/T}) 
\nonumber\\ &
 -  T C_D\int_{0}^{\infty} dp \ p^{D-2} \ln (1-e^{-p/T}) ,
 \label{V-difference}
\end{align}
where the first and second terms come from the neutral (or the diagonal) component for the gluon and ghost respectively. 
(The reason of this difference is explained below.)  However, both effective potentials give the same value for the critical ratio $M/T_d$. (Notice that the dimensionless effective potential $V_{\rm eff}(\varphi)/T^D$ is written in terms of the dimensionless quantity $M/T$.) 
This is understood as follows. 
The critical value $T_d$ is determined by the Polyakov loop average so that it separates the confined phase $L=0$ in low temperature $T<T_d$ and the  deconfined phase $L \ne 0$ in high temperature $T>T_d$. 
The Polyakov loop average $L$ is calculated as $L=\cos \frac{\varphi_{\rm min}}{2}$ using the value $\varphi_{\rm min}$ which gives the minimum of the effective potential $V_{\rm eff}$:   the derivative is equal to zero at 
 $\varphi_{\rm min}$: 
\begin{align}
 V^\prime_{\rm eff}(\varphi_{\rm min}) := \frac{\partial V_{\rm eff}(\varphi)}{\partial \varphi}|_{\varphi=\varphi_{\rm min}}=0 .
\end{align}
But, the extra terms (\ref{V-difference}) do not depend on $\varphi$, and therefore do not change the location of the minimum $\varphi_{\rm min}$. Thus we obtain the same value $\varphi_{\rm min}$, which implies the same value for the Polyakov  loop average $L=\cos \frac{\varphi_{\rm min}}{2}$ for a given ratio $T/M$. Thus we obtain the same critical value of the ratio $M/T_d$ for the two different effective potentials. 

However, the two effective potentials have the different minimum values even at the same value  $\varphi_{\rm min}$ due to the  $\varphi$-independent extra terms (\ref{V-difference}).

\begin{figure}[ptb]
\begin{center}
\includegraphics[scale=0.75]{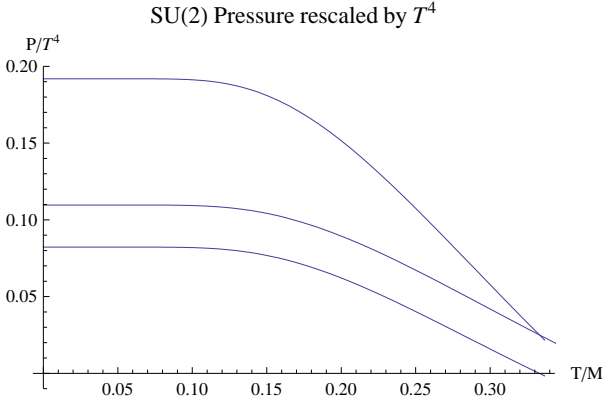}
\end{center}
\vskip -0.3cm
\caption{
The pressure $P$ (rescaled by $T^4$) as a function of $T/M$ in the low-temperature confined phase for $SU(2)$ and $D=4$.
The top line is our pressure $P^{\rm K}/T^4$, the bottom line is the RSTW pressure $P^{\rm RSTW}/T^4$ and the middle line denote the difference of the two pressures $P^{\rm K}/T^4-P^{\rm RSTW}/T^4$. 
Here the critical value is $T_d/M=0.33643$.
}
\label{fig:P-SU2-a}
\end{figure}

The pressure $P$ is defined 
through the temperature-dependent minimum value of the effective potential:
\begin{align}
 P(T) := - V_{\rm eff}(\varphi=\varphi_{\rm min}(T))
 .
\end{align}
The extra terms give the $T/M$ dependent shift for the pressure $P$. 
In our case, we have 
\begin{align}
 \frac{P^{\rm K}}{T^D}
= 
 - (D-1) F_{\hat{M}}(\varphi_{\rm min}) + F_{0}(\varphi_{\rm min}) 
 ,
\end{align}
where $F_{\hat{M}}(\varphi)$ is defined by (\ref{F-def1}) and (\ref{F-def2}). 
In the low-temperature confined phase, $\varphi_{\rm min}=\pi$ leads to the pressure: 
\begin{align}
  P^{\rm K} 
=&  -2 (D-1)  T C_D\int_{0}^{\infty} dp \ p^{D-2} \ln (1 + e^{-\sqrt{p^2+M^2}/T}  ) 
\nonumber\\ &
 + 2   T C_D\int_{0}^{\infty} dp \ p^{D-2} \ln (1 + e^{-p/T}   ) .
\end{align}
In the low-temperature limit $\hat{M}:=M/T \to \infty$, the minimum of the effective potential is given at $\varphi_{\rm min}=\pi$, which yields the positive value:
\begin{align}
 & \frac{P^{\rm K}}{T^D}  
  \to  - (D-1) F_{\infty}(\pi) + F_{0}(\pi) 
=  F_{0}(\pi)
\nonumber\\ &
\Longrightarrow  
 2 \int_{0}^{\infty} \frac{d\hat{p}}{2\pi^2} \ \hat{p}^{2} \ln (1+e^{-\hat{p}})
=  \frac{7}{360} \pi^2  \simeq 0.191909 \ (D=4) 
 ,
\end{align}
which is the behavior for the gas of free relativistic massless fermions with 2 internal degrees of freedom. 
See Fig.~\ref{fig:P-SU2-a}.
In the high-temperature limit $\hat{M}:=M/T \to 0$, the minimum of the effective potential is given at $\varphi_{\rm min}=0$, which yields the positive value:
\begin{align}
& \frac{P^{\rm K}}{T^D}  
  \to - (D-1) F_{0}(0) + F_{0}(0) 
= (2-D) F_{0}(0)
\nonumber\\ &
\Longrightarrow 
-2 \times 2\int_{0}^{\infty} \frac{d\hat{p}}{2\pi^2} \ \hat{p}^{2} \ln (1-e^{-\hat{p}})
\nonumber\\ &
= -2 \times \frac{-1}{45} \pi^2 
= \frac{2}{45}\pi^2 \simeq 0.438649 \ (D=4)
 ,
\end{align}
which is the behavior for the gas of free relativistic massless bosons with $2(D-2)$ internal degrees of freedom. 

The difference between ours and RSTW is $\varphi_{\rm min}$ independent, and 
 depends on the ratio $\hat M:=M/T$ alone for a given dimension $D$:
\begin{align}
 & \frac{P^{\rm K}}{T^D} - \frac{P^{\rm RSTW}}{T^D}     
\nonumber\\
=&  [V^{\rm RSTW}_{\rm eff}(\varphi=\varphi_{\rm min}) - V^{\rm K}_{\rm eff}(\varphi=\varphi_{\rm min})]/T^D   
\nonumber\\
=& C_D \Big[ (D-1 )  \int_{0}^{\infty} d\hat{q} \ \hat{q}^{D-2} \ln (1-e^{-\sqrt{\hat{q}^2+\hat{M}^2}}) 
\nonumber\\ & \quad\quad 
 -   \int_{0}^{\infty} d\hat{q} \ \hat{q}^{D-2} \ln (1-e^{-\hat{q}}) \Big]
 .
\end{align}
For $D=4$, the difference is given by
\begin{align}
  \frac{P^{\rm K}}{T^4} - \frac{P^{\rm RSTW}}{T^4}     
=& \frac{1}{2\pi^2} \Big[ 3  \int_{0}^{\infty} d\hat{q} \ \hat{q}^{2} \ln (1-e^{-\sqrt{\hat{q}^2+\hat{M}^2}}) 
\nonumber\\ & \quad\quad 
 -   \int_{0}^{\infty} d\hat{q} \ \hat{q}^{2} \ln (1-e^{-\hat{q}}) \Big]
 .
 \label{difference-D4}
\end{align}
In the low-temperature confined phase $T<T_d$, both pressures $P^{\rm K}/T^4$ and $P^{\rm RSTW}/T^4$ increase monotonically in $\hat{M}=M/T$ or decrease monotonically  in $T/M$. (This is not the case in the high-temperature deconfined phase.) 
The difference (\ref{difference-D4}) is monotonically increasing in $\hat{M}=M/T$ or monotonically decreasing in $T/M$ in both phases. 
In the low temperature $\hat{M}:=M/T \gg 1$,  
the difference is positive: 
\begin{align}
 \frac{P^{\rm K}}{T^4} - \frac{P^{\rm RSTW}}{T^4}     
=& \frac{1}{2\pi^2} \Big[
 -   \int_{0}^{\infty} d\hat{q} \ \hat{q}^{2} \ln (1-e^{-\hat{q}}) \Big]
\nonumber\\
=& \frac{1}{90} \pi^2  \simeq 0.109662 \ \text{for} \ D=4 
 .
\end{align}
In the high temperature $\hat{M}:=M/T \ll 1$, incidentally,  
the difference is negative: 
\begin{align}
 \frac{P^{\rm K}}{T^4} - \frac{P^{\rm RSTW}}{T^4}     
=& \frac{1}{2\pi^2} \Big[
 2 \int_{0}^{\infty} d\hat{q} \ \hat{q}^{2} \ln (1-e^{-\hat{q}}) \Big]
\nonumber\\
=& - \frac{1}{45} \pi^2  \simeq - 0.219325 \ \text{for} \ D=4 
 .
\end{align}

The RSTW pressure $P^{\rm RSTW}$ at one loop violate  slightly the positivity before reaching the critical temperature as the temperature is increased in the low-temperature confinement phase. 
Indeed,  the positivity is maximally violated $P^{\rm RSTW}/T^4 =-0.00161342$  at the critical temperature $T_d/M=0.33643$, while  our pressure is positive $P^{\rm K}/T^4 =0.0217697$ even at the critical temperature where the difference is $0.0233832$.
Thus, our pressure $P^{\rm K}$ in the initial approximation remains positive  in the low-temperature confined phase, in sharp contrast to the RSTW pressure at one loop.

This difference is understood as follows. 
Our mass term does not agree with the mass term introduced in \cite{RSTW15} even after fixing the gauge as performed in section III. 
The two mass terms have different independent degrees of freedom. 
Our mass term is written in terms of the remaining field $\mathscr{X}_\mu$ alone which are charged. 
This fact for the independent degrees of freedom for the remaining field yields the ghost fields different from those in \cite{RSTW15}. 
The authors of \cite{RSTW15} introduce the neutral ghost associated with the neutral or diagonal gauge field, in addition to the charged ghosts associated to the charged gauge fields.  
Notice that the neutral ghost gives the negative contribution to the pressure, while the charged ghosts give the positive contribution to the pressure.
However, in the gauge fixed version of our formulation, the neutral ghost is prohibited to be included and is not introduced even after fixing the gauge for the remaining fields. 
This is because the remaining field $\mathscr{X}_\mu(x)$ is required  be orthogonal to the color field $\bm{n}(x)$, namely, the defining equation (II) $\mathscr{X}_\mu(x) \cdot \bm{n}(x)=0$ in section II must be imposed, and hence the ghost fields associated with the remaining fields must be also orthogonal to the color field, which means that the ghost fields must be  charged, in other words, the neutral ghost, i.e., the component of the ghost field which is parallel to the color field must be vanishing.
(See \cite{KMS05}, section 4.6 and Appendix E for $SU(2)$, and section 5.8 and Appendix H in \cite{KKSS15} for $SU(N)$.) 
Therefore, the remaining fields and the associated ghost fields have less independent degrees of freedom than those in \cite{RSTW15}.
In fact, this difference avoids the violation of positivity of the pressure in the low-temperature confined phase.

\begin{figure}[ptb]
\begin{center}
\includegraphics[scale=0.60]{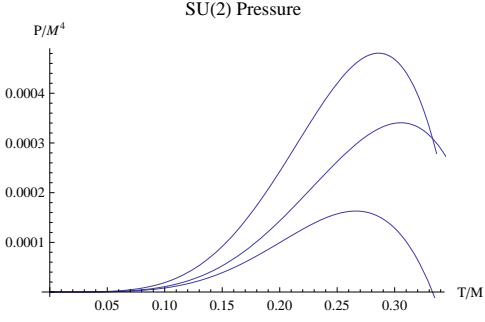}
\end{center}
\vskip -0.3cm
\caption{
The pressure $P$ (rescaled by $M^4$) as a function of $T/M$ in the low-temperature confined phase for $SU(2)$ and $D=4$.
The top line is our pressure $P^{\rm K}/M^4$, the bottom line is the RSTW pressure $P^{\rm RSTW}/M^4$ and the middle line denote the difference of the two pressures $P^{\rm K}/M^4-P^{\rm RSTW}/M^4$. 
Here the critical value is $T_d/M=0.33643$.
}
\label{fig:P-SU2-b}
\end{figure}

In view of these, our initial approximation is relatively good.
Of course, the first approximation is not enough to derive all essential aspects of the finite temperature Yang-Mills theory. 
Some results are to be improved to avoid the artifacts of the initial approximation. 
For instance, we consider the entropy density $\mathcal{S}$ defined by
\begin{align}
 \mathcal{S}(T) := \frac{dP(T)}{dT} 
 .
\end{align}
We observe that the pressure $P$ is increasing at small temperature and hence the entropy $\mathcal{S}(T)$ is positive. However, as the temperature is increased, the pressure changes its monotony and begins to decrease, indicating that the entropy becomes negative in the region $T_0 < T < T_d$ before reaching the critical temperature $T_d$, which was pointed out in \cite{RSTW15b}.
See Fig.~\ref{fig:P-SU2-b}.
Here 
$T_0/M=0.286$ for our case and $T_0/M=0.266$  for RSTW.

The positivity of the entropy density $\mathcal{S}(T)$ is equal to the  monotonic increase of the pressure $P(T)$ with respect to the temperature $T$, which means that the minimum $V_{\rm eff}(L_{\rm min})$ of the effective potential $V_{\rm eff}(L)$ is monotonically decreasing and the bottom becomes deeper as the temperature increases. The vacuum energy is further lowered by existence of more non-vanishing vacuum condensations. This suggests that $V_{\rm eff}(L)$ is insufficient to examine the minimum value near the critical temperature in the low-energy confined phase and is to be replaced by the simultaneous effective potential $V_{\rm eff}(\Phi,L)$ of $\Phi$ and $L$, since the non-vanishing temperature-dependent condensate $\Phi$ defined by (\ref{dim-2-vc}) will lower the vacuum energy to give a different temperature dependence for the pressure.  
The gauge-invariant gluonic mass $M$ obtained from $\Phi$ could be related to the glueball mass, see \cite{KOSSM06}. Taking such dynamical degrees of freedom for glueballs into consideration is expected to eliminate the artifact of considering $L$ alone to recover the positive entropy, i.e. monotonicity of the pressure.  Indeed, there exist other works suggesting that the glueball degrees of freedom reproduce the expected thermodynamic behaviors of the Yang-Mills theory, see e.g., \cite{SR12,CNP15,BEFKS12}.  
The result of the effective potential $V_{\rm eff}(\Phi, L)$ will be reported in a subsequent work.


\section{$SU(3)$ Yang-Mills theory}

\setcounter{equation}{0}

\subsection{Existence of $SU(3)$ confinement/deconfinement transition}

\begin{figure}[tb]
\begin{center}
\includegraphics[scale=0.32]{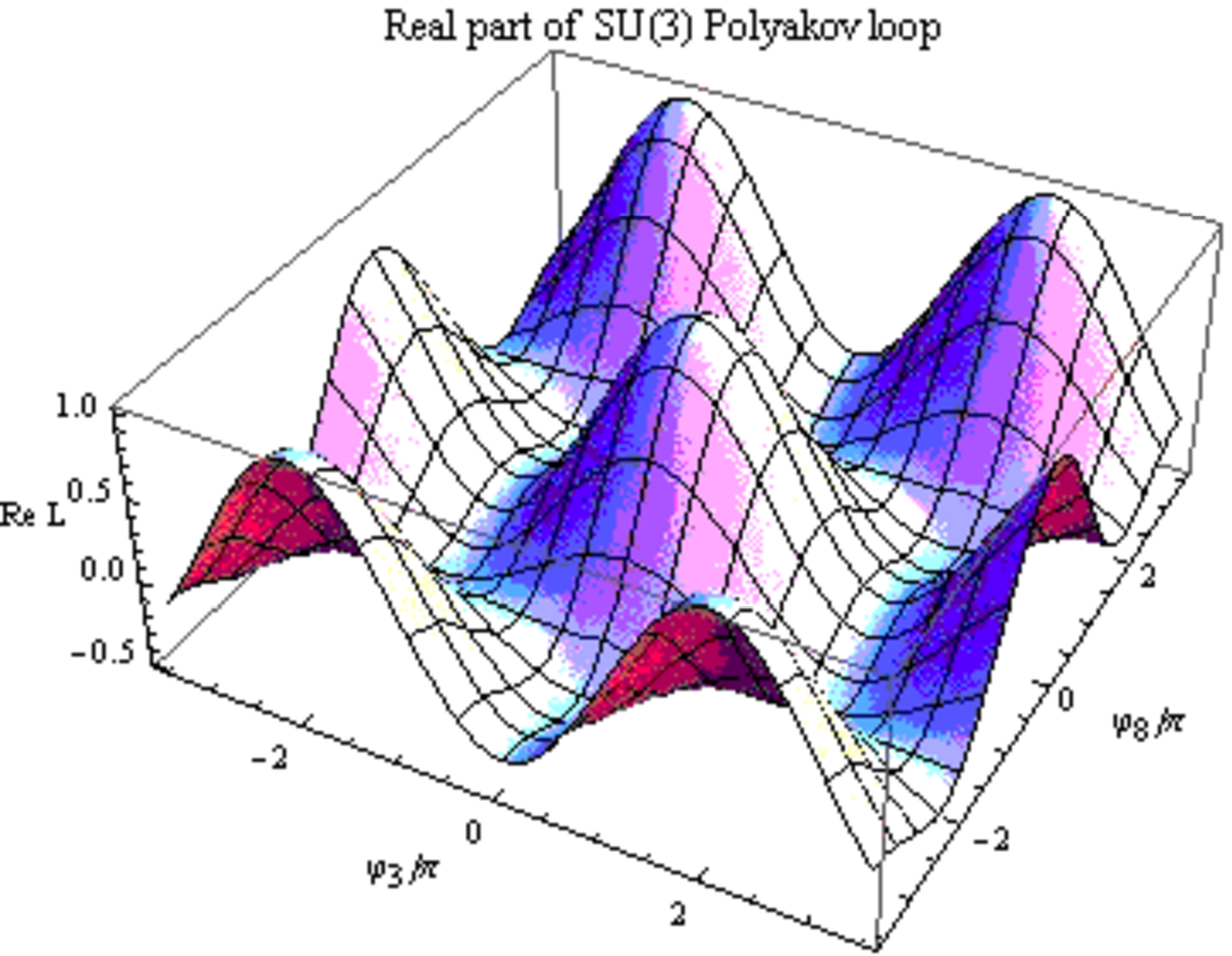}
\quad
\includegraphics[scale=0.32]{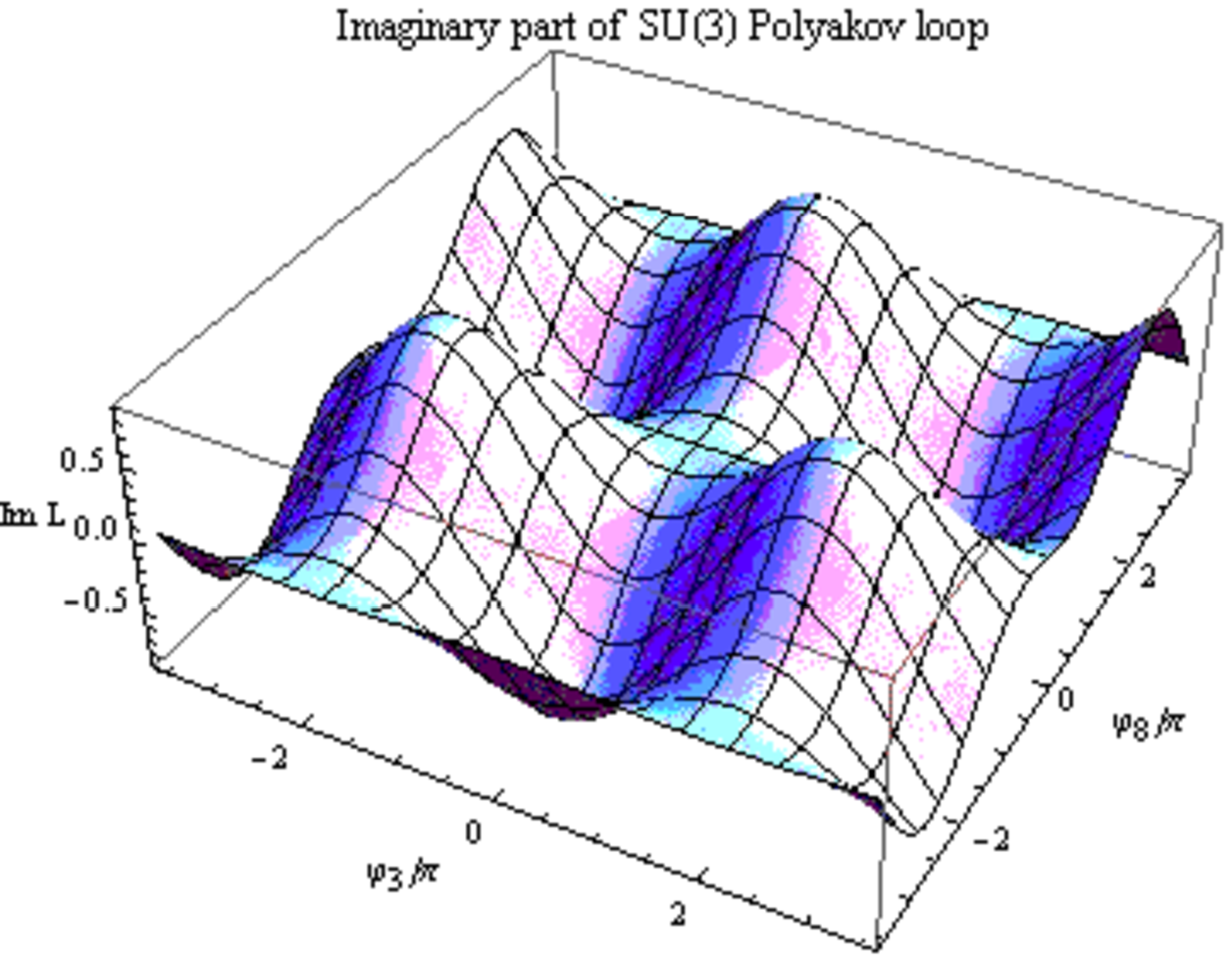}
\quad
\includegraphics[scale=0.27]{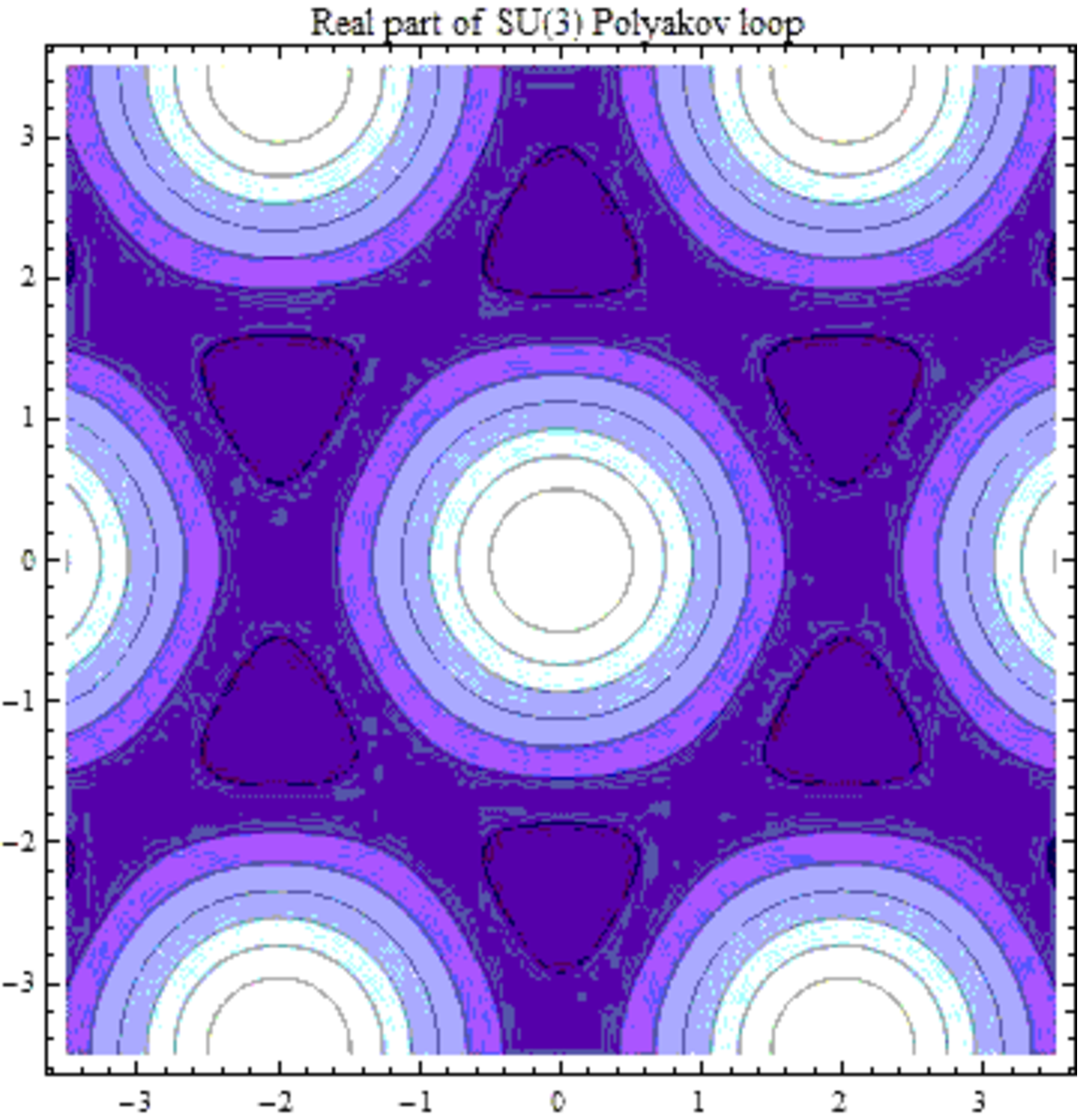}
\quad
\includegraphics[scale=0.27]{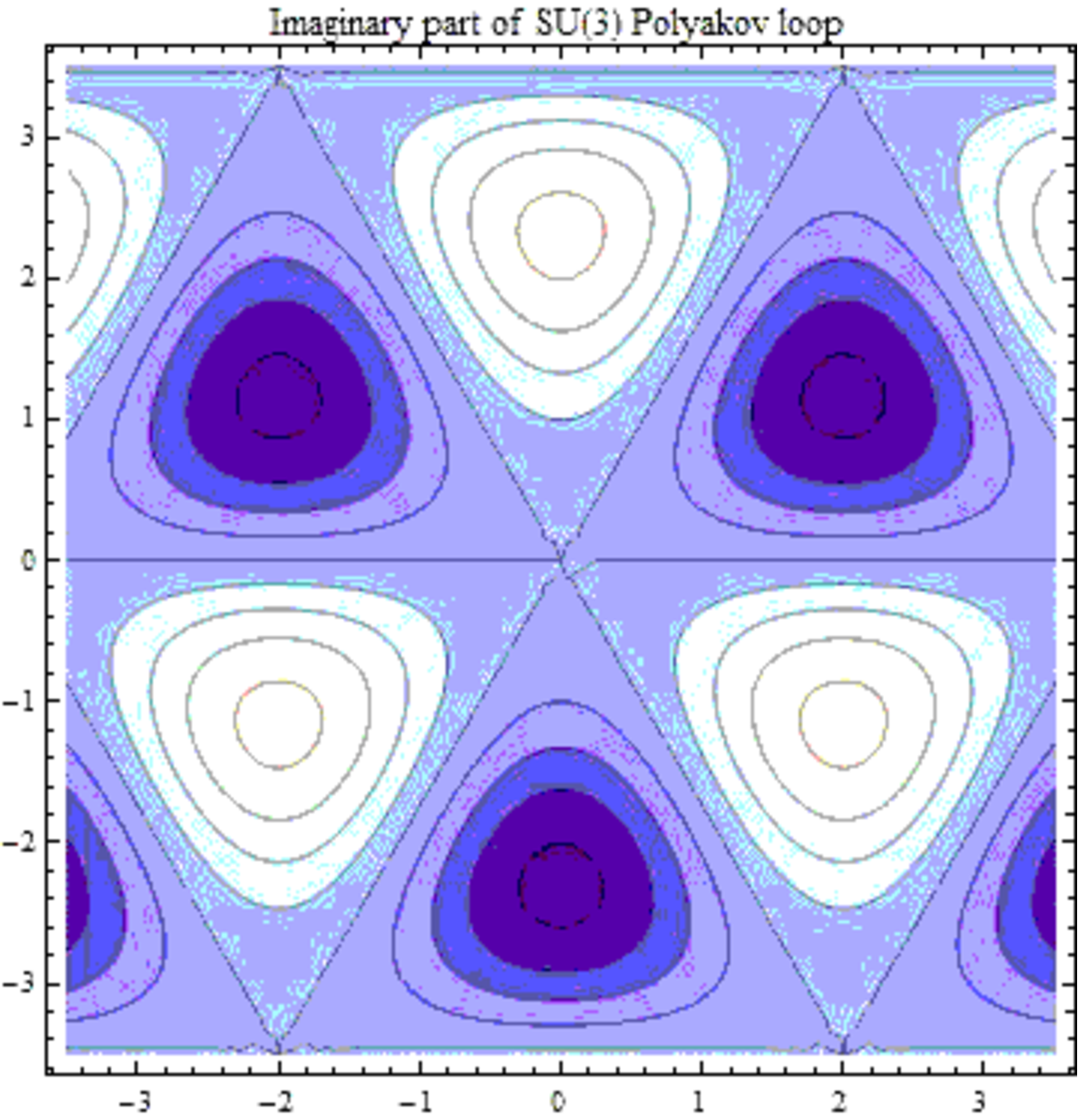}
\end{center}
\vskip -0.5cm
\caption{
3D plot and contour plot of the $SU(3)$ Polyakov loop as a function of the two angles $\varphi_3/\pi$ and $\varphi_8/\pi$: 
(Left) Real part, $Re L$,
(Right) Imaginary part, $Im L$.
}
\label{fig:V-SU3-PL}
\end{figure}

Symmetries of the $SU(3)$ Polyakov loop operator $L$ are as follows: 
See Fig.~\ref{fig:V-SU3-PL}. 
\begin{enumerate}
\item[i)] periodicity of $4\pi$ in the $\varphi_{3}$ direction and $4\sqrt{3}\pi$ in the $\varphi_{8}$ direction:
\begin{align}
  L(\varphi_{3},\varphi_{8}) 
 =& L(\varphi_{3}+4\pi,\varphi_{8}) 
=  L(\varphi_{3},\varphi_{8}+4\sqrt{3}\pi ) 
 ,
 \nonumber\\
\Longrightarrow  Re L(\varphi_{3},\varphi_{8}) 
 =& Re L(\varphi_{3}+4\pi,\varphi_{8}) 
\nonumber\\
=& Re L(\varphi_{3},\varphi_{8}+4\sqrt{3}\pi ) ,
 \nonumber\\
  Im L(\varphi_{3},\varphi_{8}) 
 =& Im L(\varphi_{3}+4\pi,\varphi_{8}) 
\nonumber\\
=& Im L(\varphi_{3},\varphi_{8}+4\sqrt{3}\pi ) ,
\end{align}

\item[ii)] reflection symmetry:
\begin{align}
 & L(\varphi_{3},\varphi_{8}) = L(-\varphi_{3},\varphi_{8}) 
 ,
 \nonumber\\
\Longrightarrow &  Re L(\varphi_{3},\varphi_{8}) =  Re L(-\varphi_{3},\varphi_{8}) , 
\nonumber\\ &
 Im L(\varphi_{3},\varphi_{8}) =  Im L(-\varphi_{3},\varphi_{8}) 
 ,
\end{align}
and
\begin{align}
& L(\varphi_{3},\varphi_{8})^* = L( \varphi_{3},-\varphi_{8}) 
 ,
 \nonumber\\
\Longrightarrow & Re L(\varphi_{3},\varphi_{8})  = Re L(\varphi_{3},-\varphi_{8}) , 
\nonumber\\ &
    Im L(\varphi_{3},\varphi_{8})  = - Im L(\varphi_{3},-\varphi_{8}) ,
\end{align}

\item[iii)]  global color symmetry:
\begin{align}
 & L(\varphi_{3}^\prime,\varphi_{8}^\prime)  = L(-\varphi_{3},-\varphi_{8}) ,
 \nonumber\\
\Longrightarrow & Re L(\varphi_{3}^\prime,\varphi_{8}^\prime)  = Re L(\varphi_{3},\varphi_{8}) , 
\nonumber\\ &
    Im L(\varphi_{3}^\prime,\varphi_{8}^\prime)  = - Im L(\varphi_{3},\varphi_{8}) ,
\end{align}
where $(\varphi_{3}^\prime,\varphi_{8}^\prime)$ is obtained from $(\varphi_{3},\varphi_{8})$ by a rotation of angle $\pm \pi/3$: 
\begin{align}
  \begin{bmatrix}
  \varphi_3^\prime \\
  \varphi_8^\prime
  \end{bmatrix} 
 =& \begin{bmatrix}
   \cos \frac{\pi}{3} & \pm \sin \frac{\pi}{3} \\
  \mp \sin \frac{\pi}{3} & \cos \frac{\pi}{3} 
   \end{bmatrix} 
  \begin{bmatrix}
  \varphi_3  \\
  \varphi_8 
  \end{bmatrix} 
 = \begin{bmatrix}
    \frac{1}{2} & \pm  \frac{\sqrt{3}}{2} \\
   \mp  \frac{\sqrt{3}}{2} & \frac{1}{2} 
   \end{bmatrix} 
  \begin{bmatrix}
  \varphi_3  \\
  \varphi_8 
  \end{bmatrix} 
 .
 \label{60-rotation}
\end{align}
The transformation (\ref{60-rotation}) is equal to 
\begin{align}
  \varphi_3^\prime =& 
\begin{cases}
\frac{1}{2} \varphi_3  +  \frac{\sqrt{3}}{2} \varphi_8 \\
\frac{1}{2} \varphi_3  -  \frac{\sqrt{3}}{2} \varphi_8    \end{cases}
 ,
\  \varphi_8^\prime 
= \begin{cases}
 -  \frac{\sqrt{3}}{2} \varphi_3 +  \frac{1}{2} \varphi_8 \\
 + \frac{\sqrt{3}}{2} \varphi_3 +  \frac{1}{2} \varphi_8
  \end{cases} ,
\end{align}
which leads to
\begin{align}
 \frac{-2}{\sqrt{3}} \varphi_8^\prime 
=& \begin{cases}
 - \left( - \varphi_3  +  \frac{1}{\sqrt{3}} \varphi_8 \right) \\
 - \left(   \varphi_3  +  \frac{1}{\sqrt{3}} \varphi_8 \right) 
  \end{cases} ,
\nonumber\\ 
   \varphi_3^\prime  +  \frac{1}{\sqrt{3}} \varphi_8^\prime 
=&  \begin{cases}
- \left( - \frac{2}{\sqrt{3}} \varphi_8 \right) \\
 - \left( - \varphi_3  +  \frac{1}{\sqrt{3}} \varphi_8 \right)
    \end{cases} ,
\nonumber\\ 
  - \varphi_3^\prime + \frac{1}{\sqrt{3}} \varphi_8^\prime 
=&  \begin{cases}
 - \left(   \varphi_3  +  \frac{1}{\sqrt{3}} \varphi_8 \right)  \\
 - \left( - \frac{2}{\sqrt{3}} \varphi_8 \right)
    \end{cases} .
\end{align}

\end{enumerate}
 It is easy to see that the Polyakov loop operator (\ref{SU(3)-PL}) respects all the symmetries i), ii) and iii).

Symmetries of the  $SU(3)$  effective potential $V_{\rm eff}(\varphi_{3},\varphi_{8})$ are as follows \cite{RSTW15}: 
\begin{enumerate}
\item[i)] periodicity of $4\pi$ in the $\varphi_{3}$ direction and $4\pi/\sqrt{3}$ in the $\varphi_{8}$ direction:
\begin{align}
  V_{\rm eff}(\varphi_{3},\varphi_{8}) 
 = V_{\rm eff}(\varphi_{3}+4\pi,\varphi_{8}) 
 = V_{\rm eff}(\varphi_{3},\varphi_{8}+4\pi/\sqrt{3}) 
 ,
\end{align}

\item[ii)]  charge conjugation invariance: 
\begin{align}
  V_{\rm eff}(\varphi_{3},\varphi_{8}) 
 = V_{\rm eff}(-\varphi_{3},-\varphi_{8}) 
 = V_{\rm eff}(-\varphi_{3},\varphi_{8}) 
 ,
\end{align}

\item[iii)] global color symmetry:
\begin{align}
  V_{\rm eff}(\varphi_{3},\varphi_{8}) 
 = V_{\rm eff}(\varphi_{3}^\prime,\varphi_{8}^\prime) 
 ,
\end{align}
where $(\varphi_{3}^\prime,\varphi_{8}^\prime)$ is obtained from $(\varphi_{3},\varphi_{8})$ by a rotation of angle $\pm \pi/3$: 
\begin{align}
  \begin{bmatrix}
  \varphi_3^\prime \\
  \varphi_8^\prime
  \end{bmatrix} 
 =& \begin{bmatrix}
   \cos \frac{\pi}{3} & \pm \sin \frac{\pi}{3} \\
  \mp \sin \frac{\pi}{3} & \cos \frac{\pi}{3} 
   \end{bmatrix} 
  \begin{bmatrix}
  \varphi_3  \\
  \varphi_8 
  \end{bmatrix} 
 = \begin{bmatrix}
    \frac{1}{2} & \pm  \frac{\sqrt{3}}{2} \\
   \mp  \frac{\sqrt{3}}{2} & \frac{1}{2} 
   \end{bmatrix} 
  \begin{bmatrix}
  \varphi_3  \\
  \varphi_8 
  \end{bmatrix} 
 .
 \label{60-rotation2}
\end{align}
The transformation (\ref{60-rotation2}) is equal to 
\begin{align}
  \varphi_3^\prime =& 
\begin{cases}
\frac{1}{2} \varphi_3  +  \frac{\sqrt{3}}{2} \varphi_8 \\
\frac{1}{2} \varphi_3  -  \frac{\sqrt{3}}{2} \varphi_8    \end{cases}
 ,
\quad  \varphi_8^\prime 
= \begin{cases}
 -  \frac{\sqrt{3}}{2} \varphi_3 +  \frac{1}{2} \varphi_8 \\
 +  \frac{\sqrt{3}}{2} \varphi_3 +  \frac{1}{2} \varphi_8 
  \end{cases} ,
\end{align}
which leads to
\begin{align}
  \frac{1}{2} \varphi_3^\prime  +  \frac{\sqrt{3}}{2} \varphi_8^\prime 
=&  \begin{cases}
-   \frac{1}{2} \varphi_3  +  \frac{\sqrt{3}}{2} \varphi_8  \\
  \varphi_3  
    \end{cases} ,
\nonumber\\ 
  - \frac{1}{2} \varphi_3^\prime  +  \frac{\sqrt{3}}{2} \varphi_8^\prime 
=&  \begin{cases}
 - \varphi_3 \\
   \frac{1}{2} \varphi_3  +  \frac{\sqrt{3}}{2} \varphi_8  
    \end{cases} .
\end{align}

\end{enumerate}

\begin{figure}[tb]
\begin{center}
\includegraphics[scale=0.57]{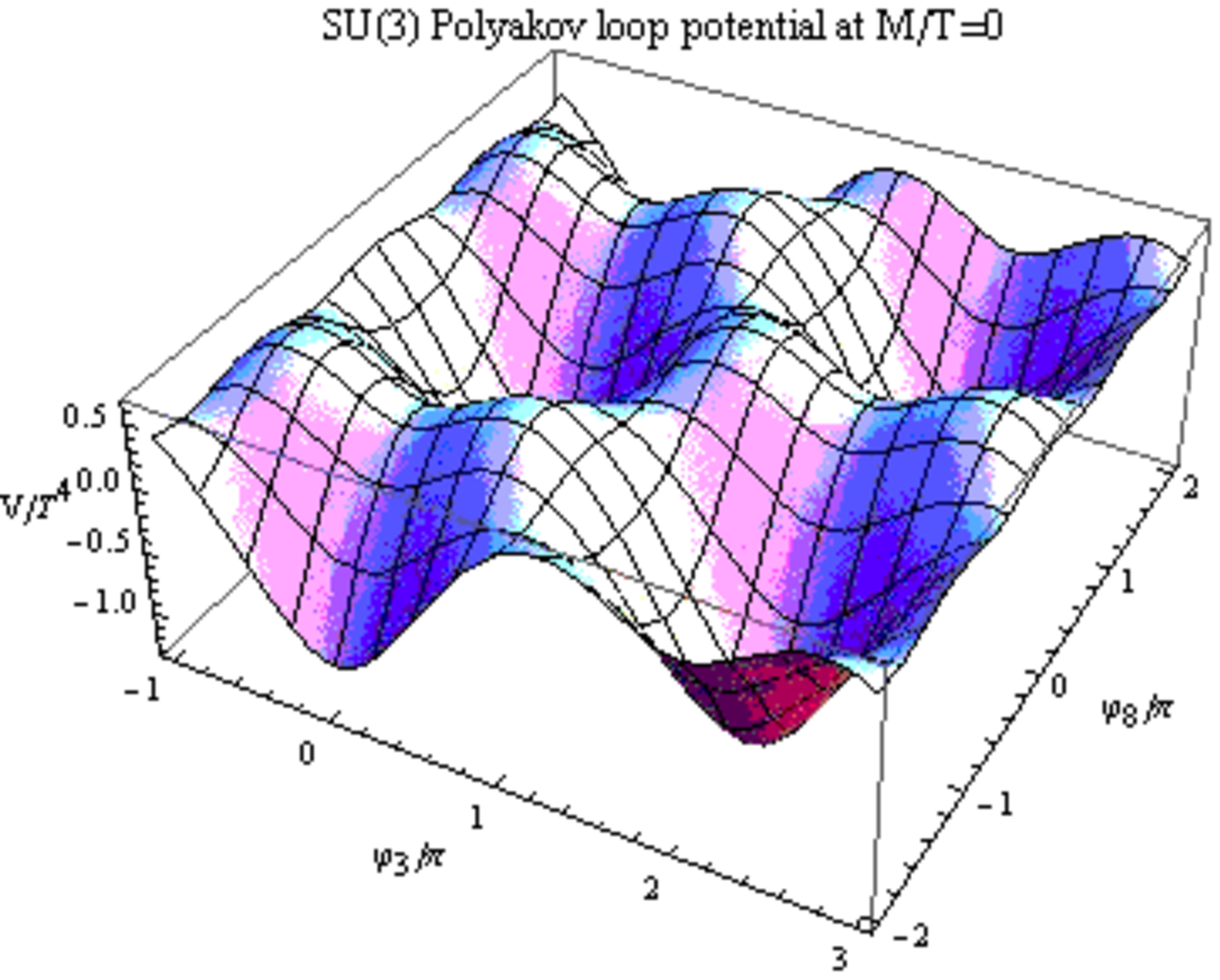}
\quad
\includegraphics[scale=0.57]{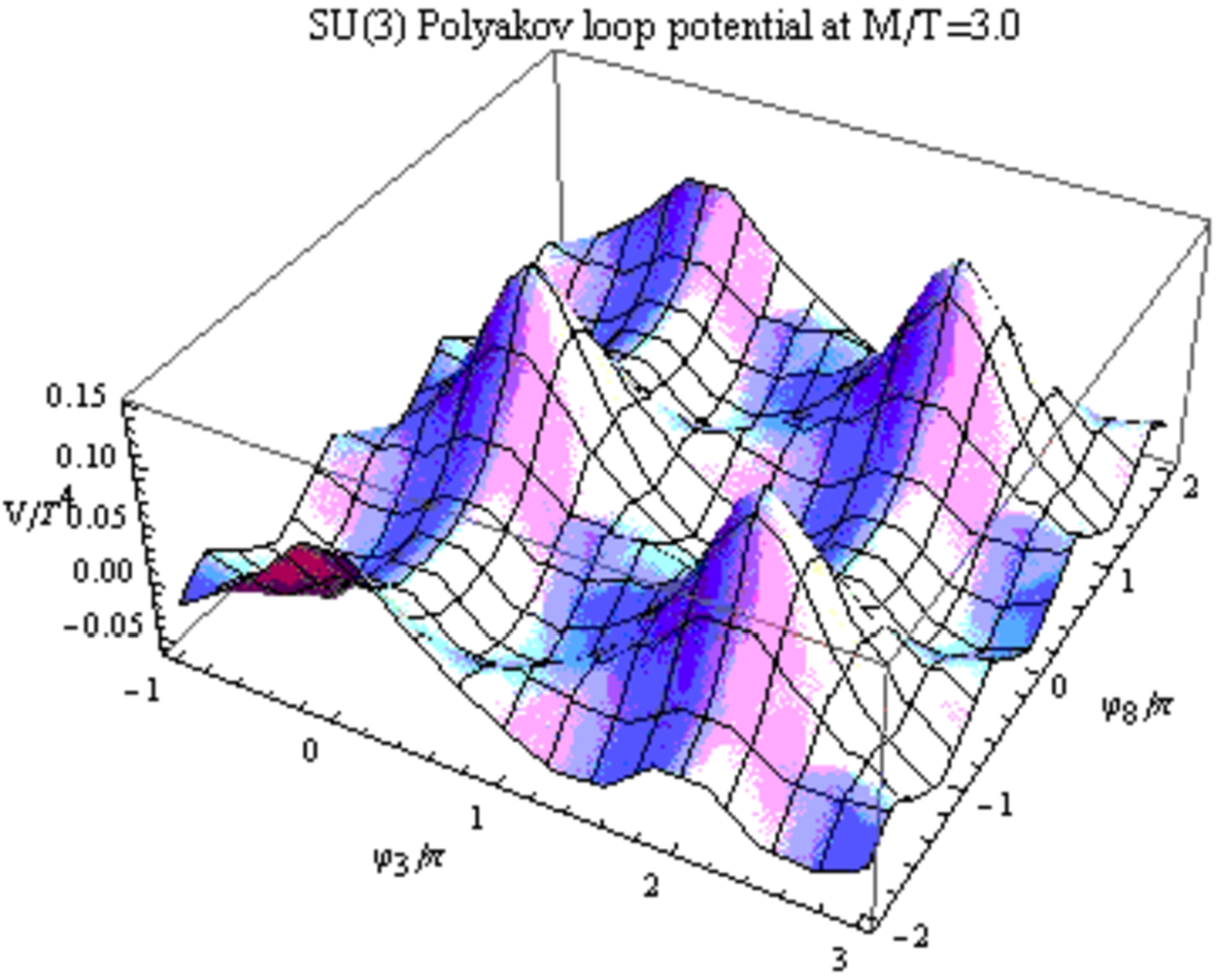}
\end{center}
\vskip -0.5cm
\caption{
Plot of the $D=4$ effective potential $\hat{V}$ of the $SU(3)$ Polyakov loop as a function of the two angles $\varphi_3/\pi$ and $\varphi_8/\pi$ 
(Left) at $\hat{M}:=M/T = 0$,
(Right) at $\hat{M}:=M/T = 3.0$.
}
\label{fig:V-SU3-ab}
\end{figure}

\begin{figure}[hptb]
\begin{center}
\includegraphics[scale=0.43]{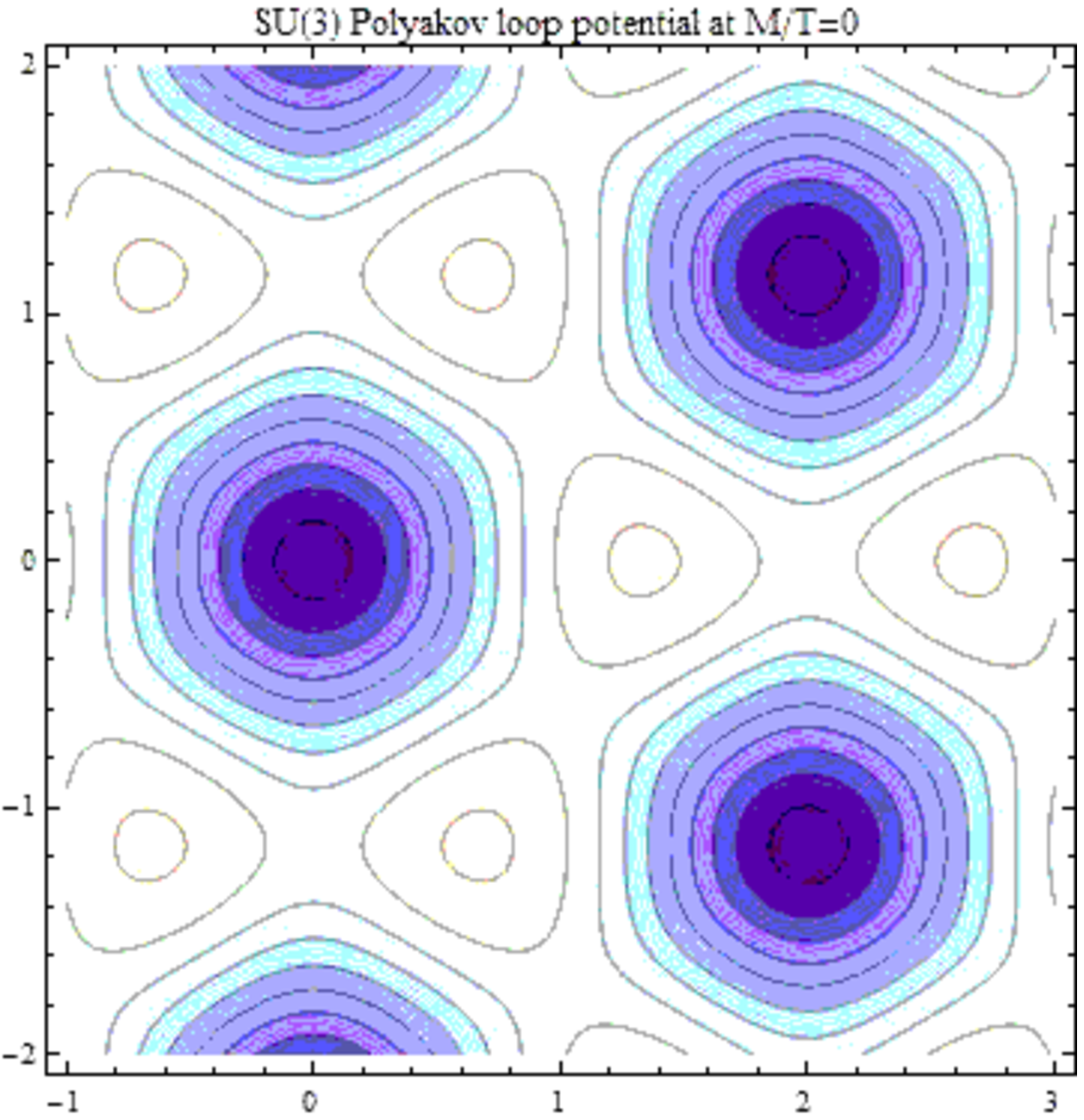}
\quad
\includegraphics[scale=0.43]{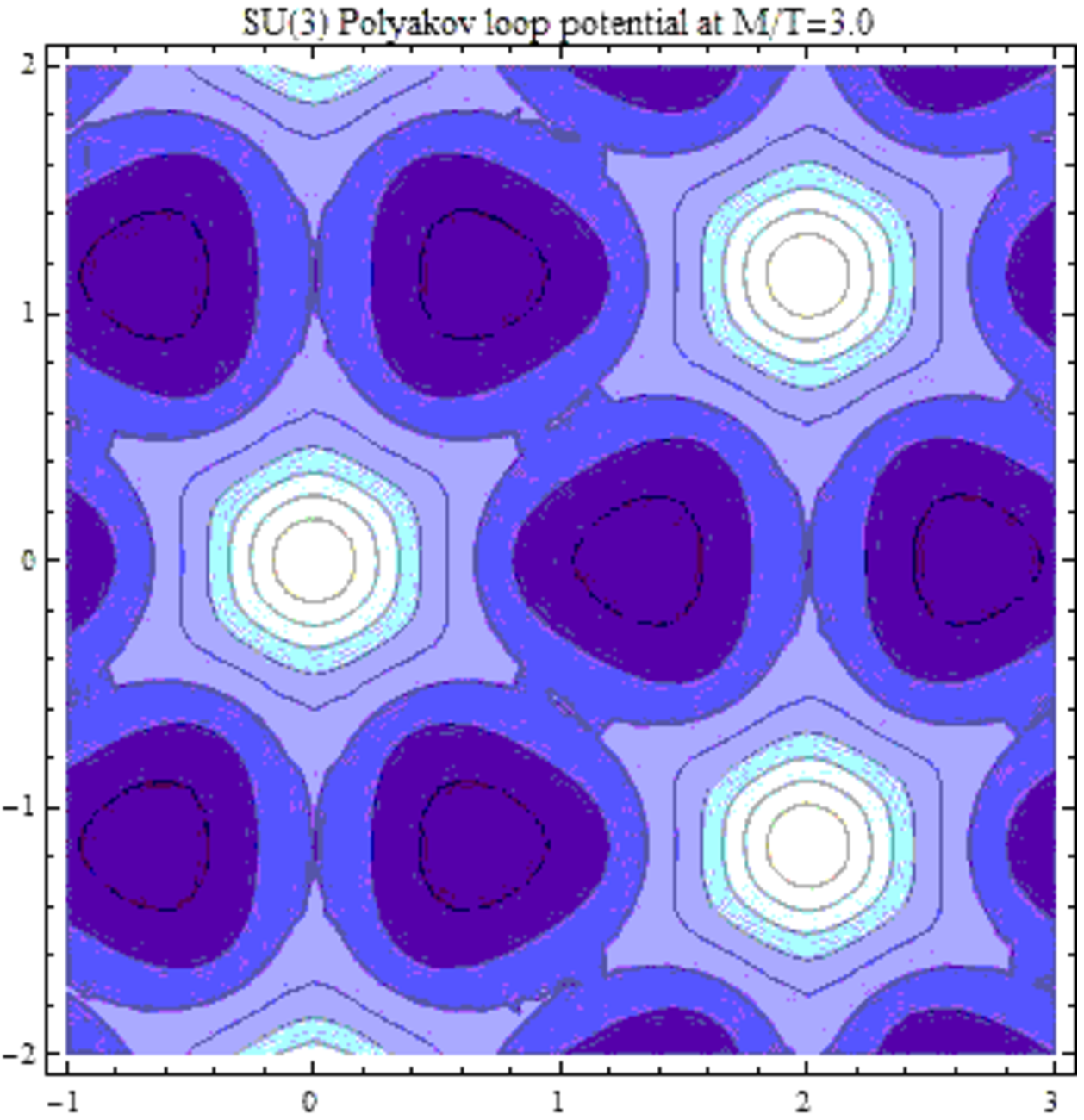}
\end{center}
\vskip -0.5cm
\caption{
Contour Plot of the $D=4$ effective potential $\hat{V}$ of the $SU(3)$ Polyakov loop as a function of the two angles $\varphi_3/\pi$ and $\varphi_8/\pi$ 
(Left) at $\hat{M}:=M/T = 0$,
(Right) at $\hat{M}:=M/T = 3.0$.
}
\label{fig:V-SU3-CPab}
\end{figure}

From the above symmetries, it is sufficient to consider the background-field effective potential $V_{\rm eff}(\varphi_{3},\varphi_{8})$ in an equilateral triangle $OAB$ of side length $4\pi/\sqrt{3}$ with the three vertices at $O:(0,0)$, $A:(2\pi,  2\pi/\sqrt{3})$ and $B:(2\pi, - 2\pi/\sqrt{3})$ in the $(\varphi_3, \varphi_8)$ plane.
The effective potential $V_{\rm eff}(\varphi_{3},\varphi_{8})$ is invariant under rotations which leave this equilateral triangle invariant. 
The vertices $O, A, B$ of the triangle and its center $G$ located at $(4\pi/3,0)$ are always extrema of the effective  potential $V_{\rm eff}(\varphi_{3},\varphi_{8})$.
Therefore, we consider the effective potential at the points $O, A, B$, and $G$ and their vicinity. 
See Fig.~\ref{fig:V-SU3-ab} and Fig.~\ref{fig:V-SU3-CPab}.

At high temperature  $M/T \ll 1$, 
the minima of the effective potential are realized at the points $O,A,B$ (the vertices of the triangle) which correspond  to the non-vanishing Polyakov loop average $L=1,e^{-i\frac23 \pi},e^{i\frac23 \pi}$ respectively, i.e., deconfinement. Choosing one of them spontaneously breaks the $Z_3$ center symmetry. The  effective  potential takes the maximum at the center point $G$ which corresponds to the vanishing Polyakov loop average $L=0$.

At low temperature $M/T \gg 1$, 
the minimum of the  effective potential is realized at the center point $G$ with vanishing Polyakov loop average $L=0$, i.e, confinement.
The points $O,A,B$ (the vertices of the triangle) are the  maxima which correspond to the non-vanishing Polyakov loop average $L=1,e^{-i\frac23 \pi},e^{i\frac23 \pi}$ respectively.

The above statements are summarized in the following table.
\begin{center}
\begin{tabular}{l||c|c||c||c|c} \hline\hline
 &   $\varphi_3$ & $\varphi_8$ & $L$ &  $ M/T \ll 1$ &  $ M/T \gg 1$ \\  \hline\hline
O &   0 &  0 & 1 & min & max \\  \hline 
A &    $2\pi$  & $\frac{2\pi}{\sqrt{3}}$  &  $e^{-i \frac23 \pi}=-\frac12 - i \frac{\sqrt{3}}{2}$ &  min &  max  \\  \hline 
B &    $2\pi$  & $-\frac{2\pi}{\sqrt{3}}$  &  $e^{+i\frac23 \pi}=-\frac12 + i \frac{\sqrt{3}}{2}$ &  min &  max  \\  \hline 
G &    $\frac43 \pi$  & 0  &  0 & max & min  \\  \hline 
\hline
\end{tabular}
\label{table:eigenvalue1}
\end{center}

By representing  the trace explicitly as the sum over the Matsubara frequencies and the integration over the spatial momentum, we obtain the effective potential $V_{\rm eff}(\varphi_{3},\varphi_{8})$   in terms of the two angles $\varphi_{3}$ and $\varphi_{8}$:
\begin{align}
   V_{\rm eff}(\varphi_{3},\varphi_{8}) 
 =& 
 \frac{D-1}{2} T \sum_{n \in \mathbb{Z}}  \sum_{\pm \vec{\alpha}^{(i)} }  \int \frac{d^{D-1}p}{(2\pi)^{D-1}}   
\nonumber\\ & 
\times \ln [(\omega_n + T \vec{\alpha}^{(i)} \cdot \vec{\varphi} )^2+\bm{p}^2 + M^2 ] 
\nonumber\\
&-  \frac{1}{2} T \sum_{n \in \mathbb{Z}}  \sum_{\pm \vec{\alpha}^{(i)} }  \int \frac{d^{D-1}p}{(2\pi)^{D-1}}   
\nonumber\\ & 
\times \ln [(\omega_n + T \vec{\alpha}^{(i)} \cdot \vec{\varphi} )^2+\bm{p}^2  ] 
 .
\end{align}
The  effective potential $V_{\rm eff}(\varphi_{3},\varphi_{8})$ is explicitly written as
\begin{align}
  & V_{\rm eff}(\varphi_{3},\varphi_{8}) 
\nonumber\\
 =& 
 \frac{D-1}{2} T \sum_{n \in \mathbb{Z}, \pm}  \int \frac{d^{D-1}p}{(2\pi)^{D-1}}   
\Biggr[ 
\ln [(\omega_n \pm T\varphi_{3})^2+\bm{p}^2 + M^2 ] 
\nonumber\\
&+  
\ln \left\{ \left[\omega_n \pm T \left(\frac{1}{2}\varphi_{3} + \frac{\sqrt{3}}{2}\varphi_{8} \right) \right]^2+\bm{p}^2 + M^2  \right\} 
\nonumber\\
&+  
\ln \left\{ \left[\omega_n \pm T \left(\frac{1}{2}\varphi_{3} - \frac{\sqrt{3}}{2}\varphi_{8} \right) \right]^2+\bm{p}^2 + M^2  \right\} 
\Biggr]
\nonumber\\
&-  \frac{1}{2} T \sum_{n \in \mathbb{Z}, \pm}  \int \frac{d^{D-1}p}{(2\pi)^{D-1}}   
\Biggr[ 
\ln [(\omega_n \pm T\varphi_{3})^2+\bm{p}^2  ] 
\nonumber\\
&+  
\ln \left\{ \left[\omega_n \pm T \left(\frac{1}{2}\varphi_{3} + \frac{\sqrt{3}}{2}\varphi_{8} \right) \right]^2+\bm{p}^2    \right\} 
\nonumber\\
&+  
\ln \left\{ \left[\omega_n \pm T \left(\frac{1}{2}\varphi_{3} - \frac{\sqrt{3}}{2}\varphi_{8} \right) \right]^2+\bm{p}^2    \right\} 
\Biggr]
 .
\end{align}

After performing the sum over the Matsubara frequencies $\omega_n=2\pi T n$, thus, we obtain the effective potential for the Polyakov loop average as
\begin{align}
 & V_{\rm eff}(\varphi_{3},\varphi_{8})/T^D 
\nonumber\\
=& (D-1) \Biggr[  F_{\hat{M}}(\varphi_{3}) + F_{\hat{M}} \left( \frac{1}{2}\varphi_{3} + \frac{\sqrt{3}}{2}\varphi_{8} \right) 
\nonumber\\&
 +  F_{\hat{M}} \left( \frac{1}{2}\varphi_{3} - \frac{\sqrt{3}}{2}\varphi_{8} \right)  \Biggr] 
\nonumber\\&
  - \Biggr[ F_{0}(\varphi_{3}) +  F_{0} \left( \frac{1}{2}\varphi_{3} + \frac{\sqrt{3}}{2}\varphi_{8} \right) + F_{0} \left( \frac{1}{2}\varphi_{3} - \frac{\sqrt{3}}{2}\varphi_{8} \right)     \Biggr]
 .
\end{align}

At high temperature, $\hat{M}=M/T \ll 1$, the mass $M$ is neglected and the effective potential reduces to
\begin{align}
  V_{\rm eff}^{\rm High}(\varphi_{3},\varphi_{8})/T^D 
& \simeq (D-2) \Biggr[ F_{0}(\varphi_{3}) + F_{0} \left( \frac{1}{2}\varphi_{3} + \frac{\sqrt{3}}{2}\varphi_{8} \right) 
\nonumber\\ & 
+ F_{0} \left( \frac{1}{2}\varphi_{3} - \frac{\sqrt{3}}{2}\varphi_{8} \right) \Biggr]  
 .
\end{align}
For $D=4$,  the effective potential reproduces the well-known $SU(3)$ Weiss potential \cite{Weiss81}. 
This potential has degenerate minima on the vertices of the basic equilateral triangle, leading to a deconfined phase with the spontaneously broken $Z(3)$ center symmetry. 

At low temperature, $\hat{M}=M/T \gg 1$, on the other hand, $F_{\hat{M}}(\varphi)$ is exponentially surpressed $F_{\hat{M}}(\varphi) \ll 1$ and the effective potential reduces to 
\begin{align}
  & V_{\rm eff}^{\rm Low}(\varphi_{3},\varphi_{8})/T^D 
  \nonumber\\  
\simeq & -\left[ F_{0}(\varphi_{3}) + F_{0} \left( \frac{1}{2}\varphi_{3} + \frac{\sqrt{3}}{2}\varphi_{8}  \right) + F_{0} \left( \frac{1}{2}\varphi_{3} - \frac{\sqrt{3}}{2}\varphi_{8} \right) \right]  
 .
\end{align}
The effective potential at the sufficiently low temperature is reversed to the Weiss potential at sufficiently high temperature: 
\begin{align}
  \hat{V}_{\rm eff}^{\rm Low}(\varphi_{3},\varphi_{8}) \simeq 
 - (D-2)^{-1} \hat{V}_{\rm eff}^{\rm High}(\varphi_{3},\varphi_{8})
 .
\end{align}
Therefore, the  effective potential has the absolute minimum at the center $G$ of the triangle leading to a $Z_3$ center symmetric confining phase. 
Thus there must exist a phase transition at a certain critical value of $T_d/M$ between the high-temperature deconfined phase and the low-temperature confined phase.

\subsection{$SU(3)$  Critical temperature and order of the phase transition}

\begin{figure}[ptb]
\begin{center}
\includegraphics[scale=0.67]{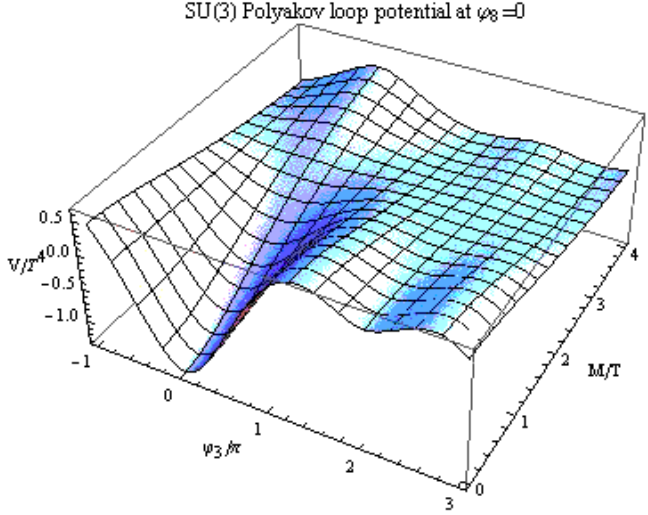}
\end{center}
\vskip -0.3cm
\caption{
The $D=4$ effective potential $\hat{V}$ of the $SU(3)$ Polyakov loop at $\varphi_8=0$ as a function of an angle $\varphi_3/\pi \in [-1, 3)$  for various values of $\hat{M}:=M/T$.
}
\label{fig:V-SU3-c}
\end{figure}

The absolute minimum of $V_{\rm eff}(\varphi_{3},\varphi_{8})$ lies on the $\varphi_8=0$ axis up to the discrete rotations by the angle $\pi/3$ for all temperature: 
\begin{align}
  V_{\rm eff}(\varphi_{3},0)/T^D 
=& (D-1) \left[F_{\hat{M}}(\varphi_{3}) + 2 F_{\hat{M}} \left(\frac{\varphi_{3}}{2}\right)  \right] 
\nonumber\\ &
 - \left[ F_{0}(\varphi_{3}) + 2 F_{0} \left(\frac{\varphi_{3} }{2} \right)   \right]
 ,
\end{align}
where
\begin{align}
&  F_{\hat{M}}(\varphi_{3}) + 2 F_{\hat{M}}\left(\frac{\varphi_{3}}{2}\right)
\nonumber\\
=& \int \frac{d^{D-1}\hat{p}}{(2\pi)^{D-1}} \left[ f_{\hat{M}}(\hat{p}^2, \varphi_{3}) + 2 f_{\hat{M}} \left(\hat{p}^2, \frac{\varphi_{3} }{2} \right) \right]
 .
\end{align}
Fig.~\ref{fig:V-SU3-c} is the plot of the Polyakov-loop effective potential $\hat{V}_0(\varphi_{3},0; \hat{M}):=V_{\rm eff}(\varphi_{3},0)/T^D$ at $\varphi_8=0$  as a function of $\varphi_3$ for various values of $M/T$ in $D=4$ dimensions. 

The integrand $f_{\hat{M}}(\hat{p}^2, \varphi_{3})$ is expanded into the power series  in  $\varphi_{3}$ about $\varphi_3=4\pi/3$ at which $L=0$: defining $\sigma :=\varphi_{3} - 4\pi/3$
\begin{align}
  f_{\hat{M}}(\hat{p}^2, \varphi_{3})
=& c^{(0)}_{\hat{M}}(\hat{p}) + c^{(1)}_{\hat{M}}(\hat{p}) \sigma + c^{(2)}_{\hat{M}}(\hat{p}) \sigma^2 
+ c^{(3)}_{\hat{M}}(\hat{p}) \sigma^3  
\nonumber\\&
+ c^{(4)}_{\hat{M}}(\hat{p}) \sigma^4 + O(\sigma^5) 
 ,
\end{align}
where the coefficients $c^{(n)}_{\hat{M}}(\hat{p})$ are given in Appendix~\ref{Appendix:series}. 
Then the expansion of the integrand 
$
f_{\hat{M}}(\hat{p}^2, \varphi_{3}) + 2 f_{\hat{M}}(\hat{p}^2, \frac{\varphi_{3} }{2})
$
into the power series of $\sigma$ is given by
\begin{align}
&  f_{\hat{M}}(\hat{p}^2, \varphi_{3}) + 2 f_{\hat{M}}(\hat{p}^2, \frac{\varphi_{3} }{2})
\nonumber\\
=& h^{(0)}_{\hat{M}}(\hat{p}) + h^{(1)}_{\hat{M}}(\hat{p}) \sigma + h^{(2)}_{\hat{M}}(\hat{p}) \sigma^2 
+ h^{(3)}_{\hat{M}}(\hat{p}) \sigma^3  
\nonumber\\&
+ h^{(4)}_{\hat{M}}(\hat{p}) \sigma^4 + O(\sigma^5) 
 ,
\end{align}
where the coefficients $h^{(n)}_{\hat{M}}(\hat{p})$ are given in Appendix~\ref{Appendix:series}.  
It should be remarked that the linear term in $\sigma$ disappears finally $h^{(1)}_{\hat{M}}(\hat{p})=0$.
Accordingly, the effective potential has the power series expansion in  $\sigma$:
\begin{align}
 & \hat{V}_0(\varphi_{3},0; \hat{M}) 
:= V_{\rm eff,0}(\varphi_{3},0)/T^D 
\nonumber\\
=& A_{0,\hat{M}}   + \frac{A_{2,\hat{M}}}{2!} \sigma^2 + \frac{A_{3,\hat{M}}}{3!} \sigma^3 + \frac{A_{4,\hat{M}}}{4!} \sigma^4 + O(\sigma^5)
 ,
\label{eff-pot-SU3} 
\end{align}
where the coefficients $A_{n,\hat{M}}$ ($n=0,1,2,...$) are given by
\begin{align}
 A_{n,\hat{M}}
=& n! C_D \int_{0}^{\infty} d \hat{p} \ \hat{p}^{D-2} [(D-1) h^{(n)}_{\hat{M}}(\hat{p}) - h^{(n)}_{0}(\hat{p})]
 .
\end{align}
In the limit $\hat{M} \to 0$, especially, we find 
\begin{align}
 A_{n,0}
=& n! C_D (D-2) \int_{0}^{\infty} d \hat{p} \ \hat{p}^{D-2}  h^{(n)}_{0}(\hat{p}) 
 ,
\end{align}
which reduces for $D=4$ to  
\begin{align}
 A_{2,0}
=& - \frac13    
 < 0 , \ 
A_{3,0} 
=   \frac{1}{2\pi} = 0.15916...> 0 , \ 
\nonumber\\
A_{4,0}
=&  \frac{9}{4\pi^2} = 0.22797... > 0
 .
\end{align}
Note that there is no linear term of $\sigma$ in the effective potential. 
Fig.~\ref{fig:V-SU2-d} is the plot of 
$A_{2,\hat{M}}$, $A_{3,\hat{M}}$, and  $A_{4,\hat{M}}$ 
for the $SU(3)$ Polyakov loop  effective potential 
as a function of $\hat{M}:=M/T$ at $D=4$.

\begin{figure}[ptb]
\begin{center}
\includegraphics[scale=0.65]{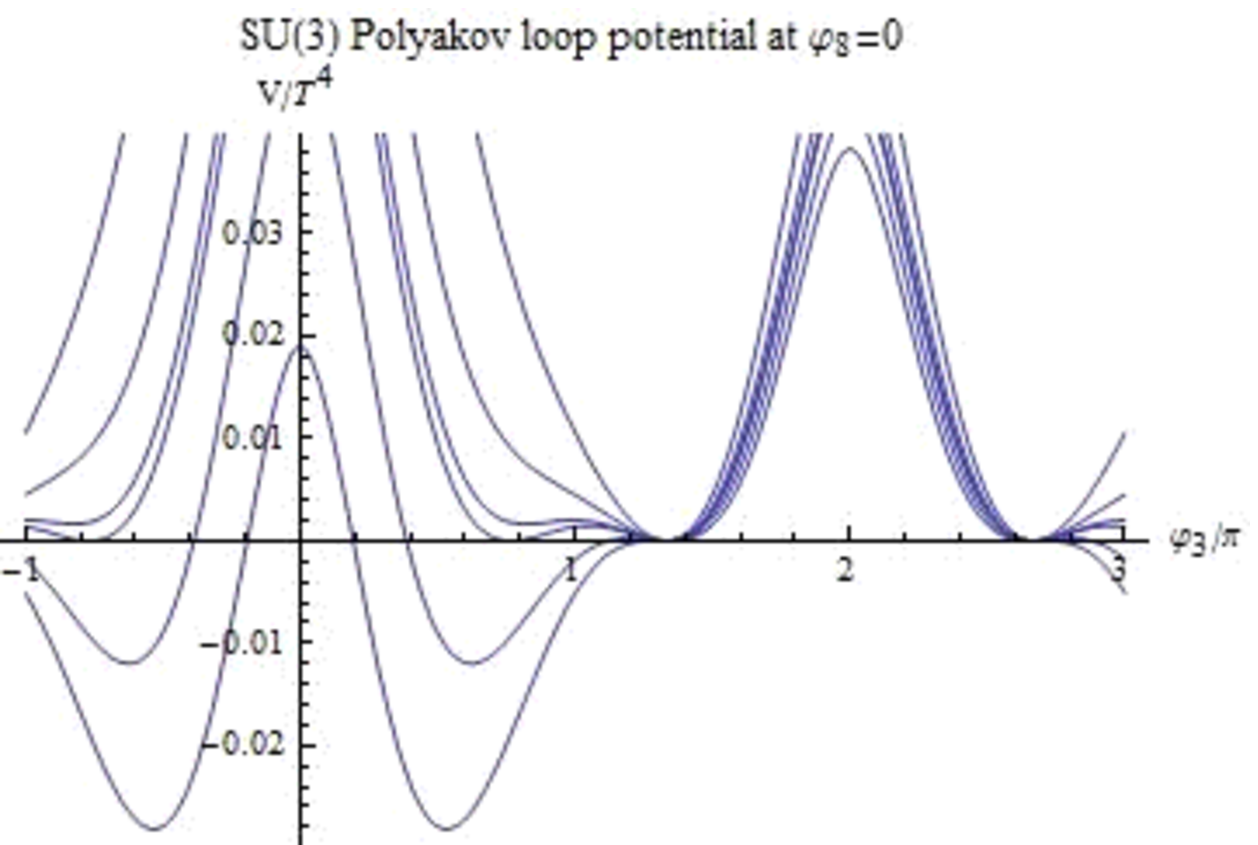}
\includegraphics[scale=0.65]{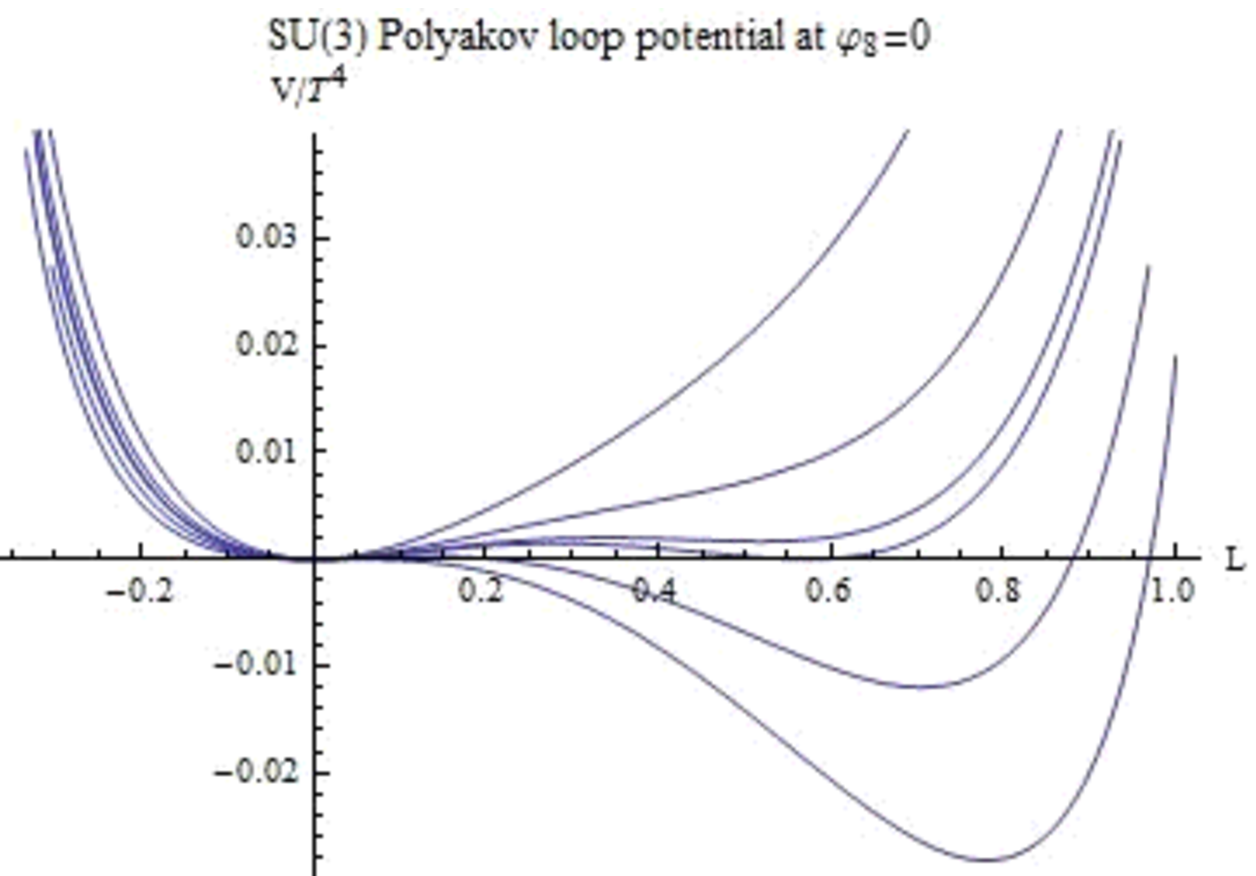}
\end{center}
\vskip -0.3cm
\caption{
The $D=4$ effective potential $\hat{V}$ of the $SU(3)$ Polyakov loop  at $\varphi_8=0$ for $\hat{M}:=M/T=2.65, 2.70, 2.75, 2.76, 2.80, 2.90$, 
(Left) as a function of an angle $\varphi_3/\pi \in [-1, 3 )$, 
(Right) as a function of the Polyakov loop average $ L  =  \frac13 \left[ 1 + 2 \cos (\frac{\varphi_{3}}{2} )  \right] \in (-1/3,1]$, normalized as $\hat{V}(L=0)=0$.
}
\label{fig:V-SU3-all}
\end{figure}

\begin{figure}[ptb]
\begin{center}
\includegraphics[scale=0.50]{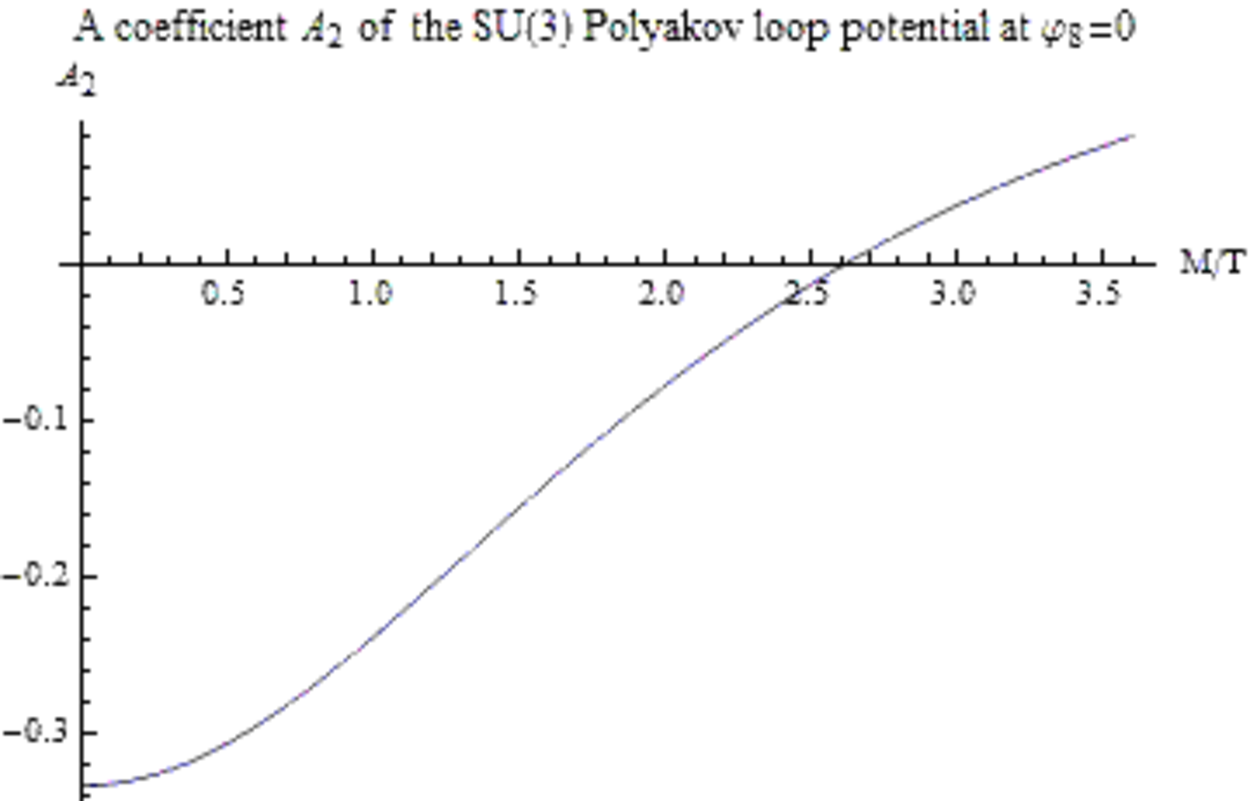}
\includegraphics[scale=0.50]{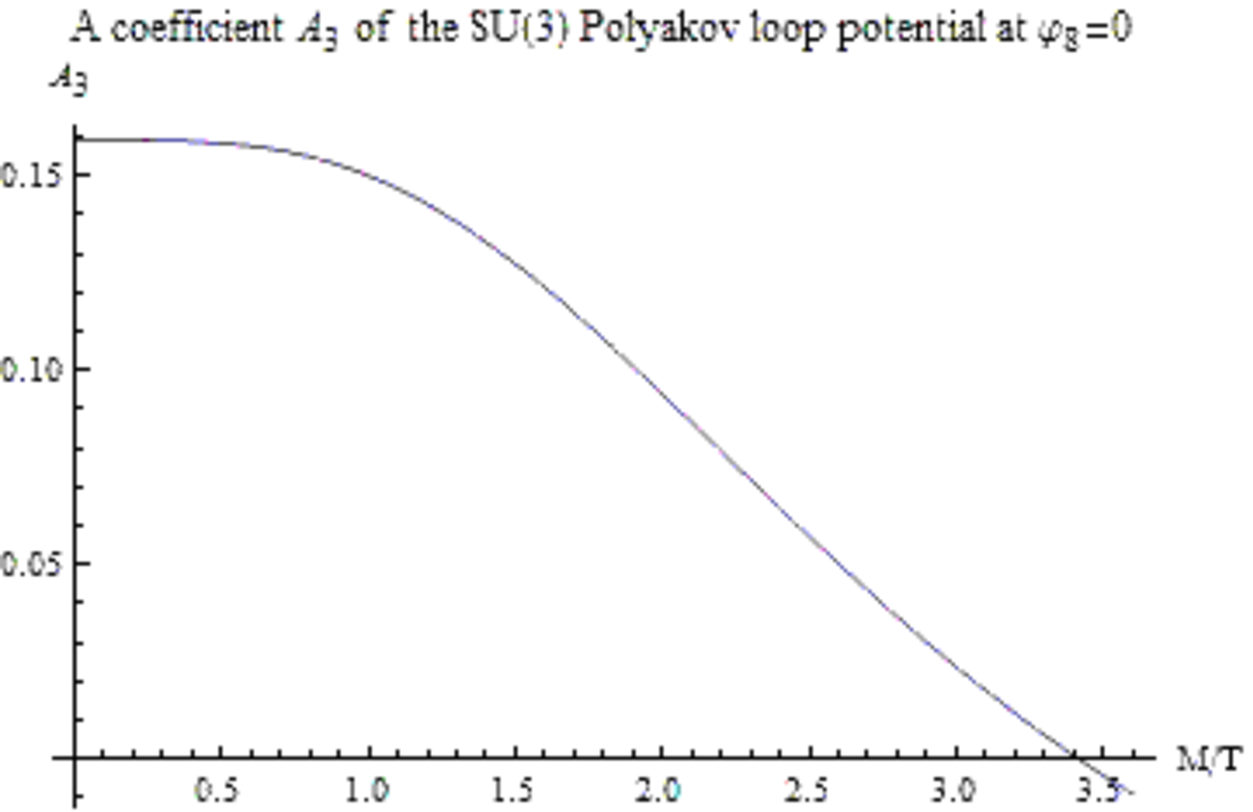}
\includegraphics[scale=0.50]{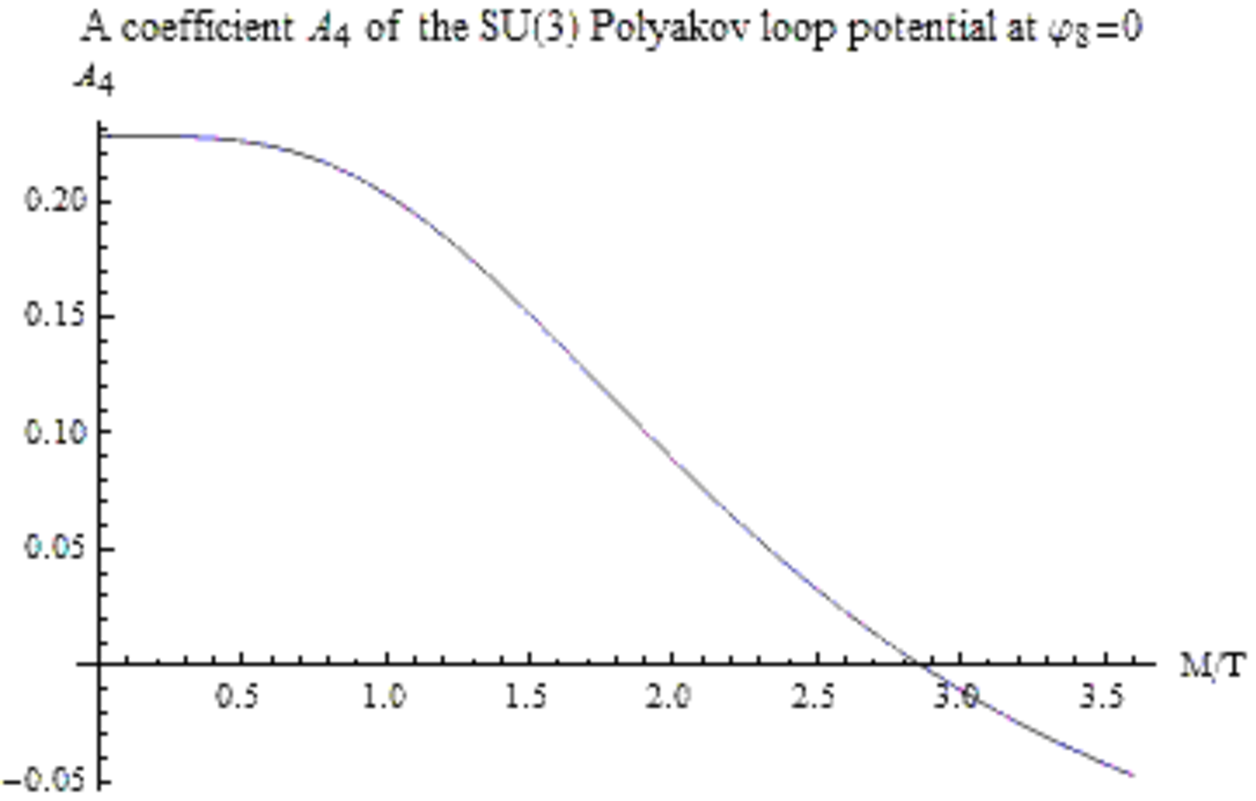}
\end{center}
\vskip -0.3cm
\caption{
The plot of 
$A_{2,\hat{M}}$,
$A_{3,\hat{M}}$,
and
$A_{4,\hat{M}}$,
for the $SU(3)$ Polyakov loop  effective potential 
as a function of $\hat{M}:=M/T$ at $D=4$. 
}
\label{fig:V-SU2-d}
\end{figure}

\begin{figure}[ptb]
\begin{center}
\includegraphics[scale=0.50]{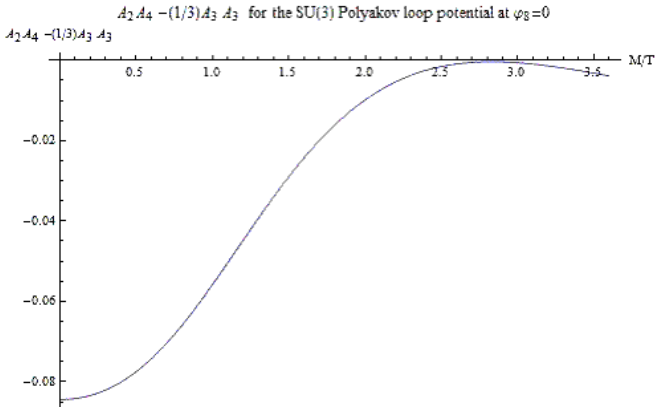}
\end{center}
\vskip -0.3cm
\caption{
The plot of 
$
A_{2,\hat{M}}A_{4,\hat{M}}  -  \frac{1}{3} (A_{3,\hat{M}})^2 
$
for the $SU(3)$ Polyakov loop  effective potential 
as a function of $\hat{M}:=M/T$ at $D=4$. 
}
\label{fig:V-SU2-d2}
\end{figure}

In order to discuss the order of the phase transition, we consider the  effective potential (the Landau function) of the form:
\begin{align}
  V(\sigma) = -h \sigma - \frac12 a \sigma^2 + \frac13 b \sigma^3 + \frac14 c \sigma^4  .\end{align}
The extrema are obtained by solving the equation:
\begin{align}
  V^\prime(\sigma) = -h   -   a \sigma  +  b \sigma^2 +  c \sigma^3 = 0 .
\end{align}
We restrict our consideration to a  case in which $h$ is negligible, i.e., $h=0$, which is the case for the $SU(3)$ Yang-Mills theory. 
The stationary points of $V(\sigma)$ at $h=0$ are given at three values of $\sigma$, i.e., $\sigma=0 , \sigma_+, \sigma_-$:
\begin{align}
  \sigma=0, \ \sigma_+:=\frac{-b + \sqrt{b^2+4ac}}{2c}, \ \sigma_-:=\frac{-b - \sqrt{b^2+4ac}}{2c}   .
\end{align}
At the stationary points $\sigma=0 , \sigma_+, \sigma_-$, the  effective potential has the values:
\begin{align}
V(\sigma=0) &= 0,
\nonumber\\
 V(\sigma_\pm) &=  - \frac{b^4+6ab^2c+6a^2c^2 \mp b(b^2+4ac)^{3/2}}{24c^3} .
\end{align}
The first order transition occurs when the two minima give  the same value of the effective potential (free energy), namely, the condition $V(\sigma=0)=V(\sigma_\pm)$ is satisfied:
\begin{align}
 a(T) = - \frac{2}{9} \frac{b(T)^2}{c(T)}  .
 \label{condition-1st}
\end{align}
at which the global minimum experiences a discontinuous jump.
This condition determines the critical temperature $T_d$. 
The first order phase transition is induced by the cubic interaction $\frac13 b \sigma^3$. 
When $b \equiv 0$, the condition (\ref{condition-1st}) reduces to $a=0$ as long as $c > 0$. This is nothing but the condition for the second order phase transition, which is indeed the $SU(2)$ Yang-Mills theory.

The confinement/deconfinement phase transition in the $SU(3)$ Yang-Mills theory described by the effective potential (\ref{eff-pot-SU3}) is of the first order.
See Fig.~\ref{fig:V-SU3-all}. 
The first order phase transition for confinement/deconfinement in the $SU(3)$ Yang-Mills theory is induced by cubic interaction $L^3$, which occurs when the condition is satisfied:
\begin{align}
 A_{2,\hat{M}}   =  \frac{1}{3} (A_{3,\hat{M}})^2/A_{4,\hat{M}}  .
 \label{1st-cond}
\end{align}
See  Fig.~\ref{fig:V-SU2-d2} for the plot of 
$
A_{2,\hat{M}}A_{4,\hat{M}}  -  \frac{1}{3} (A_{3,\hat{M}})^2 
$
for the $SU(3)$ Polyakov-loop  effective potential 
as a function of $\hat{M}:=M/T$ at $D=4$. 
This shows that the condition (\ref{1st-cond}) for the first order transition is indeed satisfied at $\hat{M}=2.75$, which is greater than $\hat{M}=2.6$ at which the second order transition $A_{2,\hat{M}}=0$ would be realized. 
In other words, when the temperature is decreased starting from the high-temperature deconfined phase, the phase transition to the low-temperature confined phase occurs at a temperature lower than the expected temperature at which the coefficient changes its signature from negative to positive.

\begin{figure}[ptb]
\begin{center}
\includegraphics[scale=0.50]{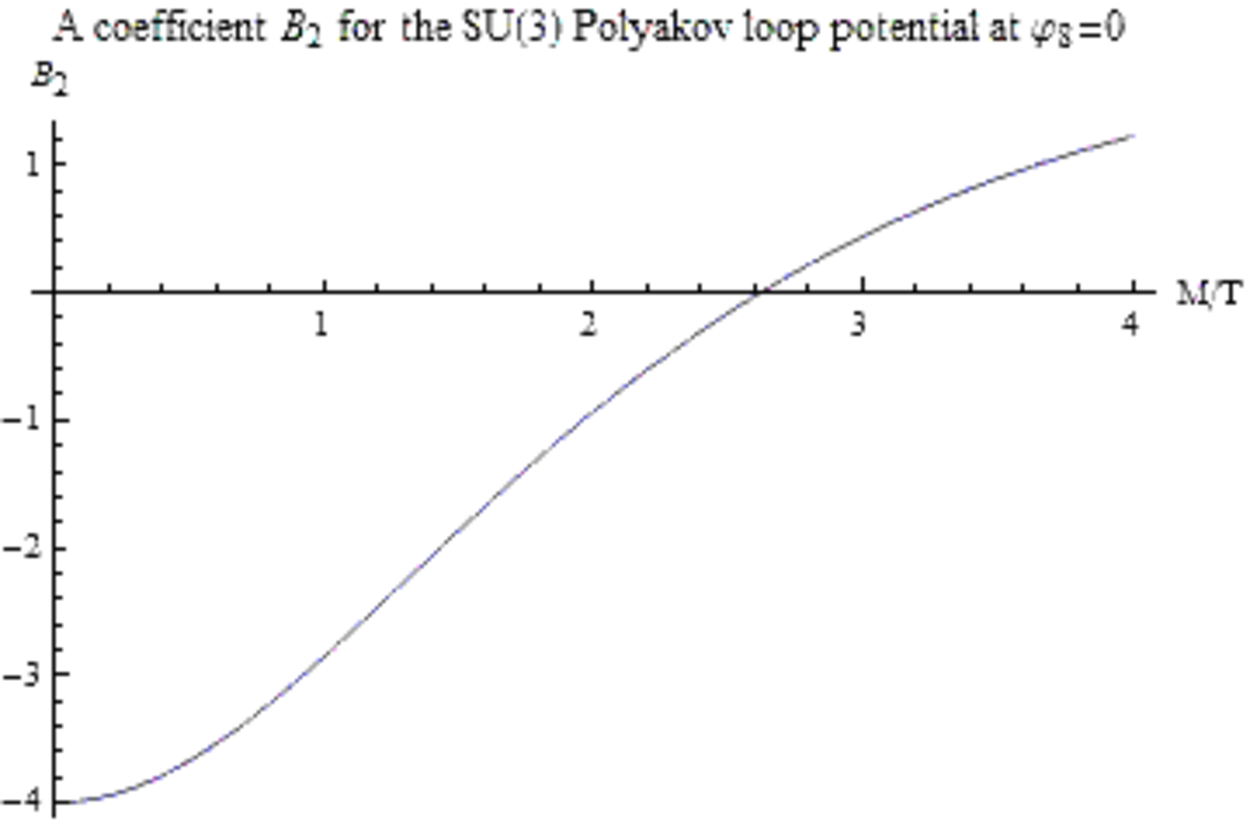}
\includegraphics[scale=0.50]{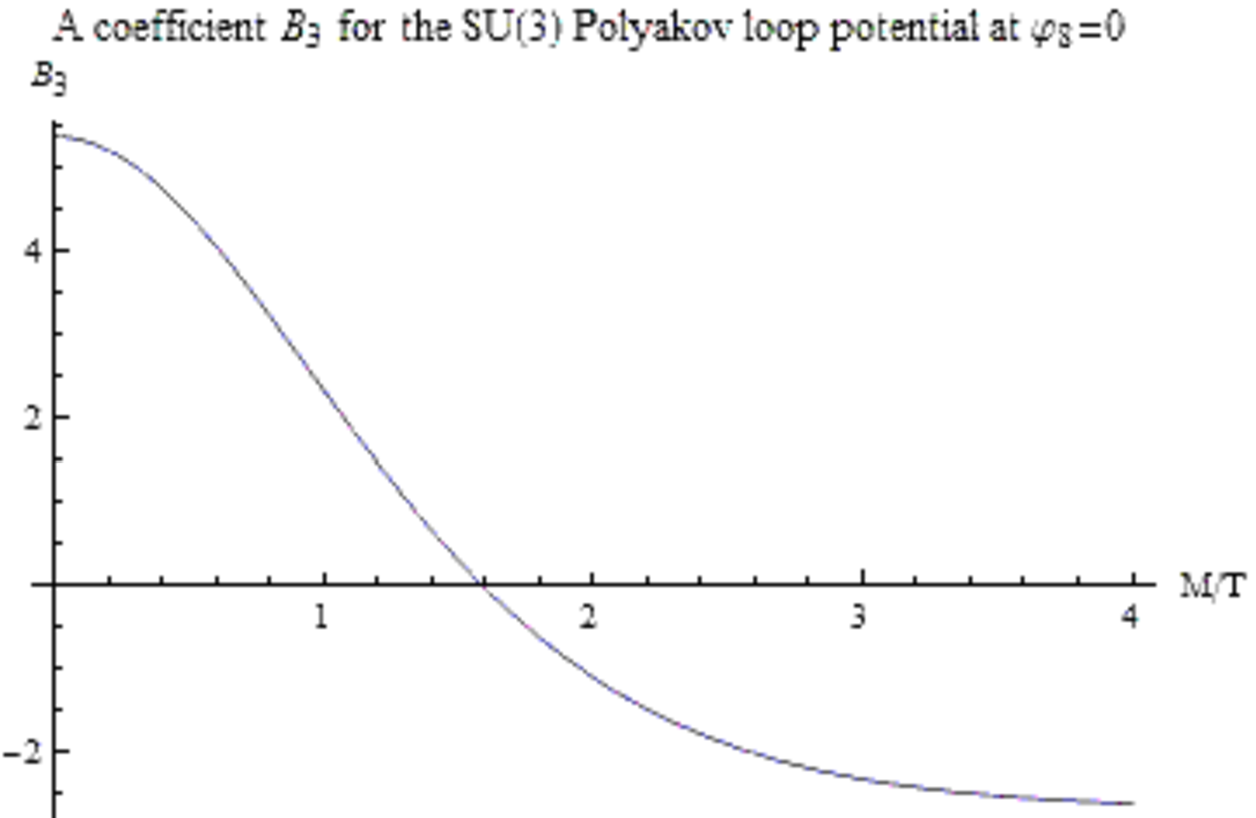}
\includegraphics[scale=0.50]{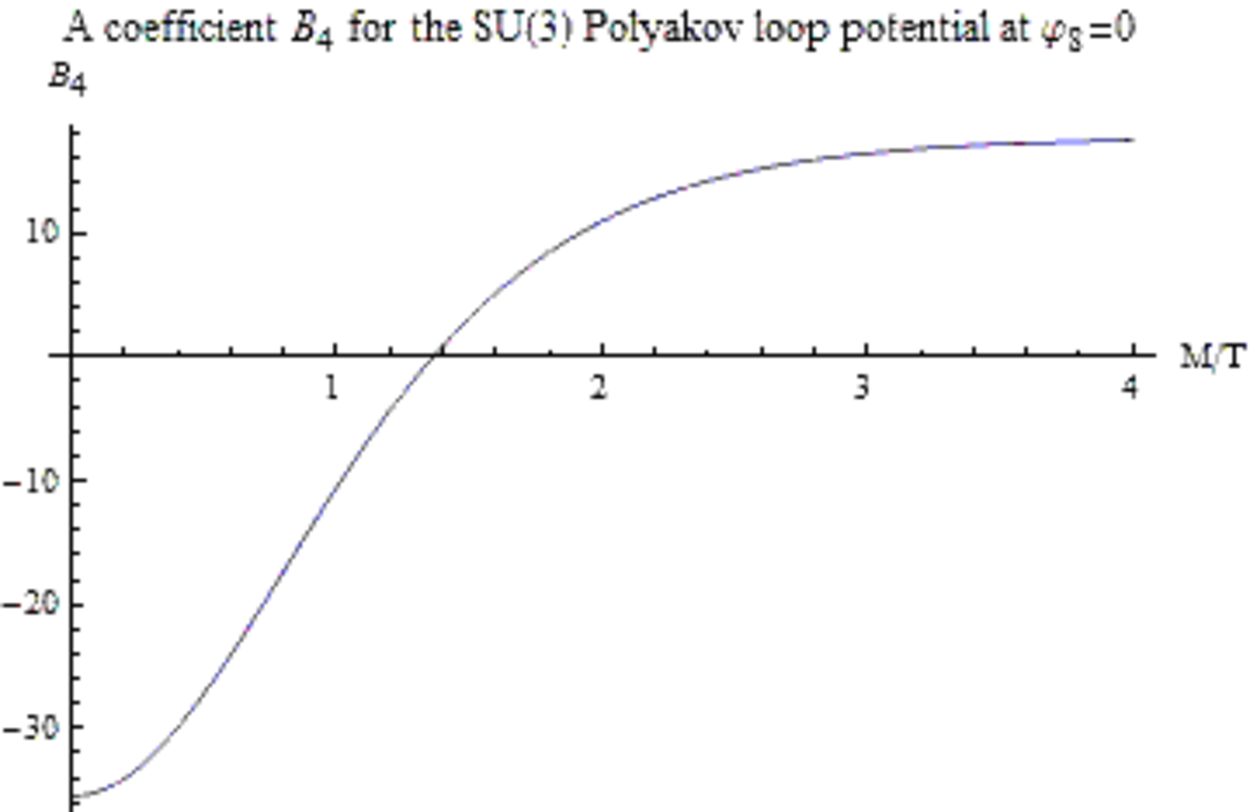}
\end{center}
\vskip -0.3cm
\caption{
The plot of 
$B_{2,\hat{M}}$,
$B_{3,\hat{M}}$,
and
$B_{4,\hat{M}}$,
for the $SU(3)$ Polyakov loop  effective potential 
as a function of $\hat{M}:=M/T$ at $D=4$. 
}
\label{fig:V-SU3-B}
\end{figure}

The transition can be observed by the effective potential directly written in terms of the gauge-invariant Polyakov loop average. 
Due to the existence of the center symmetry, the  real-valued effective potential must have the general form:
\begin{align}
 \hat{V}_0(L,0; \hat{M}) 
=&  a_{0,\hat{M}}   + \frac{a_{2,\hat{M}}}{2} L^* L + \frac{a_{3,\hat{M}}}{3} Re(L^3) 
\nonumber\\ &
+ \frac{a_{4,\hat{M}}}{4} (L^* L)^2 + O(L^5)
 ,
\end{align}
where $L$ is complex-valued in general. 
In fact, we find that the effective potential of this form reflects the $Z(3)$ center symmetry:
$\hat{V}_0(zL,0; \hat{M})=\hat{V}_0(L,0; \hat{M})$
where $z$ satisfies $zz^*=1$ and $z^3=1$.

At $\varphi_{8}=0$, the Polyakov loop operator $L$ is real-valued and given by
\begin{align}
 L  =  \frac13 \left[ 1 + 2 \cos \left(\frac{\varphi_{3}}{2} \right)  \right]  
 .  
\end{align}
At $\varphi_{8}=0$, the angle is related to the real-valued Polyakov loop operator as
\begin{align}
 \cos \left(\frac{\varphi_{3}}{2} \right) = \frac{3L-1}{2}  
 ,  
\end{align}
which yields
\begin{align}
 \cos  \varphi_{3} 
=  2 \cos ^2\left(\frac{\varphi_{3}}{2} \right) -1
= 2 \left( \frac{3L-1}{2}  \right)^2 -1 
.  
\end{align}

For $\varphi_8=0$, therefore, the  effective potential has the form:
\begin{align}
  \hat{V}_0(L,0; \hat{M}) 
=&  B_{0,\hat{M}}   + \frac{B_{2,\hat{M}}}{2!} L^2 + \frac{B_{3,\hat{M}}}{3!} L^3  
\nonumber\\ &
+ \frac{B_{4,\hat{M}}}{4!} L^4 + O(L^5)
 ,
\end{align}
If the first order transition is induced by the cubic term $\frac{B_{3,\hat{M}}}{3!} L^3$, then 
the transition from deconfinement to confinement occurs at the temperature $T_d$ at which the condition is satisfied:
\begin{align}
 B_{2,\hat{M}} = \frac{1}{3} (B_{3,\hat{M}})^2/B_{4,\hat{M}}  .
\end{align}
This condition determines the value of the ratio between the transition temperature $T_d$ and the gluonic mass $M(T)$ which may depend on temperature.
See Fig.~\ref{fig:V-SU3-B} for the plot of 
$B_{2,\hat{M}}$, $B_{3,\hat{M}}$, and $B_{4,\hat{M}}$ 
for the $SU(3)$ Polyakov loop  effective potential 
as a function of $\hat{M}:=M/T$ at $D=4$.

We find that the ratio between the transition temperature $T_d$ and the gluonic mass $M(T)$ is given for $D=4$ by  
\begin{align}
   \frac{M(T_d)}{T_d} = 2.75  \Longleftrightarrow \frac{T_d}{M(T_d)} =  0.364 
 .
\end{align}
For instance, 
\begin{align}
   M(T_d) =& 0.8 {\rm GeV} \leftrightarrow  T_d =  291 {\rm MeV} 
 .
\end{align}
This should be compared with the zero-temperature result:
\begin{align}
   M(T=0) = 0.8 \sim 1.0 {\rm GeV}  
 .
\end{align}

This result should be compared with the work \cite{RSTW15}: for $D=4$, the gluonic mass parameter $m=510{\rm MeV}$ was obtained from fits of $SU(3)$ lattice data for the gluon propagator in the Landau gauge at zero temperature, which gives the estimate on the transition temperature $T_c =185{\rm MeV}$.

\subsection{$SU(3)$ Pressure and entropy}

\begin{figure}[ptb]
\begin{center}
\includegraphics[scale=0.75]{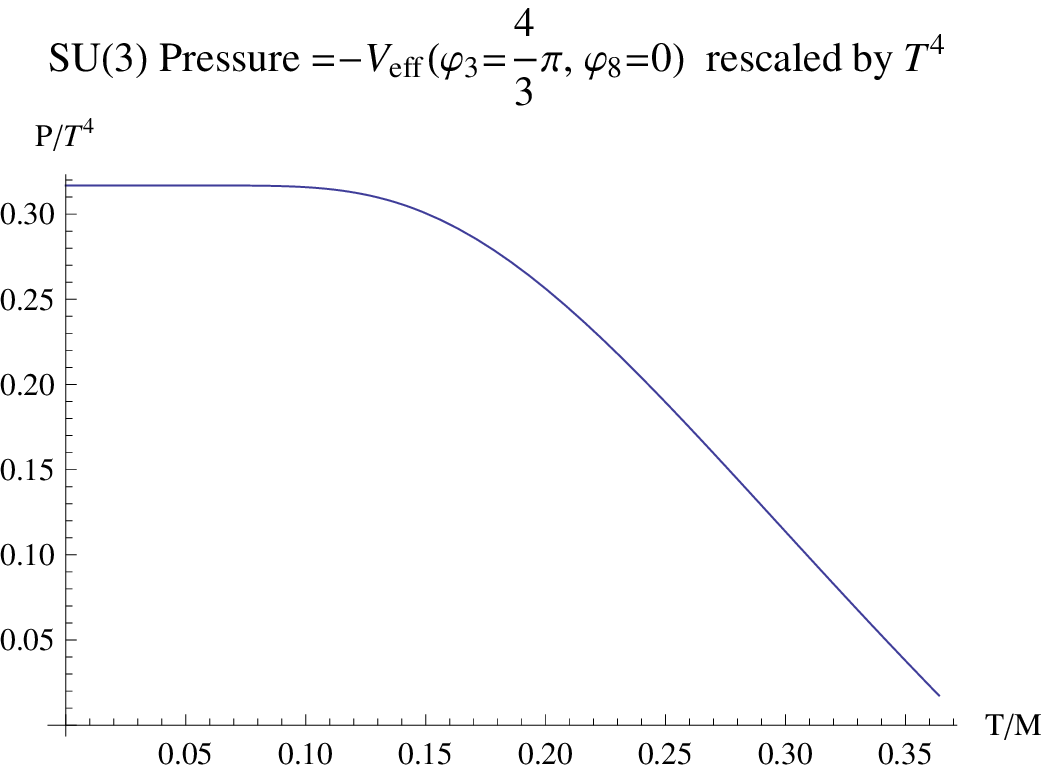}
\includegraphics[scale=0.75]{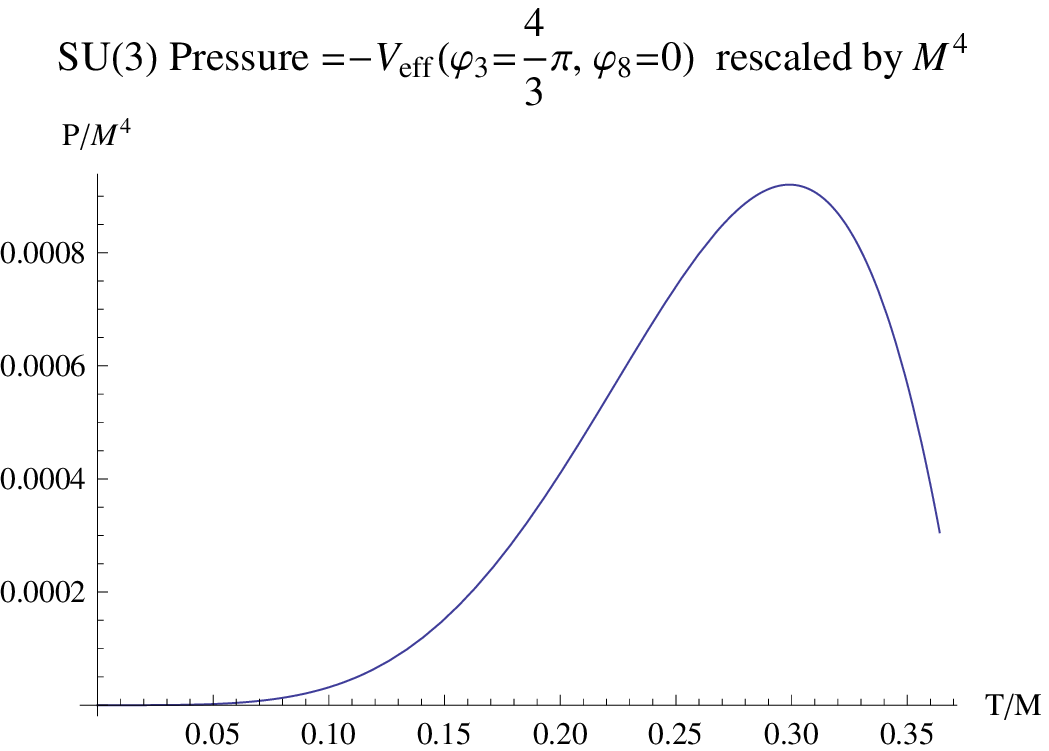}
\end{center}
\vskip -0.3cm
\caption{
The pressure $P$ as a function of $T/M$ in the low-temperature confined phase for $SU(3)$ and $D=4$. 
(Left) rescaled by $T^4$, (Right) rescaled by $M^4$.
Here the critical value is $T_d/M=0.364$.
}
\label{fig:P-SU3-a}
\end{figure}

The pressure $P$ is defined 
through the temperature-dependent minimum value of the effective potential:
\begin{align}
 P(T) := - V_{\rm eff}(\varphi=\varphi_{\rm min}(T))
 .
\end{align}
In the first approximation, we have 
\begin{align}
P/T^D =& - V_{\rm eff}(\varphi_{3}=\varphi_{3}^{\rm min},0)/T^D 
\nonumber\\
=& - (D-1) \left[F_{\hat{M}}(\varphi_{3}^{\rm min}) + 2 F_{\hat{M}} \left(\frac{\varphi_{3}^{\rm min}}{2}\right)  \right] 
\nonumber\\ &
 + \left[ F_{0}(\varphi_{3}^{\rm min}) + 2 F_{0} \left(\frac{\varphi_{3}^{\rm min} }{2} \right)   \right]
 ,
\end{align}
where $F_{\hat{M}}(\varphi)$ is defined by (\ref{F-def1}) and (\ref{F-def2}). 
See the first figure in Fig.~\ref{fig:P-SU3-a}.
In the low-temperature limit $\hat{M}:=M/T \to \infty$, the minimum of the effective potential is given at $\varphi_3^{\rm min}=\frac{4}{3}\pi$, which yields the positive value:
\begin{align}
 & P/T^D  
 \to  
   F_{0}(\frac{4}{3}\pi) + 2 F_{0} \left( \frac{2}{3}\pi \right)   
\nonumber\\ &
\Longrightarrow  \frac{13}{405} \pi^2  \simeq 0.316802  \ (D=4)  
 .
\end{align}
In the high-temperature limit $\hat{M}:=M/T \to 0$, the minimum of the effective potential is given at $\varphi_3^{\rm min}=0$, which yields the positive value:
\begin{align}
 &  P/T^D   
   \to  - (D-2) 3  F_{0}( 0)   
\nonumber\\ &
\Longrightarrow 
 -6 \times \frac{-1}{45} \pi^2 
= \frac{2}{15}\pi^2 \simeq 1.31595 \ (D=4)
 .
\end{align}

We observe that the pressure $P(T)$ remains positive at any temperature in the low-temperature confined phase $T<T_d \simeq 0.364M$. 
However, we find that the similar phenomenon to the $SU(2)$ case occurs also in the $SU(3)$ case.  
The pressure $P(T)$ is increasing at small temperature and hence the entropy $\mathcal{S}(T)$ is positive. However, as the temperature is increased, the pressure changes its monotony and begins to decrease, indicating that the entropy becomes negative in the region $T_0 < T < T_d$ before reaching the critical temperature $T_d$.
Here 
$T_0/M=0.29904$ for the $SU(3)$ case.
See the second figure in Fig.~\ref{fig:P-SU3-a}.
Therefore, we need the improvement  for $SU(3)$ case similar to the $SU(2)$ case, which will be reported in a subsequent work.


\section{Conclusion and discussion}


In this paper we have shown the existence of the confinement/deconfinement phase transition at a finite temperature $T_d$  in $SU(2)$ and $SU(3)$ Yang-Mills theories by calculating the effective potential of the Polyakov loop average.   
The key ingredient to derive the phase transition is the introduction of a dynamically generated gluonic mass $M$ in the reformulated Yang-Mills theory which allows one to introduce the gauge-invariant mass term for a specific gluonic degree of freedom. 
The transition temperature $T_d$ is estimated as the ratio to the gluonic mass $M$. The transition is continuous for $SU(2)$ and discontinuous for $SU(3)$.

The existence of the gluonic mass has been established at zero temperature and has already played the very important role in understanding quark confinement at zero temperature.
The existence of the gluonic mass across the transition temperature enables one to easily understand the occurrence of the confinement/deconfinement phase transition at finite temperature. 
Our result also confirms the well-known fact that the confinement/deconfinement phase transition signaled by the Polyakov loop average is associated to the center symmetry restoration/breaking and the infinity/finiteness of the free energy for a single quark. 
Our derivation of the transition gives also the microscopic mechanism for confinement/deconfinement. 
In this sense, the gluonic mass at finite temperature is more important and powerful than the zero temperature, since it directly yields quark confinement. 

An important point missing in our analytical study is the lack of the analytical derivation of the gluonic mass in the same framework. 
At zero temperature, such a calculation has been given in the previous work  \cite{Kondo06}. 
More detailed investigations on the gluonic mass at finite temperature will be given in a subsequent paper \cite{KS15}.

{\it Acknowledgements}\ ---
The author would like to express sincere thanks to Matthieu Tissier and Julien Serreau for valuable discussions on the related works and hospitality extended in his visit to Laboratoire APC, Universite Paris Diderot and Universite de Paris VI (Pierre et Marie Curie) from 30 March to 3 April 2015.
This work is  supported by Grant-in-Aid for Scientific Research (C) 24540252 and 15K05042 from Japan Society for the Promotion of Science (JSPS).

\appendix
\section{Integration over the fields}\label{Appendix:path-integral}

The total Lagrangian except for the FP ghost term is 
\begin{align}
   \mathscr{L}_{\rm YM} + \mathscr{L}_{\rm red} + \mathscr{L}_{\rm m}
=& -\frac{1}{4} \mathscr{F}_{\mu\nu}^A[\mathscr{V}] \mathscr{F}^{\mu\nu}{}^A[\mathscr{V}]
\nonumber\\&
- \frac{1}{2} \mathscr{X}^{\mu A}  K_{\mu\nu}^{AB}[\mathscr{V}] \mathscr{X}^{\nu B} 
\nonumber\\&
+ \mathscr{N}^A \mathscr{D}_\mu^{AB}[\mathscr{V}] \mathscr{X}^{\mu B}
\nonumber\\&
+ \frac12 \alpha \mathscr{N}^A \mathscr{N}^A 
+ O(\mathscr{X}^3) ,
\end{align}
where we have defined
\begin{align}
 K_{\mu\nu}^{AB}[\mathscr{V}]  
:=&  W_{\mu\nu}^{AB}[\mathscr{V}] - M^2 g_{\mu\nu}\delta^{AB}
\nonumber\\ 
=&  Q_{\mu\nu}^{AB}[\mathscr{V}]   + \mathscr{D}_\mu^{AC}[\mathscr{V}] \mathscr{D}_\nu^{CB}[\mathscr{V}] ,
\nonumber\\ 
Q_{\mu\nu}^{AB}[\mathscr{V}] :=&  - g_{\mu\nu} (\mathscr{D}_\rho[\mathscr{V}]\mathscr{D}^\rho[\mathscr{V}])^{AB}  
 - M^2 g_{\mu\nu}\delta^{AB}  
\nonumber\\&
+ 2g_{{}_{\rm YM}}f^{ABC} \mathscr{F}_{\mu\nu}^{C}[\mathscr{V}]  .
\end{align}
Here we have added the term $\frac12 \alpha \mathscr{N}^A \mathscr{N}^A$ to see the difference between the massless and massive cases. 
In what follows, we consider the case of $\mathscr{F}_{\mu\nu}[\mathscr{V}]=0$, which means that $Q_{\mu\nu}^{AB}$ is diagonal in $\mu, \nu$. 
Then the total Lagrangian except for the ghost term and the higher-order terms in $\mathscr{X}$ is cast into the quadratic form: 
\begin{align}
&    - \frac{1}{2} \mathscr{X}^{\mu A}  K_{\mu\nu}^{AB} \mathscr{X}^{\nu B} 
+ \mathscr{N}^A \mathscr{D}_\mu^{AB}[\mathscr{V}] \mathscr{X}^{\mu B}
\nonumber\\
=&   \frac12 
\begin{bmatrix} 
 \mathscr{X}^A_\mu & \mathscr{N}^A 
\end{bmatrix} 
\begin{bmatrix} 
 -Q_{\mu\nu}^{AB} - \mathscr{D}_\mu^{AC}  \mathscr{D}_\nu^{CB} & -\mathscr{D}_\mu^{AB} \\
 \mathscr{D}_\nu^{AB} & \alpha \delta^{AB} 
\end{bmatrix}
\begin{bmatrix} 
 \mathscr{X}^B_\nu \\
 \mathscr{N}^B 
\end{bmatrix} 
 .
\end{align}
The integration over the field $\mathscr{X}_{\mu}$ and $\mathscr{N}$ is performed by the Gaussian integration and leads to the determinant:
\begin{align}
&  \det
\begin{bmatrix} 
 -Q_{\mu\nu}^{AB}[\mathscr{V}] - \mathscr{D}_\mu^{AC}[\mathscr{V}]  \mathscr{D}_\nu^{CB}[\mathscr{V}] & -\mathscr{D}_\mu^{AB}[\mathscr{V}] \\
 \mathscr{D}_\nu^{AB}[\mathscr{V}] & \alpha \delta^{AB} 
\end{bmatrix}
\nonumber\\
=& \{ \det [(\mathscr{D}^{\rho}[\mathscr{V}] \mathscr{D}_{\rho}[\mathscr{V}] )^{AB} + M^2\delta^{AB}] \}^{D-1}
\nonumber\\ & \times
  \det [(\mathscr{D}^{\rho}[\mathscr{V}] \mathscr{D}_{\rho}[\mathscr{V}] )^{AB} + \alpha M^2 \delta^{AB}] 
 ,
\end{align}
where the determinant over the Lorentz indices $\mu, \nu$ is calculated. 
In the absence of mass $M=0$, the determinant is independent of $\alpha$. This is not the case in the presence of mass $M \ne 0$.
The reduction condition is achieved at $\alpha=0$.
Hence, we obtain the $(D-1)$ massive vector modes and one massless  scalar  mode.

The new variables have independent degrees of freedom which are less than those expected from their appearance. 
For instance, the field $\mathscr{X}_\mu^A$ ($A=1,...,{\rm dim}G=N^2-1$) has the independent components $X_\mu^a$ ($a=1,...,{\rm dim}(G/\tilde H)$).
In oder to integrate out the field variables according to the path-integral formulation, we must correctly specify independent degrees of freedom for the respective field variable. 
This was achieved by adopting an appropriate reference of frame for the target space of the field \cite{Kondo06} for $SU(2)$, and section 7.3 and Appendix J in \cite{KKSS15} for $SU(3)$. 
Consequently, we can write the Lagrangian in terms of the independent degrees of freedom as follows.
 
Thus the $SU(3)$ Yang-Mills Lagrangian density reads
\begin{align}
   \mathscr{L}_{\rm YM} 
=&  -\frac{1}{2} {\rm tr}( \mathscr{F}_{\mu\nu}[\mathscr{V}] \mathscr{F}^{\mu\nu}[\mathscr{V}] )
-    \frac12   X^{\mu a}  Q_{\mu\nu}^{ab} X^{\nu b} 
\nonumber\\& 
+ O(X^3)
 , 
\end{align}
where we have defined 
\begin{equation}
Q_{\mu\nu}^{ab}  := K^{ab}  g_{\mu\nu} 
+ 2ig  \mathscr{F}_{\mu\nu}^C [\mathscr{V}] (T_C)^{ab} ,
\label{def:Q}
\end{equation}
with 
\begin{align}
K^{ab} =& 
- \partial^\rho \partial_\rho \delta_{ab} 
-  g G_\rho^j g G^{\rho k}  f^{jac} f^{kcb}
\nonumber\\& 
+  [2 g G_\rho^j    \partial^\rho 
+   \partial^\rho  (g G_\rho^j)] f^{jab} 
  .
  \label{def:K}
\end{align}
For  $G=SU(3)$ in the \textbf{maximal option}, $K^{ab}$ is given in the usual Gell-Mann basis  by
\begin{align}
 j,k \in \{ 3,8 \}, \quad a,b,c,d  \in \{ 1,2,4,5,6,7 \} 
  .
\end{align}
For  $G=SU(3)$ in the \textbf{minimal option}, $K^{ab}$ is given in the usual Gell-Mann basis  by
\begin{align}
  j,k \in \{ 1,2,3,8 \},  \quad a,b,c,d  \in \{  4,5,6,7 \} 
  .
\end{align}
For $G=SU(2)$, in particular, we have
\begin{align}
   \mathscr{L}_{\rm YM} 
=&  -\frac{1}{2} {\rm tr}( \mathscr{F}_{\mu\nu}[\mathscr{V}] \mathscr{F}^{\mu\nu}[\mathscr{V}] )
-    \frac12   X^{\mu a}  Q_{\mu\nu}^{ab} X^{\nu b} 
\nonumber\\& 
-  \frac14 g^2 \epsilon^{ab} X_\mu^a  X_\nu^b \epsilon^{cd}  X^{\mu c}  X^{\nu d}
 , 
\nonumber\\
Q_{\mu\nu}^{ab}  :=& K^{ab}  g_{\mu\nu} 
+ 2ig  \mathscr{F}_{\mu\nu}^C [\mathscr{V}] (T_C)^{ab} 
 \quad (a,b,c,d \in \{ 1,2 \} ) ,
\end{align}
with 
\begin{align}
K^{ab} := 
[- \partial^\rho \partial_\rho + g G^{\rho} g G_\rho   ] \delta^{ab} +     [2 g G_\rho \partial^\rho + \partial^\rho  (g G_\rho)
] \epsilon^{ab} 
  ,
\end{align}
where we have used $\epsilon^{ac3} \epsilon^{cb3}=\epsilon^{ac} \epsilon^{cb}=-\delta_{ab}$.

We introduce the complex-valued field $\Phi^\pm$ defined by
\begin{equation}
 \Phi^{\pm} := \frac{1}{\sqrt{2}} ( \Phi^{1} \pm i \Phi^{2} ) \in \mathbb{C} ,
\end{equation}
and the inverse relation is given by
\begin{equation}
 \Phi^{1} = \frac{1}{\sqrt{2}} (\Phi^{+} + \Phi^{-}) , \ \Phi^{2} = \frac{1}{\sqrt{2}} ( - i \Phi^{+} + i \Phi^{-} )  .
\end{equation}
The term quadratic in $\Phi^a$ can be diagonalized by   the complex variable:
\begin{align}
& \Phi^{A} [ - (\mathscr{D}^{\rho}[\mathscr{V}] \mathscr{D}_{\rho}[\mathscr{V}] )^{AB} 
+ M^2\delta^{AB} ] \Phi^{B} 
\nonumber\\
=& \Phi^{a} [K^{ab} + M^2 \delta^{ab} ] \Phi^b
\nonumber\\
=& \Phi^+ [ -(\partial_\rho-igG_\rho)^2 + M^2 ] \Phi^- 
\nonumber\\ &
+ \Phi^- [  -(\partial_\rho+igG_\rho)^2  + M^2 ] \Phi^+ .
\end{align}
Here we have used
\begin{align}
 & P^a Q^a = P^a \delta^{ab} Q^b = P^{+} Q^{-} + P^{-} Q^{+} ,
\nonumber\\
 & P^a \epsilon^{ab} Q^b = i(P^{+} Q^{-} - P^{-} Q^{+})  
.  
\end{align}
Thus we obtain
\begin{align}
& \det [ - (\mathscr{D}^{\rho}[\mathscr{V}] \mathscr{D}_{\rho}[\mathscr{V}] )^{AB} 
+ M^2\delta^{AB} ]
\nonumber\\
=&  [ - D_{\mu}^{2} [G] + M^2 ] [ - \bar{D}_{\mu}^{2} [G] + M^2 ]  ,
\end{align}
where 
\begin{align}
  D_\mu[G] :=\partial -igG_\mu, \quad
 \bar{D}_\mu[G] :=\partial +igG_\mu .
\end{align}
This yields 
\begin{align}
& {\rm Tr} \ln   [ - (\mathscr{D}^{\rho}[\mathscr{V}] \mathscr{D}_{\rho}[\mathscr{V}] )^{AB} 
+ M^2\delta^{AB} ]
\nonumber\\
=& \ln \det [ - (\mathscr{D}^{\rho}[\mathscr{V}] \mathscr{D}_{\rho}[\mathscr{V}] )^{AB} 
+ M^2\delta^{AB} ]
\nonumber\\
=&  {\rm Tr} \ln [ - D_{\mu}^{2} [G] + M^2 ] + {\rm Tr} \ln [ - \bar{D}_{\mu}^{2} [G] + M^2 ]  .
\end{align}

\begin{widetext}
\section{Power series expansion and the coefficients}\label{Appendix:series}

The effective potential $V_{\rm eff}(\varphi_{3},\varphi_{8})$ at $\varphi_8=0$  is given by
\begin{align}
  V_{\rm eff}(\varphi_{3},0)/T^D 
=& (D-1) \left[F_{\hat{M}}(\varphi_{3}) + 2 F_{\hat{M}} \left(\frac{\varphi_{3}}{2}\right)  \right] 
 - \left[ F_{0}(\varphi_{3}) + 2 F_{0} \left(\frac{\varphi_{3} }{2} \right)   \right]
 ,
\end{align}
where
\begin{align}
 & F_{\hat{M}}(\varphi_{3}) + 2 F_{\hat{M}}(\varphi_{3}/2)
=  \int \frac{d^{D-1}\hat{p}}{(2\pi)^{D-1}} [f_{\hat{M}}(\hat{p}^2, \varphi_{3}) + 2 f_{\hat{M}}(\hat{p}^2, \varphi_{3}/2)]
\nonumber\\
 =& C_D \int_{0}^{\infty} d \hat{p} \ \hat{p}^{D-2} [f_{\hat{M}}(\hat{p}^2, \varphi_{3}) + 2 f_{\hat{M}}(\hat{p}^2, \varphi_{3}/2)]
\nonumber\\
 =& C_D \int_{0}^{\infty} d \hat{p} \ \hat{p}^{D-2} \{ \ln [1+e^{-2 \sqrt{\hat{\bm{p}}^2 +\hat{M}^2 }} - 2 e^{-  \sqrt{\hat{\bm{p}}^2 +\hat{M}^2}} \cos ( \varphi_{3})] 
+2  \ln [1+e^{-2 \sqrt{\hat{\bm{p}}^2 +\hat{M}^2 }} - 2 e^{-  \sqrt{\hat{\bm{p}}^2 +\hat{M}^2}} \cos ( \varphi_{3}/2 )] \}
 .
\end{align}
The cosine is expanded into the power series in $\varphi_{3}$ about $\varphi_3=4\pi/3$ at which $L=0$: defining $\sigma :=\varphi_{3} - 4\pi/3$
\begin{align}
  \cos \varphi_3 =& -\frac{1}{2}+\frac{1}{2} \sqrt{3} \sigma +\frac{1}{4} \sigma^2-\frac{1}{4
   \sqrt{3}} \sigma^3 -\frac{1}{48} \sigma^4+O\left( \sigma^5 \right)
 ,
\nonumber\\
  \cos \frac{\varphi_3}{2} =& -\frac{1}{2}-\frac{1}{4} \sqrt{3} \sigma +\frac{1}{16} \sigma^2+\frac{1}{32
   \sqrt{3}} \sigma^3 - \frac{1}{768} \sigma^4
+O\left( \sigma^5 \right)
 .
\end{align}
Then $f_{\hat{M}}(\hat{p}^2, \varphi_{3})$ is expanded into the power series  in $\sigma$:
\begin{align}
  f_{\hat{M}}(\hat{p}^2, \varphi_{3})
=& c^0_{\hat{M}}(\hat{p}) + c^1_{\hat{M}}(\hat{p}) \sigma + c^2_{\hat{M}}(\hat{p}) \sigma^2 
+ c^3_{\hat{M}}(\hat{p}) \sigma^3  
+ c^4_{\hat{M}}(\hat{p}) \sigma^4 + O(\sigma^5) 
 ,
\end{align}
with the coefficients: 
\begin{align}
 c^0_{\hat{M}}(\hat{p}) =&  \ln (1+e^{- \hat\epsilon_{p}}+e^{- 2\hat\epsilon_{p}}) > 0 ,
\nonumber\\
 c^1_{\hat{M}}(\hat{p}) :=& - \frac{\sqrt{3} e^{- \hat\epsilon_{p}}}{1+e^{- \hat\epsilon_{p}}+e^{- 2\hat\epsilon_{p}}} < 0 ,
\nonumber\\
 c^2_{\hat{M}}(\hat{p}) =& - \frac{ e^{- \hat\epsilon_{p}} (1+4e^{- \hat\epsilon_{p}} + e^{- 2\hat\epsilon_{p}}) }{2(1+e^{- \hat\epsilon_{p}}+e^{- 2\hat\epsilon_{p}})^2}   
 < 0 ,
\nonumber\\
 c^3_{\hat{M}}(\hat{p}) =& \frac{ e^{- \hat\epsilon_{p}} (1-e^{- \hat\epsilon_{p}} -6 e^{- 2\hat\epsilon_{p}}  -e^{-3 \hat\epsilon_{p}} + e^{-4 \hat\epsilon_{p}})}{2 \sqrt{3}(1+e^{- \hat\epsilon_{p}}+e^{- 2\hat\epsilon_{p}})^2} ,  
\nonumber\\
 c^4_{\hat{M}}(\hat{p}) =& 
 \frac{e^{-\hat\epsilon_{p}} \left( 1+12
   e^{- \hat\epsilon_{p}}-12 e^{-2 \hat\epsilon_{p}}-56 e^{-3 \hat\epsilon_{p}}-12 e^{-4
   \hat\epsilon_{p}}+12 e^{-5\hat\epsilon_{p}}+e^{-6 \hat\epsilon_{p}} \right)}{24
   \left( 1+e^{-\hat\epsilon_{p}}+e^{-2 \hat\epsilon_{p}} \right)^4} ,
\end{align}
where
\begin{align}
 \hat\epsilon_{p} :=\sqrt{\hat{\bm{p}}^2 +\hat{M}^2 } .
\end{align}
The expansion of the integrand 
$
f_{\hat{M}}(\hat{p}^2, \varphi_{3}) + 2 f_{\hat{M}}(\hat{p}^2, \varphi_{3}/2)
$
into the power series of $\sigma$ is given by

\begin{align}
    f_{\hat{M}}(\hat{p}^2, \varphi_{3}) + 2 f_{\hat{M}}(\hat{p}^2, \varphi_{3}/2)
=&  h^{(0)}_{\hat{M}}(\hat{p}) + h^{(1)}_{\hat{M}}(\hat{p}) \sigma + h^{(2)}_{\hat{M}}(\hat{p}) \sigma^2 
+ h^{(3)}_{\hat{M}}(\hat{p}) \sigma^3  
+ h^{(4)}_{\hat{M}}(\hat{p}) \sigma^4 + O(\sigma^5) 
 ,
\end{align}
with the coefficients:
\begin{align}
 h^{(0)}_{\hat{M}}(\hat{p}) =&  3 c^{(0)}_{\hat{M}}(\hat{p}) 
=  3\ln (1+e^{- \hat\epsilon_{p}}+e^{- 2\hat\epsilon_{p}}) > 0 ,
\nonumber\\
 h^{(1)}_{\hat{M}}(\hat{p}) =& 0 ,
\nonumber\\
 h^{(2)}_{\hat{M}}(\hat{p}) =& \frac32 c^{(2)}_{\hat{M}}(\hat{p}) 
= - \frac{3}{4} \frac{ e^{- \hat\epsilon_{p}} (1+4e^{- \hat\epsilon_{p}} + e^{- 2\hat\epsilon_{p}}) }{ (1+e^{- \hat\epsilon_{p}}+e^{- 2\hat\epsilon_{p}})^2}   
 < 0
\nonumber\\
 h^{(3)}_{\hat{M}}(\hat{p}) 
=& \frac{3}{4} c^{(3)}_{\hat{M}}(\hat{p}) 
=  \frac{\sqrt{3}}{8}  \frac{ e^{- \hat\epsilon_{p}} (1-e^{- \hat\epsilon_{p}} -6 e^{- 2\hat\epsilon_{p}}  -e^{-3 \hat\epsilon_{p}} + e^{-4 \hat\epsilon_{p}})}{ (1+e^{- \hat\epsilon_{p}}+e^{- 2\hat\epsilon_{p}})^3}
\nonumber\\
 h^{(4)}_{\hat{M}}(\hat{p}) 
=& \frac{9}{8} c^{(4)}_{\hat{M}}(\hat{p}) 
=   \frac{3}{64} \frac{ e^{-\hat\epsilon_{p}} \left( 1+12
   e^{- \hat\epsilon_{p}}-12 e^{-2 \hat\epsilon_{p}}-56 e^{-3 \hat\epsilon_{p}}-12 e^{-4
   \hat\epsilon_{p}}+12 e^{-5\hat\epsilon_{p}}+e^{-6 \hat\epsilon_{p}} \right)}{
   \left( 1+e^{-\hat\epsilon_{p}}+e^{-2 \hat\epsilon_{p}} \right)^4}
 .
\end{align}
\end{widetext}



\begin{thebibliography}{99}
\bibitem{QCDtexts}   
 K. Yagi, T. Hatsuda and Y. Miake,
{\em Quark-Gluon Plasma} 
(Cambridge Univ. press, Cambridge, 2005).
\\
J.B. Kogut and M.A. Stephanov,   
{\em The phases of quantum chromodynamics: From confinement to extreme environments} 
(Cambridge Univ. press, Cambridge, 2004).


\bibitem{QCDreviews}
J.O. Andersen, W.R. Naylor, and A. Tranberg,  
arXiv:1411.7176 [hep-ph] 
\\
K. Fukushima and C. Sasaki,
Prog. Part. Nucl. Phys. {\bf 72}, 99--154   (2013).
arXiv:1301.6377 [hep-ph] 
\\
K. Fukushima,
J. Phys. G{\bf 39},  013101 (2012). 
arXiv:1108.2939 [hep-ph] 
\\
K. Fukushima and T. Hatsuda,
Rept. Prog. Phys. {\bf 74},  014001 (2011). 
arXiv:1005.4814 [hep-ph]
\\
O. Philipsen,
Prog. Part. Nucl. Phys. {\bf 70}, 55--107 (2013).  
arXiv:1207.5999 [hep-lat]
 

\bibitem{Kondo10}
K.-I. Kondo,
Phys. Rev. D{\bf 82}, 065024 (2010).  
arXiv:1005.0314 [hep-th] 


\bibitem{KKSS15}
K.-I. Kondo, S. Kato, A. Shibata and T. Shinohara,
Phys. Rept. {\bf 579}, 1--226 (2015).
arXiv:1409.1599 [hep-th].


\bibitem{Cho80}
  Y.M. Cho,
Phys. Rev. D 21, 1080--1088 (1980).
Y.M. Cho,
Phys. Rev. D {\bf 23}, 2415--2426 (1981). 


\bibitem{DG79}
  Y.S. Duan and M.L. Ge, 
Sinica Sci., {\bf 11}, 1072 (1979). 


\bibitem{FN99} 
  L. Faddeev and A.J. Niemi,
Phys. Rev. Lett. {\bf 82}, 1624--1627 
(1999).
[hep-th/9807069] 


\bibitem{Shabanov99}
  S.V. Shabanov,
Phys. Lett. B{\bf 458}, 322--330 (1999).
[hep-th/9903223] 
\\
  S.V. Shabanov, 
Phys. Lett. B{\bf 463}, 263--272 (1999).
[hep-th/9907182] 


\bibitem{Cho80c}
Y.M. Cho, 
Unpublished preprint, 
MPI-PAE/PTh 14/80 (1980).
\\
Y.M. Cho,
Phys. Rev. Lett. {\bf 44}, 1115--1118 (1980).


\bibitem{FN99a} 
L. Faddeev and A.J. Niemi,
Phys. Lett. B{\bf 449}, 214--218 (1999).
[hep-th/9812090] 
\\
L. Faddeev and A.J. Niemi,
Phys. Lett. B{\bf 464}, 90--93 (1999).
[hep-th/9907180] 
\\
T.A. Bolokhov and L.D. Faddeev,
Theoretical and Mathematical Physics, {\bf 139}, 679--692 (2004).


\bibitem{KMS05}
  K.-I. Kondo, T. Murakami and T. Shinohara,
Eur. Phys. J. C {\bf 42}, 475--481 (2005).
[hep-th/0504198] 


\bibitem{KMS06}
  K.-I. Kondo, T. Murakami and T. Shinohara,
Prog. Theor. Phys. {\bf 115}, 201--216 (2006). 
[hep-th/0504107]  


\bibitem{Kondo06}
  K.-I. Kondo,
Phys. Rev. D{\bf 74},  125003 (2006).  
[hep-th/0609166]  


\bibitem{KSM08}
K.-I. Kondo, T. Shinohara, and T. Murakami,
Prog. Theor. Phys. {\bf 120}, 1-50  (2008). 
arXiv:0803.0176 [hep-th] 


\bibitem{Wetterich93}
C. Wetterich, 
Phys. Lett. B{\bf 301}, 90--94 (1993).


\bibitem{FRG}
Y. Igarashi, K. Itoh and H. Sonoda,
Prog. Theor. Phys. Suppl. {\bf 181}, 1--166 (2009).  
\\
J.M. Pawlowski,
Annals Phys. 322, 2831--2915 (2007).  
[hep-th/0512261] 
\\
H. Gies, 
hep-ph/0611146. 
\\
J. Berges, N. Tetradis and C. Wetterich,  
Phys. Rept. {\bf 363}, 223--386 (2002). 
[hep-ph/0005122]   
 

\bibitem{WRG}
 K.G. Wilson and John B. Kogut,
Phys. Rept. {\bf 12}, 75--200 (1974). 


\bibitem{HRCW08}
T. Hell, S. R\"ossner, M. Cristoforetti and W. Weise, 
Phys. Rev. D{\bf 79}, 014022 (2009).
arXiv:0810.1099  


\bibitem{SFR06}
C. Sasaki, B. Friman and K. Redlich, 
Phys. Rev. D{\bf 75}, 074013 (2007). 
[hep-ph/0611147] 


\bibitem{BBRV07}
D. Blaschke, M. Buballa, A.E. Radzhabov and M.K. Volkov,
Yad.Fiz. {\bf 71}, 2012--2018 (2008), Phys. Atom. Nucl.{\bf 71}, 1981--1987 (2008). 
arXiv:0705.0384 [hep-ph]  
\\
D. Gomez Dumm, D.B. Blaschke, A.G. Grunfeld, and N.N. Scoccola,    
Phys. Rev. D{\bf 73}, 114019 (2006). 
[hep-ph/0512218] 


\bibitem{Fukushima04}
K. Fukushima, 
Phys. Lett. B{\bf 591}, 277--284 (2004). 
[hep-ph/0310121] 


\bibitem{HRCW10}
T. Hell, S. Rossner, M. Cristoforetti, W. Weise,
Phys.Rev. D81, 074034  (2010). 
arXiv:0911.3510 [hep-ph]


\bibitem{HKW11}
T. Hell, K. Kashiwa, and W. Weise,
Phys. Rev. D{\bf 83}, 114008 (2011). 
e-Print: arXiv:1104.0572 [hep-ph]

 
\bibitem{RBBV11}
A.E. Radzhabov, D. Blaschke, M. Buballa, M.K. Volkov,
Phys. Rev. D83, 116004 (2011).  
arXiv:1012.0664 [hep-ph]


\bibitem{SSKY10}
Y. Sakai, T. Sasaki, H. Kouno, M. Yahiro,
Phys. Rev. D{\bf 82},  076003 (2010). 
arXiv:1006.3648 [hep-ph] 
\\
Y. Sakai, T. Sasaki, H. Kouno, M. Yahiro,
J. Phys. G{\bf 39},  035004 (2012). 
arXiv:1104.2394 [hep-ph] 


\bibitem{BS83}
C. Borgs, and E. Seiler,
Commun. Math. Phys. {\bf 91}, 329--380 (1983).  
Nucl. Phys. B{\bf 215}, 125--135 (1983).  


\bibitem{Polyakov78}
A.M. Polyakov, 
Phys. Lett. B {\bf 72}, 477--480 (1978).


\bibitem{MP08}
F. Marhauser and J.M. Pawlowski,
arXiv:0812.1144 [hep-ph].


\bibitem{BGP10}
J. Braun, H. Gies and J.M. Pawlowski, 
Phys. Lett. B{\bf 684}, 262
 (2010). 
arXiv:0708.2413 [hep-th], 


\bibitem{BEGP10}
J. Braun, A. Eichhorn, H. Gies, J.M. Pawlowski,
Eur. Phys. J. C\textbf{70},  689--702 (2010). 
arXiv:1007.2619 [hep-ph] 


\bibitem{FP13}
L. Fister and J.M. Pawlowski, 
Phys. Rev. D{\bf 88}, 045010 (2013). 
arXiv:1301.4163 [hep-ph],


\bibitem{HPS11}
T.K. Herbst, J.M. Pawlowski, and Bernd-Jochen Schaefer,
Phys. Lett. B\textbf{696},  58--67 (2011). 
arXiv:1008.0081 [hep-ph] 


\bibitem{HSBBPSB13}
L.M. Haas, R. Stiele, J. Braun, J.M. Pawlowski, J. Schaffner-Bielich,
Phys.Rev. D{\bf 87}, 076004 (2013). 
arXiv:1302.1993 [hep-ph] 


\bibitem{tHooft81}
  G. 't Hooft,
  Nucl.Phys. B{\bf 190} [FS3], 455--478 (1981).


\bibitem{EI82}
  Z.F. Ezawa and A. Iwazaki,
  Phys. Rev. D{\bf 25}, 2681--2689 (1982).


\bibitem{SY90}
  T. Suzuki and I. Yotsuyanagi,
  Phys. Rev. D{\bf 42}, 4257--4260 (1990).


\bibitem{KLSW87}
  A. Kronfeld, M. Laursen, G. Schierholz and U.-J. Wiese,
  Phys.Lett. B {\bf 198}, 516--520 (1987).  


\bibitem{SNW94}
J.D. Stack, S.D. Neiman and R. Wensley,
Phys. Rev. D{\bf 50}, 3399--3405 (1994).
[hep-lat/9404014] 
H.~Shiba and T.~Suzuki,
Phys.Lett.B{\bf 333}, 461--466 (1994).
[hep-lat/9404015]  


\bibitem{dualsuper}
  Y. Nambu,
  Phys. Rev. D{\bf 10}, 4262--4268 (1974).
\\
G. 't Hooft,
  in: High Energy Physics, edited by A. Zichichi 
(Editorice Compositori, Bologna, 1975).
\\
S. Mandelstam,
 Phys. Report{\bf 23}, 245--249 (1976).
\\
A.M. Polyakov,
  Phys. Lett. B{\bf 59}, 82--84 (1975).
  Nucl. Phys. B{\bf 120}, 429--458 (1977).


\bibitem{AS99}
  K. Amemiya and H. Suganuma,
Phys. Rev. D{\bf 60}, 114509 (1999).
[hep-lat/9811035] 


\bibitem{BCGMP03}
  V.G. Bornyakov, M.N. Chernodub, F.V. Gubarev, S.M. Morozov and M.I. Polikarpov, 
 Phys. Lett. B{\bf 559}, 214--222 (2003).
  [hep-lat/0302002]  


\bibitem{KKMSSI05}
  S. Kato, K.-I. Kondo, T. Murakami, A. Shibata, T. Shinohara and S. Ito,
Phys. Lett. B {\bf 632}, 326--332
 (2006).
[hep-lat/0509069]  


\bibitem{IKKMSS06}
  S. Ito, S. Kato, K.-I. Kondo, T. Murakami, A. Shibata and T. Shinohara,  
Phys. Lett. B {\bf 645}, 67--74  (2007).  
[hep-lat/0604016]  


\bibitem{SKKMSI07}
A. Shibata, S. Kato, K.-I. Kondo, T. Murakami, T. Shinohara and  S. Ito,
Phys. Lett. B{\bf 653}, 101--108  (2007). 
arXiv:0706.2529 [hep-lat] 


\bibitem{Shibata-lattice2007}
A. Shibata, S. Kato, K.-I. Kondo, T. Murakami, T. Shinohara, and S. Ito, 
POS(LATTICE-2007)331, 
arXiv:0710.3221 [hep-lat] 
 

\bibitem{SKKS13}
A. Shibata, K.-I. Kondo, S. Kato and T. Shinohara, 
Phys. Rev. D{\bf 87}, 054011 (2013).
arXiv:1212.6512 [hep-lat] 


\bibitem{Kondo14}
K.-I. Kondo,
Phys. Rev. D{\bf 89}, 105013 (2014).  
arXiv:1309.2337 [hep-th] 


\bibitem{Kondo04}
K.-I. Kondo,
Phys. Lett. B{\bf 600}, 287--296 (2004). 
e-Print: hep-th/0404252 
\\
K.-I. Kondo,
Int. J. Mod. Phys. A{\bf 20}, 4609--4614 (2005). 
e-Print: hep-th/0410024 


\bibitem{Kondo01}
K.-I. Kondo,
Phys. Lett. B{\bf 514}, 335--345 (2001). 
[hep-th/0105299] 
\\
K.-I. Kondo,
Phys. Lett. B{\bf 572}, 210--215 (2003). 
[hep-th/0306195]  


\bibitem{EGP11}
A. Eichhorn, H. Gies, and J. M. Pawlowski,
Phys. Rev. D{\bf 83}, 045014  (2011), Erratum-ibid. D{\bf 83}, 069903  (2011).
arXiv:1010.2153 [hep-ph] 


\bibitem{CSME00}
G.W. Carter, O. Scavenius, I.N. Mishustin, P.J. Ellis,
Phys.Rev. C{\bf 61}, 045206 (2000). 
e-Print: nucl-th/9812014


\bibitem{Svetitsky86}
B. Svetitsky, 
Phys. Rept. {\bf 132}, 1--53 (1986).  


\bibitem{KS15}
K.-I. Kondo and A. Shibata, 
in preparation.
\\
A. Shibata, K.-I. Kondo, S. Kato, and T. Shinohara,
PoS LATTICE2014 (2015) 340. 
arXiv:1501.06271 [hep-lat]  


\bibitem{LP13}
B. Lucini and M. Panero,
Phys. Rept. {\bf 526}, 93--163 (2013). 
arXiv:1210.4997 [hep-th]  


\bibitem{Weiss81}
N. Weiss,
Phys. Rev. D{\bf 24}, 475--480 (1981).
\\
N. Weiss,
Phys.Rev. D{\bf 25}, 2667--2672  (1982). 


\bibitem{GPY81}
D.J. Gross, R.D. Pisarski, L.G. Yaffe,
Rev. Mod. Phys. {\bf 53}, 43-- (1981). 


\bibitem{RSTW15}
U. Reinosa, J. Serreau, M. Tissier, and N. Wschebor,
Phys. Lett. B{\bf 742}, 61--68 (2015). 
arXiv:1407.6469 [hep-ph] 


\bibitem{RSTW15b}
U. Reinosa, J. Serreau, M. Tissier, and N. Wschebor,
Phys. Rev. D{\bf 91}, 045035 (2015).  
arXiv:1412.5672 [hep-th] 


\bibitem{TW11}
M. Tissier and N. Wschebor,  
Phys. Rev. D\textbf{84}, 045018 (2011). 
arXiv:1105.2475 [hep-th] 
\\
M. Tissier and N. Wschebor,  
Phys. Rev. D\textbf{82}, 101701 (2010). 
arXiv:1004.1607 [hep-ph]  
\\
J. Serreau and M. Tissier,
Phys. Lett. B\textbf{712},  97--103 (2012).
arXiv:1202.3432 [hep-th]  


\bibitem{SR12}
C. Sasaki and K. Redlich,
Phys.Rev. D86 (2012) 014007 
arXiv:1204.4330 [hep-ph] 


\bibitem{FK13}
K. Fukushima and K. Kashiwa,
Phys. Lett. B{\bf 723}, 360--364 (2013).  
arXiv:1206.0685 [hep-ph] 


\bibitem{RH13}
H. Reinhardt and J. Heffner,
Phys. Rev. D{\bf 88},  045024 (2013). 
arXiv:1304.2980 [hep-th]
\\
H. Reinhardt and J. Heffner, 
Phys. Lett. B{\bf 718}, 672--677 (2012). 
arXiv:1210.1742 [hep-th]


\bibitem{Fischer09}
C.S. Fischer, J. Luecker, J.A. Mueller,
Phys. Lett. B{\bf 702}, 438--441 (2011). 
arXiv:1104.1564 [hep-ph] 
\\
C.S. Fischer, A. Maas, J.A. Muller,
Eur. Phys. J. C{\bf 68}, 165--181 (2010).  
arXiv:1003.1960 [hep-ph]
\\
C.S. Fischer and J.A. Mueller,
Phys. Rev. D{\bf 80},  074029 (2009). 
arXiv:0908.0007 [hep-ph]
\\
C.S. Fischer,
Phys. Rev. Lett.{\bf 103}, 052003 (2009). 
arXiv:0904.2700 [hep-ph]


\bibitem{KondoIV}
K.-I. Kondo,
Phys. Rev. D {\bf 58}, 105016 (1998).
[hep-th/9805153] 


\bibitem{KT00b}
K.-I. Kondo and Y. Taira,
Mod. Phys. Lett. A {\bf 15}, 367--377 (2000); 
[hep-th/9906129] 


\bibitem{KT00}
K.-I. Kondo and Y. Taira,
Prog. Theor. Phys. {\bf 104}, 1189--1265 (2000).
[hep-th/9911242] 




\bibitem{Kondo08}
K.-I. Kondo,
Phys. Rev. D {\bf 77}, 085029 (2008).
arXiv:0801.1274 [hep-th] 


\bibitem{Kondo08b}
K.-I. Kondo, 
J. Phys. G: Nucl. Part. Phys. {\bf 35}, 085001  (2008).
arXiv:0802.3829 [hep-th] 


\bibitem{KG67}
T. Kunimasa and T. Goto,
Prog. Theor. Phys. {\bf 37}, 452--464 (1967).


A. A. Slavnov, 
Teor. Mat. Fiz. {\bf 10}, 305--328 (1972), [transl. Theor. Math. Phys. {\bf 10} (1972) 201--217.]


A. A. Slavnov and L. D. Faddeev, 
Teor. Mat. Fiz. {\bf 3}, 18--23 (1970), [transl. Theor. Math.
Phys. {\bf 3}, 312--316 (1970)].


R. Delbourgo, S. Twisk, and G. Thompson,  
Int. J. Mod. Phys. A{\bf 3}, 435--449 (1988).


\bibitem{KOSSM06}
 K.-I. Kondo, A. Ono, A. Shibata, T. Shinohara and T. Murakami,
J. Phys. A: Math. Gen. {\bf 39}, 13767--13782 (2006). 
[hep-th/0604006] 


\bibitem{CNP15}
M. Caselle, A. Nada, and M. Panero,
JHEP 1507 (2015) 143 
arXiv:1505.01106 [hep-lat]
\\
M. Caselle, A. Nada, and M. Panero,
arXiv:1509.06905 [hep-lat] 


\bibitem{BEFKS12}
Sz. Borsanyi, G. Endrodi, Z. Fodor, S.D. Katz, and K.K. Szabo,
JHEP 1207 (2012) 056, 
e-Print: arXiv:1204.6184 [hep-lat]


\bibitem{Kondo11}
K.-I. Kondo,
Phys. Rev. D{\bf 84}, 061702 (2011).  
arXiv:1103.3829 [hep-th] 


\bibitem{Kondo12}
K.-I. Kondo, 
Phys. Rev. D{\bf 87}, 025008 (2013).   
arXiv:1208.3521 [hep-th]
\\
K.-I. Kondo, K. Suzuki, H. Fukamachi, S. Nishino, and T. Shinohara,
Phys. Rev. D\textbf{87}, 025017 (2013). 
arXiv:1209.3994 [hep-th]. 


\bibitem{Litim00}
D.F. Litim,
[hep-th/0103195], 
Phys. Rev. D{\bf 64}, 105007 (2001). 
D.F. Litim,
[hep-th/0005245], 
Phys. Lett. B{\bf 486},  92--99 (2000). 



\end{thebibliography}
\end{document}